\documentclass[11pt]{article}
\usepackage{jheppub} 
\usepackage[normalem]{ulem}
\usepackage[T1]{fontenc}
\usepackage{slashed,amsmath,amsfonts,amssymb,bbm}
\usepackage{mathrsfs,wasysym,array,booktabs}
\usepackage{units}
\usepackage[dvipsnames,usenames]{xcolor}
\usepackage{graphicx}
\usepackage{feynmp}
\usepackage[compat=1.0.0]{tikz-feynman}
\usepackage{bbold}
\usepackage{dashrule}

\RequirePackage[colorlinks=true,urlcolor=blue,anchorcolor=blue,citecolor=blue,filecolor=blue,linkcolor=blue,menucolor=blue,linktocpage=true,pdfproducer=medialab,pagebackref=true]{hyperref}
 \hypersetup{
 colorlinks   = true, 
 urlcolor     = PineGreen, 
 linkcolor    = blue, 
  citecolor   = blue 
}

\newcommand{\blue}[1]{\color{blue}#1\color{black}}

\usepackage{cleveref}

\renewcommand{\a}{\alpha}

\renewcommand{\d}{\delta}

\newcommand{\g}{\gamma}
\renewcommand{\to}{\rightarrow}
\newcommand{\tpdf}[2]{\texorpdfstring{#1}{#2}}

\newcommand{\Lag}{\mathscr{L}}
\newcommand{\de}{\partial}

\newcommand{\hc}{\text{h.c.}}

\renewcommand{\O}{\mathbf{O}}

\newcommand{\mf}{\text{f}}
\newcommand{\mfp}{\text{f'}}

\renewcommand{\c}{\mathbf{c}}
\newcommand{\x}{\mathbf{x}}
\DeclareMathOperator{\tr}{Tr}
\newcommand{\beq}{\begin{equation}}
\newcommand{\eeq}{\end{equation}}
\newcommand{\bea}{\begin{eqnarray}}
\newcommand{\eea}{\end{eqnarray}}
\renewcommand{\[}{\begin{equation}}
\renewcommand{\]}{\end{equation}}

\graphicspath{/..}
\subheader{\rm\blue{IFT-UAM/CSIC-21-82}}

\title{ One-loop corrections to ALP couplings}
\author[a]{ J.~Bonilla,}
\author[b]{I.~Brivio,}
\author[a]{M.B.~Gavela,} 
\author[c,d]{V.~Sanz}

\affiliation[a]{Departamento de F\'isica Te\'orica, Universidad Aut\'onoma de Madrid, and Instituto de F\'isica Te\'orica IFT-UAM/CSIC, 
Cantoblanco, E-28049, Madrid, Spain}
\affiliation[b]{Institut f\"ur Theoretische Physik, Universit\"at Heidelberg, Philosophenweg 16,\\ D-69120 Heidelberg, Germany}
\affiliation[c]{Instituto de F{\' i}sica Corpuscular (IFIC), Universidad de Valencia-CSIC, E-46980 Valencia, Spain}
\affiliation[d]{Department of Physics and Astronomy, University of Sussex, Brighton BN1 9QH, UK}
\emailAdd{jesus.bonilla@uam.es}
\emailAdd{brivio@thphys.uni-heidelberg.de}
\emailAdd{belen.gavela@uam.es}
\emailAdd{veronica.sanz@uv.es}

\abstract{ 
The plethora of increasingly precise experiments which hunt for axion-like particles (ALPs), as well as their widely different energy reach, call for the theoretical understanding of ALP couplings at loop-level.
We derive the one-loop contributions to ALP-SM effective couplings, 
including finite corrections. 
The  complete leading-order --dimension five-- effective linear Lagrangian is 
considered.  
The ALP is left off-shell, which is of particular impact on LHC and accelerator 
searches of 
ALP couplings to 
$\gamma\gamma$, $ZZ$, $Z\gamma$, $WW$, gluons and fermions. All  results are obtained 
in the covariant $R_\xi$ gauge.
A few phenomenological consequences are also explored as illustration,  with flavour diagonal channels in the case of fermions: in particular, we explore constraints on the coupling of the ALP to top quarks, that can be extracted from LHC data, from astrophysical  sources and from Dark Matter direct detection experiments  such as PandaX, LUX and XENON1T.  
Furthermore, we  clarify  the relation between alternative ALP bases, the role of gauge anomalous couplings and their interface with chirality-conserving and chirality-flip fermion interactions, and we briefly discuss renormalization group aspects.

}

\begin{document}

\maketitle

\section{Introduction}

The field of axions and axion-like particles (ALPs) is undergoing a phase of spectacular development, both theoretical and experimental. This should come as no surprise. 
No firm signal of new physics has shown up yet at colliders or elsewhere, which transforms the fine-tuning issues of the Standard Model of particle physics (SM) in most pressing ones, and also impacts on the  dark matter (DM) quest.  
The silence of data is calling for a rerouting guided by fundamental issues such as the strong CP problem, as well as  for an open-minded approach to hunt for the generic tell-tale of global hidden symmetries: derivative couplings, as in the case of axions and ALPs.

Indeed,  axions appear in dynamical solutions to the strong CP problem as the pseudo Goldstone-bosons (pGB) of a global chiral $U(1)$ symmetry~\cite{
Peccei:1977hh, Peccei:1977ur, Weinberg:1977ma,Wilczek:1977pj}. 
Theories of pGBs extend well beyond those true axions, though.  They appear in a plethora of beyond the SM (BSM) constructions,  typically  as SM scalar singlets, and often receive the  generic name of  ALPs (in particular when gauge anomalous couplings are present in addition to pure derivative ones). 
Paradigmatic examples of  pGBs  physics include:
i) theories with extra dimensions, because the Wilson line around a compact dimension behaves as a 4-dimensional axion; ii) dynamical explanations to the smallness of neutrino masses, with the Majoron~\cite{Gelmini:1980re} as a pGB of a hidden $U(1)$ lepton symmetry (the Majoron and the axion could even be identified~\cite{Langacker:1986rj,Ballesteros:2016xej}); iv) string theory models, which tend to have a plethora of hidden $U(1)$'s and axions~\cite{Cicoli:2013ana}; iv) dynamical flavour theories (``axiflavons''~\cite{Wilczek:1982rv,Calibbi:2016hwq,Ema:2016ops}), to cite just a few examples.  
 As a wonderful byproduct, axions and a variety of ALPs are often  excellent candidates to account for DM.

The landscape of experimental searches for axions and/or ALPs is undergoing a flourishing period, covering orders of magnitude in energy scale and using very different techniques. In particular, 
the couplings of ALPs  to heavy SM bosons are under increasing experimental scrutiny~\cite{Jaeckel:2015jla,Mimasu:2014nea,Bauer:2017ris,Bauer:2018uxu,Frugiuele:2018coc,Brivio:2017ije,Craig:2018kne,Gavela:2019cmq,Ebadi:2019gij}. 
Indeed, because of electroweak gauge invariance   they are generically expected at the same level as the photonic interactions. Through the ensemble of ALP bosonic couplings, ALP  scales ranging from hundreds of GeV to  several TeV  
 are within the reach of  the LHC and of future collider experiments, favored by the prospects of increasing energy and precision. 
In addition, the impact of ALP electroweak couplings on  flavour rare decays is already setting impressive constraints on the ALP parameter space~\cite{Izaguirre:2016dfi,Freytsis:2009ct} (for ALP masses below 5 GeV), offering 
a complementary  window of high-precision.

A model-independent approach to the search for a true axion or an ALP --both denoted here as $a$-- is that of effective Lagrangians, with the tower of effective operators weighted down by its BSM scale $f_a$. The parameter space is then simply defined by the mass vs. scale $\{m_a, f_a\}$ plane, with  $m_a \ll f_a$ and the model-dependence encoded in the arbitrary operator coefficients. The couplings are mainly derivative --proportional to the ALP momentum-- as befits pGBs, plus anomalous couplings to gauge field strengths. The practical difference between a canonical QCD axion~\cite{Weinberg:1977ma,Wilczek:1977pj} which solves the strong CP problem and generic ALPs is that for the latter  
$f_a$ and 
$m_a$ are  treated as independent parameters.
 The exploration of the ALP parameter space is thus free from the stringent phenomenological constraints which hold for the canonical QCD axion.\footnote{ The anomalous coupling to gluons is necessarily present for axions that solve the strong CP problem. For true axions, the precise relation between $m_a$ and $f_a$
depends on the characteristics of the strong interacting sector of the theory: QCD in the case of the canonical axion, and an enlarged confining sector for true axions  which are either heavier~\cite{Rubakov:1997vp,Berezhiani:2000gh,Gianfagna:2004je,Hsu:2004mf,Hook:2014cda,Fukuda:2015ana,Chiang:2016eav,Dimopoulos:2016lvn,Gherghetta:2016fhp,Kobakhidze:2016rwh,Agrawal:2017ksf,Agrawal:2017evu,Gaillard:2018xgk,Buen-Abad:2019uoc,Hook:2019qoh,Csaki:2019vte,Gherghetta:2020ofz} or lighter~\cite{Hook:2018jle,DiLuzio:2021pxd,DiLuzio:2021gos} than the canonical QCD axion.}  For the purpose of this work, the difference between a true axion and an ALP is of no consequence and the name ALP will be used indistinctly.

We explore at one-loop order all possible  CP-even operators coupling one pseudoscalar ALP to SM fields: to the gluon, the photon, $W^\pm$, $Z$, the Higgs particle and to fermions,
at next-to leading order (NLO) of the linear effective field theory (EFT) formulation, i.e. mass dimension five operators.
The approach is  in the same spirit as the usual SMEFT theory, but including the ALP $a$ 
as an additional low-energy active field. 
The necessity to address these interactions at loop-level stems,  on one side, from the high precision experimentally achieved in certain channels, and on the other from the very different energy scales explored by different experiments. Motivated by the latter, updated studies of the renormalization group evolution of the ALP effective Lagrangian have already appeared very recently~\cite{Chala:2020wvs,Bauer:2020jbp}.     

We provide here the complete one-loop corrections, i.e.  divergent and finite contributions,  
for an off-shell  ALP and on-shell SM fields.  Previously,  those corrections had been worked out only for the contributions  to the axion-photon-photon coupling $g_{a\gamma\gamma}$ and to the axion leptonic coupling (in certain limits), for an on-shell ALP~\cite{Bauer:2017ris}.    Recently,  fermionic contributions to $g_{a\gamma Z}$ have also appeared~\cite{Bauer:2020jbp} for an on-shell ALP.  The physical impact of our results will be presented as contributions to the set of measurable CP-even interactions $\{g_{a\gamma\gamma},\,g_{aWW},\,  g_{aZZ}\,,g_{a\gamma Z}\,,g_{agg}\, , c_{\text{f}}  \}$, 
where the first five denote ALP anomalous couplings to gauge bosons and $\text{f}$ denotes a generic fermion, with the SM fields on-shell. All our computations are performed in the covariant $R_\xi$ gauge.   
The only restriction on fermions is that flavour diagonal channels are computed, disregarding generation mixing. Neutrino masses are disregarded as well.

Furthermore, the constraints that gauge invariance imposes on the complete set of ALP couplings will be discussed, showing how the one-loop corrections modify the tree-level gauge invariance relations which relate physical channels. The results impact in particular the variety of LHC and collider ALP searches.

We will also clarify  the  one-loop impact of 
ALP-fermion couplings on gauge anomalous ALP interactions. This will allow to elucidate ongoing discussions in the literature on the relation between different types of complete and non-redundant bases of operators. Some aspects of the RG running above the electroweak scale will be briefly discussed as well. 
  
The structure of the paper can be easily inferred from the Table of Contents.

\section{Effective Lagrangian}
\label{secEffLag}
The formulation of  the  CP-even ALP effective Lagrangian at next-to-leading order (NLO) of the linear expansion, i.e. up to $\mathcal{O} (1/f_a)$ couplings of mass dimension five, is discussed next 
assuming the field $a$  to be a pseudoscalar.  A complete basis of  independent ALP operators --bosonic plus fermionic-- is considered,
 and its relation to other complete bases and to the purely bosonic one is also clarified.

In addition to ALP kinetic energy and mass terms, any 
ALP EFT is  defined by an ensemble of
effective operators which are invariant under the shift symmetry $a \rightarrow a +c$ where $c$ is a constant (i.e. purely derivative ALP couplings, as it would befit Goldstone bosons) plus 
ALP-gauge couplings resulting from the axial anomaly of the form 
$a X_{\mu \nu} {\tilde{X}}^{\mu\nu}$, where $X_{\mu \nu}$ denotes a generic SM gauge field strength and $\tilde{X}^{\mu\nu}$ its dual $\tilde{X}^{\mu\nu}\equiv \frac12 \epsilon^{\mu\nu\rho\sigma}X_{\rho\sigma}$ with $\varepsilon^{0123}=1$.\footnote{We do not 
consider other shift-invariant ALP couplings to gauge fields which have been recently argued to be independent in some BSM theories~\cite{Bonnefoy:2020gyh}.}

\subsection{Complete and non-redundant bases}\label{sec.lagrangian}
A complete  and non-redundant ALP effective Lagrangian 
is given at $\mathcal{O} (1/f_a)$  by
\begin{equation}
 \Lag_{ALP} = \Lag_{SM} + \Lag_a^{\rm total}\,,
\end{equation}
where $\Lag_{SM}$ denotes the SM Lagrangian, 
\begin{equation}\label{eq.LSM}
\begin{aligned}
\Lag_{SM} &= -\frac14 W_{\mu\nu}^\a W^{\a\mu\nu}-\frac14 B_{\mu\nu} B^{\mu\nu}-\frac14 G_{\mu\nu}^a G^{a\mu\nu}
+D_\mu\Phi^\dag D^\mu\Phi
+\sum_{\text{f}} \bar{\text{f}}\, i\slashed{D}\,  \text{f}
\\&
-\left[\bar Q_L Y_d \Phi d_R+\bar Q_L Y_u \tilde \Phi u_R + \bar L_L Y_e \Phi e_R+\hc\right]
-V(\Phi^\dag\Phi)\,.
\end{aligned}
\end{equation}
Here, the index $\text{f}$ runs over the chiral fermion fields $\text{f}=\{Q_L, u_R, d_R, L_L, e_R\}$ which  are  vectors in three-dimensional flavour space, $Y_\text{f}$ denote $n_g\times n_g$  Yukawa matrices in flavour space, where $n_g$ denotes the number of fermion generations, $\Phi$ is the Higgs doublet with $\tilde \Phi = i\sigma^2\Phi^*$,   and $V(\Phi^\dag\Phi)$ is the Higgs potential. 
In this equation, $G_{\mu\nu}$, $W_{\mu\nu}$ and $B_{\mu\nu}$  denote respectively the $SU(3)_c$,  $ SU(2)_L$ and $U(1)_Y$ gauge field strengths.
Neutrino masses are disregarded here and all through this work; no right-handed neutrino fields will be considered.  

All possible shift-invariant fermionic coupling of mass dimension five are contained in the set
\begin{align}
\label{ferm}
\O_u&\equiv \dfrac{\de_\mu a}{f_a} \, \left(\bar{u}_R \gamma^\mu u_R \right)\,,
\qquad
\O_d \equiv \dfrac{\de_\mu a}{f_a} \, \left(\bar{d}_R \gamma^\mu d_R \right)\,,
\qquad \O_Q \equiv \dfrac{\de_\mu a}{f_a} \, \left(\bar{Q}_L \gamma^\mu Q_L \right)\,,
\\
\label{lept}
\O_L &\equiv\dfrac{\de_\mu a}{f_a} \, \left(\bar{Q}_L \gamma^\mu Q_L \right)\,,
\qquad
\O_e \equiv \dfrac{\de_\mu a}{f_a} \, \left(\bar{e}_R \gamma^\mu e_R \right)\,
\end{align}
 in a compact notation in which each of these terms is a $n_g\times n_g$ matrix in flavour space, with flavour indices $\{i,j\}$  left implicit, e.g. $\O_u\equiv \{ \O_u^{ij}  =  {\de_\mu a}/{f_a} \, \big(\bar{u}_R^i \gamma_\mu  u_R^j \big)\}$. The question is how many of those fermionic couplings can be included in a a complete and non-redundant basis of ALP operators.

The most general CP-conserving ALP effective Lagrangian  $\Lag_a^\text{total}$, including {\it bosonic and fermionic} ALP couplings~\cite{Georgi:1986df, Choi:1986zw}, admits many possible choices of basis.  A {\it complete and non-redundant basis} --to be used in this paper-- is that defined  by the Lagrangian
\begin{equation}
 \Lag_a^\text{total}\,= \frac12 \de_\mu a\de^\mu a + \frac{m_a^2}{2}a^2+\,c_{\tilde{W}}\O_{\tilde{W}}+c_{\tilde{B}}\O_{\tilde{B}}+c_{\tilde{G}}\O_{\tilde{G}}+ \sum_{\text{f}=u,\,d,\,e}\,\text{\bf{c}}_{\text{f}}\, \O_{\text{f}}  + \sum_{\text{f}=Q,\,L}\,\text{\bf{c}}_{\text{f}}\, \slashed{\O}_{\text{f}} \,,
\label{general-NLOLag-lin}
\end{equation}
where  the effective operators are as given in Table~\ref{tab.basis}, and the coefficients $\c_{\text{f}}$ are $n_g\times n_g$ hermitian tensors; 
 in addition, because of the assumption of CP conservation, they will obey $\c_{\text{f }}= \c_{\text{f }}^T$.   The  convention to be used for the $\c \, \O$ products is the popular one in which their implicit flavour indices $\{i,j\}$ are {\it not} contracted  as a matrix product, but as follows:
\begin{equation}
\c \,\O\equiv \sum_{i,j} {(\c)}_{ij} \, {\O}^{ij}\,.
\label{convention}
\end{equation}
 Note that the fermionic basis is chosen here to include all possible right-handed currents, while --in order to avoid redundancies-- one of the quark operators made out of left-handed currents  has been excluded (see $\slashed{\O}_Q$) together with all diagonal elements of the leptonic operators made out of left-handed currents ($\slashed{\O}_L$), as indicated in short-hand notation, i.e. 
\begin{align}
 \slashed{\O}_Q \equiv \{{\O}_Q^{ij}  &=  \dfrac{\de_\mu a}{f_a} \, \big(\bar{Q}_L^i \gamma_\mu  Q_L^j \big) \quad \text{where}\,\, i,j\ne 1,1\}\\
 \slashed{\O}_L \equiv \{{\O}_L^{ij}  &=  \dfrac{\de_\mu a} {f_a} \, \big(\bar{L}_L^i \gamma_\mu  L_L^j \big) \quad \text{where}\,\, i\ne j \}\,.
\end{align}
The exclusion of the $(\slashed{\O}_Q)_{11}$ element can be replaced by that of any other of the diagonal elements of $\slashed{\O}_Q$.

It follows that the most general CP-conserving ALP Lagrangian is described by a total of 
\begin{equation}
\label{dof}
3 (\text{bosonic}) + [ n_g(5n_g+3)/2 -1 ] (\text{fermionic}) = 2+n_g(5n_g+3)/2 
\end{equation}
independent couplings, i.e. 6 couplings in the case of just one generation, and 29 couplings for $n_g=3$. 
\begin{table}[t]\centering
\renewcommand{\arraystretch}{2}
\begin{tabular}{|*3{>{$}c<{$}@{  }>{$}l<{$}}|}
\toprule
\O_{\tilde W} & = -\dfrac{a}{f_a}W_{\mu\nu}^\a\tilde W^{\a\mu\nu}
&
\O_{\tilde B} & = -\dfrac{a}{f_a}B_{\mu\nu}  \tilde B^{\mu\nu}
&
\O_{\tilde G} & = -\dfrac{a}{f_a}G_{\mu\nu}^a\tilde G^{a\mu\nu}
\\
\O_u & = \dfrac{\de_\mu a}{f_a} \, \left(\bar{u}_R\gamma^\mu u_R\right) 
&
\O_d & = \dfrac{\de_\mu a}{f_a} \, \left(\bar{d}_R\gamma^\mu d_R \right)
&
\O_e & = \dfrac{\de_\mu a}{f_a} \, \left(\bar{e}_R\gamma^\mu e_R\right)
\\
\slashed{\O}_Q & = \dfrac{\de_\mu a}{f_a} \, \left(\bar{Q}_L\gamma^\mu Q_L\right)_{i,j \ne 1,1}
&
\slashed{\O}_L & = \dfrac{\de_\mu a}{f_a} \, \left(\bar{L}_L\gamma^\mu L_L\right)_{i \ne j}
& &
\\
\bottomrule
\end{tabular}
\caption{ A complete and non-redundant basis of bosonic+fermionic operators, in the presence of quark mixing.
 Each fermionic structure is a $n_g\times n_g$  matrix in flavour space, with flavour indices $\{i,j\}$ left implicit except in the operators  on the last row (which become redundant-- for $n_g=1$).   
  For the anomalous terms,   a ``hatted'' renaming will be used when convenient, $\hat\O_{\tilde X} \equiv \a_X/4\pi \,\O_{\tilde X}$, see text.}
\label{tab.basis}
\end{table}

The key point to identify redundancies, and the origin of the different  number of degrees of freedom for quarks and leptons, is related to baryon  and lepton number conservation. Classically, with neutrino masses disregarded (only the SM left-handed neutrino fields are considered),   lepton number $L_i$  is  
separately conserved  for each generation $i$ (i.e. $L_e$, $L_{\mu}$ and $L_{\tau}$  for $n_g=3$), 
while for quarks with all generations mixed only the total baryon number $B$ is. In consequence, $n_g$ leptonic diagonal couplings become redundant, in contrast to just one for quarks.  Indeed, the ALP coupling to the baryonic and leptonic currents reads (see App.~\ref{relations-flip-non-flip}) 
\begin{align}
\label{eq.B}
 \frac{\de_\mu a}{f_a}\, J^\mu_{B}&= \tr \left[\frac{\O_{Q} + \O_u + \O_d}{3} \right] = \frac{n_g}{32\pi^2}\left(g^{2}\O_{\tilde W }-g^{\prime 2} \O_{\tilde B}\right) \,,
 \\
 \label{eq.Li}
 \frac{\de_\mu a}{f_a}\, J^\mu_{L_i}&= \left[\O_{L }+ \O_{e}\right]^{ii} = \frac{1}{32\pi^2}\left(g^{2}\O_{\tilde W }-g^{\prime 2} \O_{\tilde B}\right)\,,
\end{align}
 where in the last equation there is no sum over the $i$ index, and the right-hand side of these equations stems from the fermion rotations involved.  These relations provide one constraint on diagonal quark operators and $n_g$ constraints on diagonal leptonic operators, which reduce in consequence the number of independent degrees of freedom.  

Eqs.~(\ref{eq.B}) and (\ref{eq.Li}) also illustrate that the ALP coupling to  the $B+L$ current $J^\mu_{B+L}$ is anomalous, where $L$ denotes total lepton number $L=\sum_i L_i$,   which is precisely why that coupling  can be traded by purely derivative operators.\footnote{This is analogous to how the  Peccei-Quinn current, precisely because it is anomalous, allows to rotate away the $\bar\theta$ terms which combine fermion mass and anomalous gauge terms.} The $B-L$ current  $J^\mu_{B-L}$ is instead exactly conserved,
\begin{align}
\label{eq.B+L-1}
 \frac{\de_\mu a}{f_a}\, J^\mu_{B+L}= \tr \left[\frac{\O_{Q} + \O_u + \O_d}{3} + \O_L + \O_e\right] &= \frac{n_g}{16\pi^2}\left(g^{2}\O_{\tilde W }-g^{\prime 2} \O_{\tilde B}\right) \,,
 \\
 \label{eq.B-L}
 \frac{\de_\mu a}{f_a}\, J^\mu_{B-L}=\tr\left[\frac{\O_{Q} + \O_u + \O_d}{3} - \O_L - \O_e\right] &= 0\,.
\end{align}

The role of the left-handed and right-handed ALP operators in Table~\ref{tab.basis}.  can be exchanged. 
 For completeness, we discuss in the next subsection  other fair choices of shift-invariant fermionic operators --e.g. containing all possible left-handed currents.

\paragraph {A frequent redefinition.}  Often in the literature~\cite{Bauer:2017ris,Irastorza:2018dyq,Bauer:2018uxu,Bauer:2020jbp,MartinCamalich:2020dfe,Galda:2021hbr} the normalization used for the ALP coupling to gauge anomalous currents differs slightly from that in Tab.~\ref{tab.basis}. 
We will denote with a hat (``hat basis'') that variant:
\begin{align}
\label{Ohats}
 \hat\O_{\tilde B} &\equiv \frac{\a_1}{4\pi} \O_{\tilde B}\,,
 &
 \hat\O_{\tilde W} &\equiv \frac{\a_2}{4\pi} \O_{\tilde W}\,,
 &
 \hat\O_{\tilde G} &\equiv \frac{\a_s}{4\pi} \O_{\tilde G}\,,
\end{align}
where $\a_1=g^{\prime 2}/4\pi$, $\a_2= g^2/4\pi$ and $\a_s= g_s^2/4\pi$ denote respectively the $SU(3)_c$,  $ SU(2)_L$ and $U(1)$ fine structure constants.  
The corresponding Wilson coefficients of the ALP anomalous  gauge couplings  are simply related by
\begin{align}
 c_{\tilde B} &=  \hat{c}_{\tilde B}\,\frac{\alpha_1}{4\pi}\,,  
 &
 c_{\tilde W} &=  \hat{c}_{\tilde W}\,\frac{\alpha_2}{4\pi}\,, 
 &
 c_{\tilde G} &=  \hat{c}_{\tilde G}\,\frac{\alpha_s}{4\pi}\,.  
\label{chats}
\end{align}

\subsection{ Alternative complete basis} \label{alternative-basis}
Many  choices of complete basis other than that in Eq.~(\ref{general-NLOLag-lin}) and Table~\ref{tab.basis} are possible, as far as the total number of independent couplings is consistently maintained.  Several examples have been proposed in the literature.

\paragraph {Chirality-conserving fermionic alternatives.} 
A valid option is to include in the basis {\it all} possible  operators made out of left-handed fields, including all diagonal couplings, i.e.   all $n_g\times (n_g+1)/2$ operators 
  $\O_Q$ and all  $n_g\times (n_g+1)/2$ operators $\O_L$, see  Eqs.~(\ref{ferm}) and (\ref{lept}).  
 With respect to the choice in Table~\ref{tab.basis}, and still maintaining in the basis the three anomalous couplings, this would require --to avoid redundancies-- to drop all flavour diagonal leptonic  operators in $\O_{\text{e}}$ (i.e. replace  ${\O}_{e}\to \slashed{\O}_{e}\equiv{\de_\mu a}/{f_a} \, \left(\bar{e}_R\gamma^\mu e_R\right)_{i \ne j}$, plus one of the flavour-diagonal ones in $\O_{\text{f}={u}}$ or $\O_{\text{f}={d}}$. Several other intermediate exchange patterns are legitimate as far as the number of degrees of freedom is consistently maintained. 
 
 It is also valid to omit from the basis some of the anomalous bosonic operators, substituting them for flavour-diagonal fermionic couplings.
 Indeed, Eqs.~(\ref{eq.B}) and~(\ref{eq.Li}) show that a complete and non-redundant basis would result for instance from substituting  $\slashed{\O}_Q$ in Table~\ref{tab.basis}  by the whole set $\O_Q$ together with the omission of either  $\O_{\tilde W }$ or $\O_{\tilde B}$, or other similar tradings involving the lepton sector.

\vspace{0.3cm}

\paragraph{The case $\mathbf{n_g=1}$.} In the simplified case of one generation,  the operators $\slashed{\O}_{\{Q,L\}}$ in the basis in Table~\ref{tab.basis} are absent and pure right-handed operators suffice in addition to the three anomalous ones. That is, for just one generation the set of operators $\{\O_{\tilde W},\O_{\tilde B},\O_{\tilde G}, \O_u, \O_d\, \O_e\}$ in Table~\ref{tab.basis} suffices to form a complete basis of linearly independent operators,
 unlike for $n_g>1$. 
Indeed, in the one-generation case the following relations hold
\begin{align}
\label{eq.Q_to_ud_anomalous}
\O_Q
&=
- \left[
 \O_u
+\O_d 
\right]  + \frac{3}{32\pi^2}\left(g^{2}\O_{\tilde W }-g^{\prime 2} \O_{\tilde B}\right)\,,
\\
\label{eq.L_to_e_anomalous}
\O_L
&=
- \O_e 
+ \frac{1}{32\pi^2}\left(g^{2}\O_{\tilde W }-g^{\prime 2} \O_{\tilde B}\right)\,,
\end{align}
 which demonstrate that it would be redundant to consider any element of $\O_Q$ and $\O_L$ in addition to all possible operators made out of right-handed currents plus the three anomalous couplings.  

 \paragraph{No flavour mixing, $\text{CKM}\mathbf{=\mathbb{1}}$.} When $n_g> 1$ but CKM flavour mixing is disregarded, the quark sector mirrors what is described above for the lepton sector. There will be then $n_g$ quark baryon charges independently conserved, each of them obeying separately an equation alike to Eq.~(\ref{eq.Li}), instead of only the combined one  Eq.~(\ref{eq.B}). In consequence,  $n_g$ constraints follow on the diagonal elements of the quark sector, and all  
   diagonal elements of $\O_Q$ become redundant (assuming that the complete set of right-handed quark currents is retained in the basis together with the anomalous operators).  
  In other words, when  CKM mixing is disregarded, a complete and non-redundant basis is given by that in Table~\ref{tab.basis} albeit with the redefinition
  \begin{equation}
  \label{slashedOQ-noCKM}
  \slashed{\O}_Q \equiv \{{\O}_Q^{ij}  =  \dfrac{\de_\mu a} {f_a} \, \big(\bar{Q}_L^i \gamma_\mu  Q_L^j \big) \quad \text{where}\,\, i\ne j \}
  \end{equation}
We will use this simplified framework in the one-loop computations  in Sec.~\ref{complete-one-loop-main}.

\paragraph{On the use of chirality-flip fermionic operators.}  
Chirality-flip fermion currents are sometimes used to describe the ALP Lagrangian,  together with the three anomalous gauge couplings. That is, some or all of the chirality-conserving  fermionic structures in  Table~\ref{tab.basis} are traded by chirality-flip ones, i.e. 
\begin{align}
\O_{u\Phi}  \equiv i\, \frac{a}{f_a} \bar{Q}_L\, \widetilde{\Phi}\, u_{{R}}\,, \quad
\O_{d\Phi} \equiv i\, \frac{a}{f_a} \bar{Q}_L\, \Phi\, d_{{R}}\,, \quad
\O_{e\Phi}  \equiv i\, \frac{a}{f_a} \bar{L}_L\, \Phi\, e_{{R}}\,.
\label{ferm-Yuk-1}
\end{align}
Although this is possible if done with care, it could be misleading. The point is that, in all generality, the operators in Eq.~(\ref{ferm-Yuk-1}) do not belong to the ALP Lagrangian in the sense that they are not invariant {\it per se} under the required shift symmetry $a\rightarrow a +c$ 
(which in the ALP paradigm is assumed to be broken only by  gauge anomalous currents).

Only in some particular cases the chirality-flip couplings are tradable for generic chirality-preserving ones (plus redefinitions of the $c_{\tilde{X}}$ anomalous coefficients). For instance, this is the  case for just one fermion generation or when the EFT respects Minimal Flavour Violation (MFV)\footnote{This requires  the coefficients of the chirality-flip operators to be proportional to the corresponding Yukawa matrices.}. 
Otherwise, it suffices to note here that  the number of degrees of freedom of  a hermitian coefficient matrix (as for chirality-preserving operators)  differs in general from that of a general $n_g\times n_g$ matrix (as for chirality-flip ones).  In the CP-even case,  any complete and non-redundant basis made out of purely shift-invariant fermionic operators spans $n_g(5n_g+3)/2-1$ degrees of freedom --see Eq.~(\ref{dof}), which differs from the $3n_g^2$ independent parameters of the chirality-flip set $\{\O_{u\Phi}, \O_{d\Phi}, \O_{e\Phi}\}$  in Eq.~(\ref{ferm-Yuk-1}). 
The precise combinations of chirality-flip structures which are equivalent to shift-invariant ALP couplings (plus anomalous gauge couplings) are identified in App.~\ref{relations-flip-non-flip}, see also Ref.~\cite{Chala:2020wvs}.

\paragraph{Trading anomalous operators by fermionic ones.} Anomalous gauge couplings are intrinsically non shift-invariant.
Chirality-flip structures will thus necessarily appear  if  anomalous operators were to be traded by purely fermionic ones.
 It is shown in App.~\ref{app.removing_anomalous} how  each of the operators $\O_{\tilde B}$, $\O_{\tilde W}$ and $\O_{\tilde G}$ can be  traded by a combination of purely fermionic structures which necessarily includes chirality-flip terms.  
 
Let us consider here as illustration the situation when only one anomalous gauge coupling is removed from the complete Lagrangian.  Eq.~(\ref{eq.B+L-1}) showed that $\O_{\tilde W}$ can be removed  
without  introducing any chirality-flip operator if $\O_{\tilde B}$ is maintained, and viceversa, as a consequence of $B+L$ being an anomalous global symmetry of the SM.  For instance, in our basis in  Table~\ref{tab.basis} it would suffice to replace either $\O_{\tilde W}$ or $\O_{\tilde B}$ by a trace of chirality-conserving fermionic structures defined in Eqs.~(\ref{ferm}) and~(\ref{lept}). 
In contrast,  
the  combination of $\O_{\tilde B}$ and $\O_{\tilde W}$ with opposite sign to  that in Eq.~(\ref{eq.B+L-1})  does not correspond to an anomalous current, and thus requires chirality-flip structures when traded by fermionic currents, namely 
\begin{align}
\label{eq.B-L-1}
\frac{n_g}{16\pi^2}\left(g^{\prime 2} \O_{\tilde B} + g^2 \O_{\tilde W}\right) &= 2
 \tr \O_{L}  - 2 (Y_e\O_{e\Phi} + \hc)
 \\ 
 &= \frac23\tr \left(\O_{Q} - 2\O_u + 4\O_d \right)  + 2 \left( Y_d\O_{d\Phi} - Y_u\O_{u\Phi}+ \hc \right)\,.
\end{align}

Analogously, and as expected from  the non-perturbative nature of  $a\,G_{\mu\nu}\tilde{G^{\mu\nu}}$ and the fact that this term may induce a potential for the ALP field,\footnote{In fact, it is well-known that  $\O_{\tilde G}$ generates a scalar potential for the QCD axion~\cite{Weinberg:1977ma,Wilczek:1977pj}.}  
it is not possible to remove $\O_{\tilde G}$ altogether in favour of another anomalous coupling plus purely chirality-conserving (and thus shift-invariant) terms.   For instance, some  alternative equivalences of interest are 
\begin{align}
\O_{\tilde G} &= -
 \frac{32\pi^2}{n_g g_s^2} \left[ \tr \O_d+ \left( Y_d\O_{d\Phi} + \hc\right)
 \right] - \frac{2}{3}\frac{g^{\prime 2}}{g_s^2}\O_{\tilde B}\,,
 \\
\O_{\tilde G} &= -
 \frac{32\pi^2}{n_g g_s^2}\left[ \tr \O_u+ \left(Y_u\O_{u\Phi} + \hc\right)
 \right] - \frac{8}{3}\frac{g^{\prime 2}}{g_s^2}\O_{\tilde B}\,,
 \\
 \O_{\tilde G} &=
 \frac{32 \pi^2}{3 n_g g_s^2} \left[ \tr \left( \O_u - 4 \O_d \right) + ( Y_u\O_{u\Phi} - 4Y_d\O_{d\Phi} + \hc)\right]\,.
\end{align}

An interesting question in the chirality-flip vs. chirality conserving arena is that of the one-loop ($\mathcal{O} (\alpha_X)$) impact of fermionic operators on anomalous ALP-couplings. The results allow to understand which combinations of 
chirality-flip operators discussed are exactly equivalent to purely derivative fermionic ones. That this happens at all could seem paradoxical from the quantum loop perspective, as chirality-flip operators  will exclusively induce at one-loop corrections proportional to fermion masses squared, while derivative chirality-conserving operators contribute in addition a finite and mass independent term, which is the contribution from the chiral anomaly of the fermionic currents. Nevertheless, the relations above among both type of fermionic structures --see also App.~\ref{relations-flip-non-flip}-- are precisely such that the matching holds at any order. An illustrative example of the one-loop matching of chirality flip and chirality conserving contributions can be found at the end of Sec.~\ref{app.Oaphi}.

\subsection{Purely bosonic basis}
\label{sec-pure-bosonic}
\begin{table}[t]\centering
\renewcommand{\arraystretch}{2}
\begin{tabular}{|*3{>{$}c<{$}@{ = }>{$}l<{$}}|}
\toprule
\O_{\tilde W} & -\dfrac{a}{f_a}W_{\mu\nu}^\a\tilde W^{\a\mu\nu}
&
\O_{\tilde B} & -\dfrac{a}{f_a}B_{\mu\nu}  \tilde B^{\mu\nu}
&
\O_{\tilde G} & -\dfrac{a}{f_a}G_{\mu\nu}^a\tilde G^{a\mu\nu}
\\
\O_{a\Phi} & \dfrac{\de^\mu a}{f_a} \left(\Phi^\dag i\overleftrightarrow{D_\mu}\Phi\right)
&\multicolumn{4}{l|}{}
\\\bottomrule
\end{tabular}
\caption{Purely bosonic operator basis.}
\label{tab.basis_bosonic}
\end{table}
The addition of an ALP  to the SM interactions is an enlargement of the scalar sector of the low-energy theory. In some contexts, it may 
be pertinent to focus exclusively on the bosonic Lagrangian.

The most general and complete  {\it purely bosonic} effective ALP Lagrangian describing CP-even  couplings at NLO is extraordinarily simple. It contains just four linearly independent effective operators~\cite{Georgi:1986df,Choi:1986zw,Salvio:2013iaa,Brivio:2017ije}:
\begin{equation}\label{deltaLbosonic-lin}
\Lag_{a}^{\text{bosonic}} =
c_{\tilde W} \O_{\tilde W}
+c_{\tilde B} \O_{\tilde B}
+c_{\tilde G} \O_{\tilde G}
+c_{a\Phi} \O_{a\Phi}\,,
\end{equation}
where $c_{a\Phi}$ is a real constant and 
\begin{equation}
\O_{a\Phi} \equiv \frac{\de^\mu a}{f_a}(\Phi^\dag i\overleftrightarrow{D_\mu}\Phi)\,,
\label{eq.bos_operators_phi}
\end{equation}
being $\Phi^\dag i\overleftrightarrow{D}_\mu\Phi = i\Phi^\dag (D_\mu\Phi)-i(D_\mu\Phi^\dag)\Phi$.  The purely bosonic basis  is summarized in Table~\ref{tab.basis_bosonic}.  
The operator $\O_{a\Phi}$ is equivalent to a precise linear combination of the fermionic operators in 
Table~\ref{tab.basis}: 
\begin{equation}
 \O_{a\Phi}=  \tr \left(  \O_e + \O_d -  \O_u\right)\,,
 \label{Axion-fermion-from-bosonic-1}
\end{equation}
and  it would have thus been redundant to add it to the set in Table~\ref{tab.basis}.  The direct impact of $\O_{a\Phi}$ is to induce a kinetic mixing between $a$ and the would-be Goldstone boson eaten by the $Z$ boson. This mixing is cumbersome to work with, and it can be removed via a Higgs field redefinition of the form $\Phi \to \Phi \, e^{i c_{a\Phi} a/f_a}$~\cite{Georgi:1986df,Gavela:2014vra,Brivio:2017ije}, which is equivalent to the application of the Higgs EOM. This delivers chirality-flip operators that can next be turned via the fermionic EOM into the chirality-conserving combination in Eq.~\eqref{Axion-fermion-from-bosonic-1}.  
Note that no trace of anomalous gauge couplings remains in the final expression Eq.~(\ref{Axion-fermion-from-bosonic-1}) in spite of the  fermion rotations involved, as expected for a purely bosonic ALP interaction.   A comprehensive discussion 
of how the anomalous terms that {\it a priori} could be induced by fermion rotations cancel each other for this operator can be found in Appendix~\ref{app.Oaphi}. 

Finally,  note that $\O_{a\Phi}$ could be kept as one of the operators of a complete and non-redundant basis at the expense of some other coupling. Eq.~(\ref{Axion-fermion-from-bosonic-1}) shows that it could be included  at the price of omitting any of the diagonal operators of the right-handed set $\{ \O_e,\, \O_d,\,  \O_u\}$. Another possibility --among many-- is for $\O_{a\Phi}$ to replace certain flavour-diagonal fermionic couplings of the left-handed set $\{\O_Q,\O_L\}$,  as indicated by the identity (see Appendix~\ref{app.Oaphi})
\begin{align}
 \O_{a\Phi} &=
 - \tr \left( \O_L +  \O_Q + 2 \O_u \right) + \frac{1}{8\pi^2}\left(g^{2}\O_{\tilde W} - g^{\prime 2} \O_{\tilde B}\right) n_g\,.
\end{align}
This equation also suggests yet another alternative: to include $\O_{a\Phi}$ in the complete and non-redundant basis at the expense of omitting either $\O_{\tilde W}$ or $\O_{\tilde B}$. The exact expression of the degrees of freedom which may be replaced by $\O_{a\Phi}$ is to be analyzed for each possible basis.

\subsection{Phenomenological parameters}\label{sec.pheno_parameters}

The ALP EFT  presented above in terms of $SU(3)_c\times SU(2)_L \times U(1)_Y$ gauge invariant operators  leads to multiple experimental signals. 
The ultimate goal is to detect or constraint from data the set of fundamental independent variables
\begin{equation}
\{c_{\tilde{W}}\,,c_{\tilde{B}}\,,c_{\tilde{G}}\,,\text{\bf{c}}_{\text{f}}\}\,,
\end{equation}
which are to be treated as free Lagrangian parameters.

The three anomalous gauge couplings,  
$\O_{\tilde G}$, 
$\O_{\tilde W}$ and $\O_{\tilde B}$,  induce five distinct physical interactions with gluons, photons, $W$ and $Z$ bosons, which are customarily codified  as  
\begin{equation} \label{coupboson2}
    \Lag_a \supset - \frac{1}{4}g_{agg}\,a\,G_{\mu\nu} \widetilde{G}^{\mu\nu}  -\frac{1}{4} g_{a \gamma \gamma} a F_{\mu \nu} \tilde{F}^{\mu \nu} - \frac{1}{4} g_{a \gamma Z} a F_{\mu \nu} \tilde{Z}^{\mu \nu} - \frac{1}{4} g_{a Z Z} a Z_{\mu \nu} \tilde{Z}^{\mu \nu} - \frac{1}{2} g_{a W W} a W^+_{\mu \nu} \tilde{W}^{- \mu \nu} \,,
\end{equation}
where 
\begin{align}
\label{gagluon}
g_{agg} & \equiv\frac{4}{ f_{a}}\,c_{\tilde{G}} \,, \quad
g_{a\gamma\gamma} \equiv \frac{4}{f_{a}} \, \big(s_w^2 \, c_{\tilde{W}}+ c_w^2 \, c_{\tilde{B}} \big)\,,\\
g_{aWW} &\equiv \frac{4}{ f_a}  \, c_{\tilde{W}} \,,\quad
g_{aZZ} \equiv \frac{4}{ f_a} \,({c_w^2}\, c_{\tilde{W}}+ {s_w^2}\, c_{\tilde{B}}) \label{gaZZ}\,,
\\ 
g_{a\gamma Z} &\equiv \frac{8}{ f_a}  \,s_w c_w ( c_{\tilde{W}}-  c_{\tilde{B}} )\,,
\label{gXX0}
\end{align}
where  $s_w$ and $c_w$ denote respectively the sine and cosine of the Weinberg mixing angle, given at tree-level by
\begin{equation}\label{cosWeinberg}
c_w \equiv  \frac{M_W}{M_Z}\,.
\end{equation}
It follows that the two independent electroweak anomalous couplings may source 
four independent measurable quantities,
\begin{equation}
\{c_{\tilde{W}},  c_{\tilde{B}}\}\,\longrightarrow\,\{g_{a\gamma\gamma}, g_{aWW}, g_{aZZ} , g_{a\gamma Z}\}\,,
\label{only2indep}
\end{equation}
a fact that allows to overconstrain the electroweak gauge sector of the parameter space.  In other words, electroweak gauge invariance imposes at tree-level the constraints
\begin{equation}
 \label{relationcouplingc2}
    \begin{aligned}
      &  g_{aWW}= g_{a\gamma\gamma} + \frac{c_w}{2 s_w} g_{a \gamma Z} \,, \\
& g_{a ZZ} = g_{a \gamma \gamma} + \frac{c_w^2 - s_w^2}{2 c_w s_w} g_{a \gamma Z} \,.
    \end{aligned}
\end{equation}
From the experimental point of view these two expressions are quite useful, since they can be used to set constraints on one coupling  based on the constraints on other couplings, barring fine-tuned cancellations. For example, $g_{a \gamma \gamma}$ is strongly constrained from multiple experiments, while $g_{a ZZ}$ is harder to measure directly. Nevertheless, applying Eq.~(\ref{relationcouplingc2}) one can translate the constraints on $g_{a \gamma \gamma}$ into constraints on $g_{a Z Z}$ that are stronger than those extracted from direct searches of the latter. This approach has already led  to cross-relations among different measurements, resulting in a noticeable reduction of parameter space allowed by present data~\cite{Bauer:2017ris,Alonso-Alvarez:2018irt}.
It is thus relevant  
from the phenomenological point of view to determine how the relations in Eq.~(\ref{relationcouplingc2}) are modified when one-loop corrections are taken into account. We will address this task in Sec.~\ref{sec-gauge-inv-one-loop}.

It is also convenient for later use to consider the following combination of the couplings in Eq.~(\ref{gaZZ}) and (\ref{gXX0}), which corresponds to the $aB_{\mu\nu}{\tilde{B}}_{\mu\nu}$ coupling:
\beq 
g_{aBB}\equiv\frac{4}{ f_{a}}\,c_{\tilde{B}}= g_{aWW} -\frac{1}{2s_w c_w}g_{a \gamma Z}\,.
\label{gaBB}
\eeq
Furthermore, in the cases in which the hatted basis of gauge invariant operators  in Eq.~(\ref{chats}) is preferred as description,  the corresponding phenomenological parameters ${\hat {g}}_{iXX}$ follow trivially form the substitution $\{c_{i} \longrightarrow  {\hat{c}}_{i}\,,g_{iXX} \longrightarrow  {\hat{g}}_{iXX}\}$
in Eqs.~(\ref{gagluon})-(\ref{gXX0}), i.e. 
 \begin{equation}
\hat{g}_{aGG}  \equiv\frac{4}{ f_{a}}\,\hat{c}_{\tilde{G}}\,, \qquad
\hat{g}_{aWW}  \equiv\frac{4}{ f_{a}}\,\hat{c}_{\tilde{W}}\,, \qquad
\hat{g}_{aBB} 
 \equiv\frac{4}{ f_{a}}\,\hat{c}_{\tilde{B}}\,, 
\label{gXX0-II}
\end{equation}
with  the relation between $c_i$ and ${\hat{c}}_{i}$ as discussed in Eq.~(\ref{chats}).

In all cases, the data on fermion EFT couplings  can be directly expressed in terms of the EFT Lagrangian parameter matrix $\text{\bf{c}}_{\text{f}}$ corresponding to the complete basis in  Table~\ref{tab.basis}.  For practical purposes, a simplified notation can be useful when considering flavour-diagonal transitions. The latter are proportional only to the axial part of the fermionic derivative couplings, i.e. the coupling has Lorentz structure R$-$L.  
For instance, a general --basis independent-- definition of phenomenological flavour-diagonal couplings can be written as 
\begin{align}
c_u & \equiv (\c_u -\c_Q)^{11}\,,\quad  
&
c_c & \equiv (\c_u -\c_Q)^{22}\,, \quad 
&
c_t & \equiv (\c_u -\c_Q)^{33}\,, \label{updiagonal}
\\
c_d & \equiv (\c_d -U^\dagger \c_Q U)^{11} \,, \quad 
&
c_s & \equiv (\c_d -U^\dagger \c_Q U)^{22}\,, \quad 
& 
c_b & \equiv (\c_d -U^\dagger \c_Q U)^{33} \,, \label{downdiagonal}
\\
c_e & \equiv (\c_e -\c_L)^{11}\,,\quad 
&
c_\mu & \equiv (\c_e -\c_L)^{22}\,, \quad 
& 
c_\tau & \equiv (\c_e -\c_L)^{33}\,, \label{leptdiagonal}
\end{align}
where $U=U_{\text{CKM}}$ is the CKM mixing matrix. This notation simplifies further in the particular complete basis in Table~\ref{tab.basis} in which {\it de facto} $(\c_Q)^{11}=0$ and $(\c_L)^{i=j}=0$ , e.g. 
\begin{equation}
c_u= (\c_u)^{11}\,,\quad c_e= (\c_e)^{11}\,,\quad c_\mu= (\c_e)^{22}\,,\quad c_\tau= (\c_e)^{33}\,.
\end{equation}

\section{Non-renormalization theorems}
The renormalization group (RG) properties of the ALP effective coupling have received considerable attention lately.

\subsection*{Above the electroweak scale} 
\begin{figure}[h]
\centering
\includegraphics{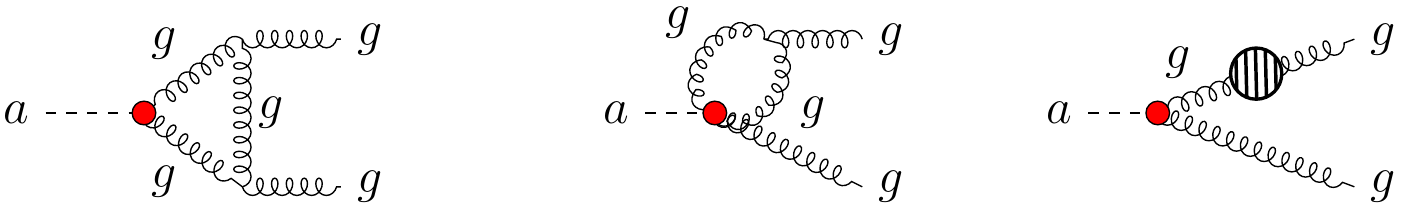}
\caption{One-loop diagrams which renormalize the effective $a\, G {\tilde G}$ interaction. The blob in the last diagram stands for one-loop gluon and quark contributions (a similar contribution holds for the other external gauge leg).}
\label{betaGG}
\end{figure}

\begin{figure}[h]
\centering
\includegraphics{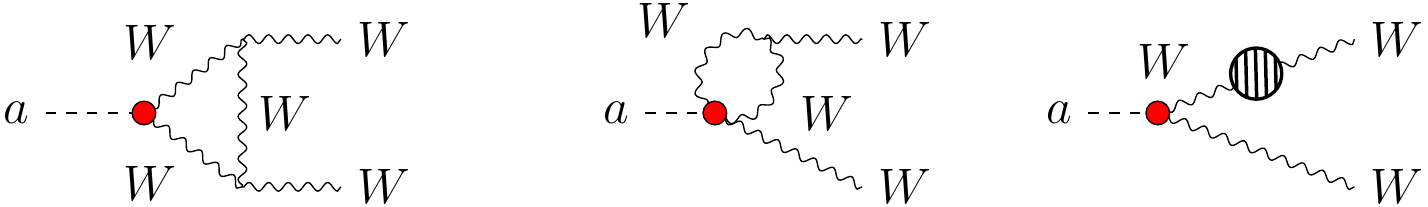}
\caption{One-loop diagrams which renormalize the effective $a\, W {\tilde W}$ interaction. The blob in the last diagram stands for one-loop $W$ and $SU(2)_L$ charged fermion contributions (a similar contribution holds for the other external gauge leg).}
\label{betaWW}
\end{figure}

\begin{figure}[h]
\centering
\includegraphics{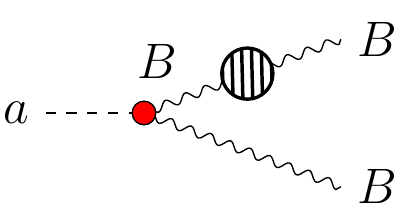}
\caption{One-loop diagrams which renormalize the effective $a\, B {\tilde B}$ interaction. The blob in the last diagram stands for one-loop fermion contributions (a similar contribution holds for the other external gauge leg).}
\label{betaBB}
\end{figure}

CP-odd anomalous gauge couplings within the SM, 
i.e. Lagrangian terms of 
the generic form  ${\alpha_X}X_{\mu\nu}  \tilde{X}^{\mu\nu}$ where $X_{\mu\nu}$ denotes a generic gauge field strength and $\alpha_X$ its fine structure coupling, are not multiplicatively renormalized at any order in perturbation theory. The reason is their topological character, which  ensures anomaly matching conditions~\cite{tHooft:1979rat}. 
Indeed the combinations  
${\alpha_1}/{2\pi}B{\tilde B} \, $,  ${\alpha_2}/{2\pi} \,W {\tilde W} $ and  ${\alpha_s}/{2\pi} \,G {\tilde G} $ 
appear in the Lagrangian multiplied by ``$\theta$'' angles which are periodic variables with periodicity $2\pi$, and cannot thus be multiplicatively renormalized~\cite{Kaplan:1988ku,Grojean:2013kd}.  This can be inferred from the fact that  a chiral rotation 
induces a contribution to the divergence of the axial current $J_\mu$ precisely of the form
\begin{equation}
\partial_\mu J_\mu \supset \frac{\alpha_X}{2\pi} X_{\mu\nu}  \tilde{X}^{\mu\nu}\,.
\end{equation}
Now, when considering ALP-SM anomalous couplings, the ratio $a/f_a$ plays the role of an effective angle. 
The non-renormalization theorems thus apply as well to  ALP couplings of the form
${\alpha_X}/({2\pi}f)\,a \, X_{\mu\nu}  \tilde{X}^{\mu\nu}$, where $2\pi f$ is the periodicity of $a$~\cite{Agrawal:2019lkr}.   In consequence, 
 no UV divergent terms can result from corrections
 to the combinations 
${\alpha_1}/{2\pi} \,\O_{\tilde W} $,  ${\alpha_2}/{2\pi} \,\O_{\tilde B} $ and ${\alpha_s}/{2\pi} \,\O_{\tilde G} $. In other words, in the {\it hat basis} of effective ALP operators --see Eqs.~(\ref{Ohats}) and (\ref{chats})--  
the $\beta$ functions for  the electroweak anomalous couplings
must vanish, 
 \begin{equation}
 \beta_{{\hat{c}}_{\tilde B}} = \frac{\text{d}}{\text{d} \log \mu} {\hat{c}}_{\tilde B} =0\,,\qquad \beta_{{\hat{c}}_{\tilde W}} = \frac{\text{d}}{\text{d} \log \mu} {\hat{c}}_{\tilde W} =0\,,\qquad \beta_{{\hat{c}}_{\tilde G}} = \frac{\text{d}}{\text{d} \log \mu} {\hat{c}}_{\tilde G} =0\,.
\end{equation}
It is easy to check these results at one-loop, from the contributions of the Feynman diagrams in  Figs.~\ref{betaGG}, \ref{betaWW} and \ref{betaBB}. 
 Correspondingly, the RG evolution of the $\{c_{\tilde{G}},c_{\tilde{W}},  c_{\tilde{B}}\}$ coefficients for the  basis in Table~\ref{tab.basis} reflects that of the $\alpha_i$ couplings,  see Eq.~(\ref{chats}),  
  \begin{align}
  \beta_{{c}_{\tilde B}} &= \frac{\text{d}}{\text{d} \log \mu} {c}_{\tilde B} =  \beta_{\alpha_1}= \left( \frac{1}{12} + \frac{10}{9} n_g \right) \frac{\a_{1} }{\pi} \, c_{\tilde B} = \frac{41}{12} \frac{\a_{1} }{\pi} \, c_{\tilde B} \,,\label{betacB}
  \\
  \beta_{{c}_{\tilde W}} &= \frac{\text{d}}{\text{d} \log \mu} {c}_{\tilde W} =\beta_{\alpha_2}= - \left( \frac{43}{12} - \frac{2}{3} n_g \right) \frac{\a_{2} }{\pi } \, c_{\tilde W} = - \frac{19}{12} \frac{\a_{2} }{\pi} \, c_{\tilde W} \,,\label{betacW}
  \\
  \beta_{{c}_{\tilde G}} &= \frac{\text{d}}{\text{d} \log \mu} {c}_{\tilde G} = \beta_{\alpha_s}=- \left( \frac{11}{2} - \frac{2}{3} n_g \right) \frac{\a_{s} }{\pi} \, c_{\tilde G} = - \frac{7}{2} \frac{\a_{s} }{\pi} \, c_{\tilde G} \,,\label{betacG}
  \end{align}
where $n_g$ is the number of generations of fermions, and $n_g=3$ has been taken on the last equalities of these equations. This results had been 
previously derived in Ref.~\cite{Chala:2020wvs}.
  
The beta functions for the ALP-fermion couplings have been previously obtained as well, using a variety of fermionic bases, and we refer the reader to the corresponding literature~\cite{Chala:2020wvs,MartinCamalich:2020dfe,Bauer:2020jbp}. The beta function for the bosonic operator $\O_{a\Phi}$ can be found  In Ref.~\cite{MartinCamalich:2020dfe},  in a  redundant basis which contemplates all possible operators.

\section{Complete one-loop contributions to ALP couplings}
\label{complete-one-loop-main}

We present here the one-loop contributions to the phenomenological ALP couplings,  including all finite corrections.  The ALP field will be left off-shell (which is of practical interest for collider and other searches away from the ALP resonance, besides adapting trivially to ALP on-shell searches), while the external SM fields will be considered on-shell. For channels with external fermions, we only provide corrections to the flavour diagonal ones. Furthermore, CKM mixing is disregarded in the loop corrections to all couplings, which means the framework depicted in Sec.~\ref{alternative-basis} for CKM$=\mathbb{1}$. That is, the complete and non-redundant basis corresponds to that in Table~\ref{tab.basis} with the proviso in Eq.~(\ref{slashedOQ-noCKM}).

The operator basis used is that defined in Eq.~(\ref{general-NLOLag-lin}) and  Table~\ref{tab.basis}.   We will trade the set of two linearly independent electroweak anomalous couplings $\{c_{\tilde{W}}, c_{\tilde{B}}\}$ for the 
set of four phenomenological couplings $\{g_{a\gamma\gamma}, g_{aWW}, g_{aZZ} , g_{a\gamma Z}\}$  in Eqs.~(\ref{coupboson2})-(\ref{gXX0}), which are in consequence linked by gauge invariance (as shown at tree-level  in Eq.~(\ref{relationcouplingc2})).  
The latter means that the final one-loop results for a given 
effective electroweak coupling $g^{\text{eff}}_{aXX}$ can be expressed in terms of just two tree-level phenomenological couplings of choice, e.g. in terms of the set $\{g_{aXX}, g_{aWW}\}$.  These can be easily transcribed back  in terms of the set $\{c_{\tilde{W}}, c_{\tilde{B}}\}$  if wished,
using  Eqs.~(\ref{gagluon})-(\ref{gXX0}) and (\ref{gaBB}).

All computations have been carried out in the covariant $R_{\xi}$-gauge, with the help of  {\tt Mathematica} packages {\tt FeynCalc} and {\tt Package-X}~\cite{Shtabovenko:2020gxv,Patel:2016fam}.  
The individual one-loop diagrams are in general $\xi$-dependent. 
The same applies to each of the one-loop corrected amplitudes in the ensemble $ \{g^{\text{eff}}_{a\gamma\gamma}, g^{\text{eff}}_{aWW}, g^{\text{eff}}_{aZZ} , g^{\text{eff}}_{a\gamma Z}\}$
resulting from directly inserting all possible  tree-level phenomenological couplings $\{g_{a\gamma\gamma}, g_{aWW}, g_{aZZ} , g_{a\gamma Z}\}$. Their $\xi$-independence (with external SM fields on-shell) becomes explicit only when the gauge invariance relations in  Eq.~(\ref{relationcouplingc2}) are applied to the electroweak radiative results, so as to reduce the parameter space.
Details of  $\xi$-dependent intermediate steps are provided in \href{https://notebookarchive.org/2021-07-9otlr9o}{NotebookArchive}.
  
\paragraph{Renormalization and measurable parameters.} We will use as renormalization framework of the electroweak sector the scheme 
in which  its four linearly independent parameters  (other than fermion Yukawa couplings), i.e. the $SU(2)_L$ and $U(1)_Y$ coupling constants ($g$ and $g'$ respectively), the Higgs vev $v$ and Higgs self-coupling denoted here $\tilde{\lambda}$, are to be traded by precisely measured input parameters as follows 
\begin{equation}
\{g, g', v, \tilde{\lambda}\} \longrightarrow  \{\alpha_{em}, M_Z, M_W, M_H\}|_{exp}
\end{equation}
where the experimental value of 
$\alpha_{em}$ is extracted from Thompson scattering (e.g. $Q^2=0$) and the values of $M_W$,
 $M_Z$ and $M_H$ are determined from their resonant peaks.\footnote{ $ \alpha_{em}=1/137.035999139(31) \, \mbox{  at } Q^2=0$,     
 $M_Z=91.1876(21) \mbox{ GeV}$, $M_W=80.379(12)$  and   $M_H=125.25(17) \mbox{ GeV}$ . }
The ALP effective operators do not contribute to these observables at one-loop and $\mathcal{O}(1/f_a)$. In consequence, the relation between the Lagrangian parameters and those four observables is not modified with respect to the SM case. In other words, at tree-level it holds that 
\begin{equation} \label{relationsZscheme}
    \begin{aligned}
   & \alpha_{em} = \frac{e^2}{4 \pi} = \frac{g^2 g'^2}{4 \pi (g^2 + g'^2)} = \frac{\a_1 \a_2}{\a_1 + \a_2} \,,  \qquad &    M_W = \frac{1}{2}\,gv  \,, \\
 &   M_Z = \frac{1}{2} \sqrt{g^2 + g'^2} v \,,  \qquad &   M_H^2 = \tilde{\lambda} v^2 \,,
    \end{aligned}
\end{equation}
a set of relations that can be easily inverted.
 All other SM observable quantities to be predicted can be expressed in terms of those four input observables plus fermion masses.
While the fermion masses of leptons have a direct physical meaning which allows simple renormalization procedures, in QCD due to confinement such a natural scale does not exist.  Alike considerations apply to the QCD coupling strength $\alpha_s$. The renormalization scale and scheme must be chosen with other criteria, based on simplicity and convergence.  There are many alternative ways proposed to deal with the infrarred behaviour of the QCD coupling constant, that is, on how to extract from observables the strength of $\alpha_s$ at a variety of scales, see for instance Ref.~\cite{Prosperi:2006hx} and Sec.~\ref{sec-gagg}.

\paragraph{One-loop corrections.} Let us briefly rename with a bar the one-loop renormalized parameters  whose values are to be identified with the experimentally inputs mentioned above, i.e. $\{\overline{\alpha}_{em},  \overline{M}_Z, \overline{M}_W, \overline{M}_H\}$. Their relation with the (unbarred) tree-level quantities can be written as 
\begin{equation} \label{relationsZscheme}
\begin{aligned}
 & {\overline{\alpha}_{em} }= \alpha_{em} +\delta \alpha_{em}\, \qquad
    \,,  \qquad &    \overline{M}^2_Z= M^2_Z +\delta M_Z^2  \,, \\
 &    \overline{M}^2_W= M_W^2 +\delta M_W^2  \,,  \qquad &    \overline{M}^2_H= M_H^2 +\delta M_H^2  \,.
\end{aligned}
\end{equation}
While the symbol $\delta$ is used here for the corrections involved in the definition of the input parameters, we will use the symbol $\Delta$ for the physical predictions, that is, for the measurable deviations with respect to the SM, that follow for any other observable.
Of particular practical interest is the Weinberg angle,  defined at tree-level in Eq.~(\ref{cosWeinberg}). Let us define a ratio $\bar{c}_w$ as
\begin{equation}\label{cwbar}
\bar{c}_w \,\equiv\, \frac{\overline{M}_W}{\overline{M}_Z}= c_w\,\left(1+   \frac{\Delta c_w}{c_w}\right)\,,
\end{equation}
where
\begin{equation}\label{Deltacwbar}
 \frac{\Delta c_w}{c_w} = - \frac{1}{2} \left( \frac{\delta M_Z^2}{M_Z^2} - \frac{\delta M_W^2}{M_W^2} \right)\,,
\end{equation}
 and  $\delta M_{V=Z,W}^2$  are computed in terms of the $Z$ and $W$ transverse self-energies as $\delta M_V^2 = \Sigma_V (q^2 = M_V^2)$, see   whose exact expressions 
can be found in App.~\ref{sec-cos-renorm}.  
 The  tree-level variables $\{ g_{a\gamma\gamma}, g_{a\gamma Z}, g_{aZZ}\}$ can now be written as a combination of the set $\{c_{\tilde{B}}, c_{\tilde{W}}  \}$ and physical boson masses,  
\begin{equation}\label{gagamma_cWcB}
 g_{a \g \g} = \frac{4}{f_a} ( c_w^2 c_{\tilde B} + s_w^2 c_{\tilde W}) = \frac{4}{f_a} ( \bar{c}_w^2 c_{\tilde B} + \bar{s}_w^2 c_{\tilde W}) + \frac{8}{f_a} \bar{c}_w^2 ( c_{\tilde W} - c_{\tilde B} ) \frac{\Delta c_w}{c_w} \,,
\end{equation}
\begin{equation} \label{gagammaZ_cWcB}
 g_{a \g Z} = \frac{8}{f_a} c_w s_w(  c_{\tilde W} - c_{\tilde B} ) = \frac{8}{f_a} \bar{c}_w \bar{s}_w(  c_{\tilde W} - c_{\tilde B} ) \left( 1 + \frac{\bar{c}_w^2 - \bar{s}_w^2}{\bar{s}_w^2} \frac{\Delta c_w}{c_w} \right) \,,
\end{equation}
\begin{equation}\label{gaZZ_cWcB}
 g_{a Z Z} = \frac{4}{f_a} ( s_w^2 c_{\tilde B} + c_w^2 c_{\tilde W}) = \frac{4}{f_a} ( \bar{s}_w^2 c_{\tilde B} + \bar{c}_w^2 c_{\tilde W}) - \frac{8}{f_a} \bar{c}_w^2 ( c_{\tilde W} - c_{\tilde B} ) \frac{\Delta c_w}{c_w} \,.
\end{equation}

\vspace{0.5cm}
We will denote below by $\{ g^{\text{eff}}_{agg}, g^{\text{eff}}_{a\gamma\gamma}, g^{\text{eff}}_{a\gamma Z}, g^{\text{eff}}_{aZZ}, g^{\text{eff}}_{aWW}, c^{\text{eff}}_{\text{f}} \} $ the physical amplitudes computed at one loop, which are to be compared with data. They will be expressed in terms of the tree-level variables $\{ g_{agg}, g_{a\gamma\gamma}, g_{a\gamma Z}, g_{aZZ},g_{aWW}, c_{\text{f}} \} $ and SM quantities.  The $\Delta c_w$ corrections shown above are to be taken into account whenever a fit  to the fundamental electroweak ALP variables $\{c_{\tilde{B}}, c_{\tilde{W}}  \}$ is attempted from data, i.e. the equalities to the right in Eqs.~(\ref{gagamma_cWcB})-(\ref{gaZZ_cWcB}) must be used in the transcription.  Aside from taking into account this proviso, the bars will be omitted from now on in all expressions.

\subsection{ ALP anomalous coupling to photons}
\label{sec-photon-couplings}

\begin{figure}[h]
\centering
\includegraphics{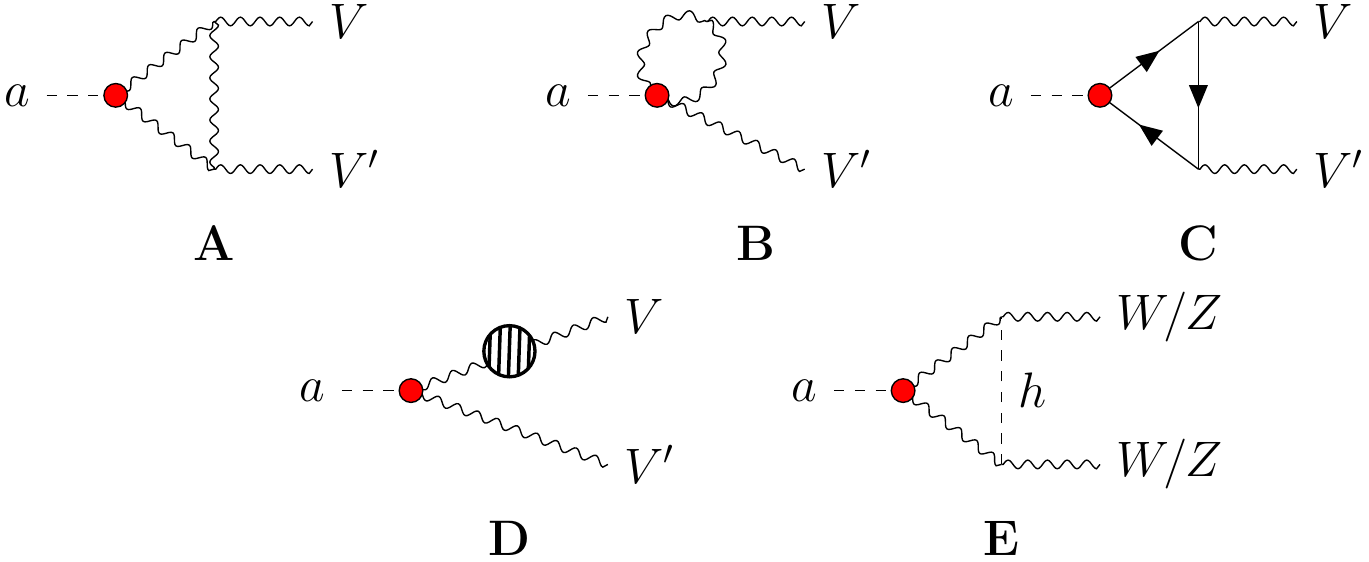}
\caption{One-loop diagrams contributing to $g_{a gg}$, $g_{a \gamma\gamma}$, $g_{a \gamma Z}$, $g_{a ZZ}$ and $g_{aWW}$  at one-loop (the corresponding diagrams with Goldstone bosons and the diagrams exchanging the gauge boson legs are left implicit), where $V$ and $V'$ are either a  gluon,  a photon,
a $Z$ boson or a $W$ boson. The last diagram only corrects insertions of the $g_{a ZZ}$ and $g_{aWW}$ couplings.}
\label{correctionsgammaZ}
\end{figure}

\begin{figure}[h]
\centering
\includegraphics{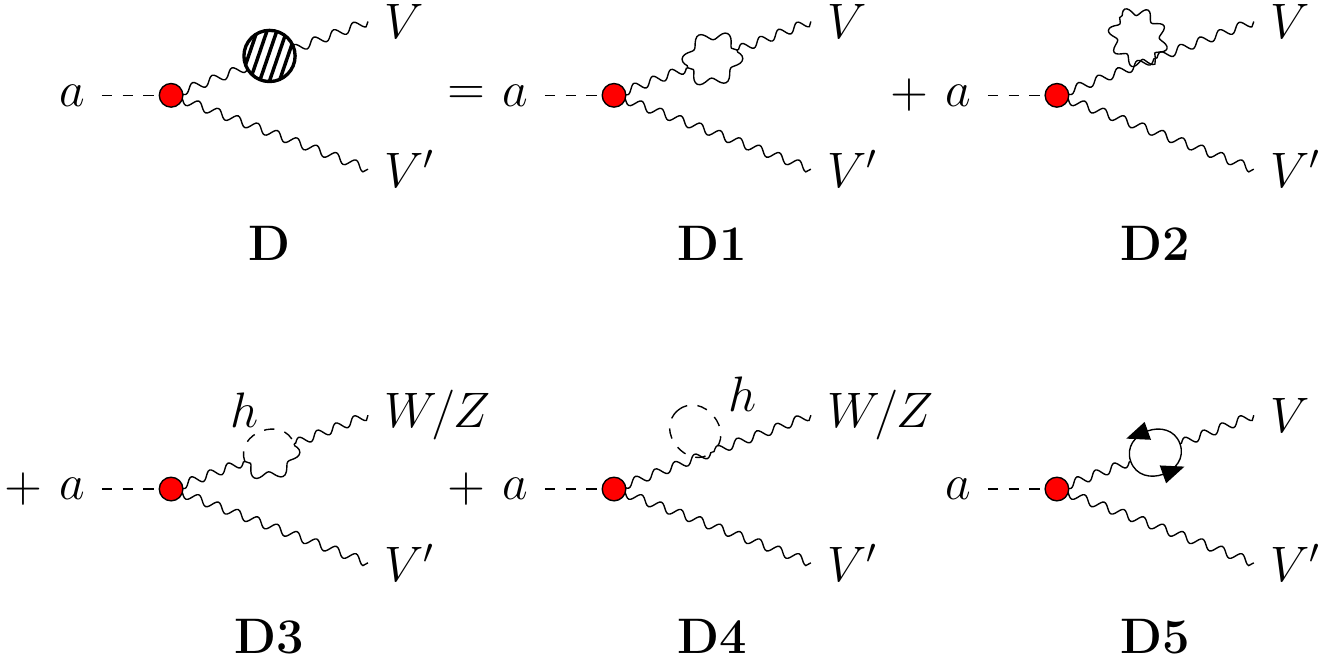}
\caption{One-loop diagrams contributing to the correction to the external gauge boson legs. Diagrams with Goldstone bosons and Higgs tadpole diagrams are included. Notice that diagrams \textbf{D3} and \textbf{D4} are only present for a $Z$ or $W$ boson external legs.}
\label{correctionexternallegs}
\end{figure}

The Feynman diagrams which induce one loop corrections to the effective anomalous ALP coupling to photons, $g_{a\gamma\gamma}$, are depicted in 
Fig.~\ref{correctionsgammaZ} {\textbf A}, {\textbf B}, {\textbf C} and {\textbf D}, with  $V=V'=\gamma$ (which implies that $W$ is the gauge boson running in the closed gauge loops, while the virtual gauge boson coupled to $a$ in diagram \textbf{D}  is either a photon or a $Z$ boson).  Among the four effective electroweak couplings, insertions of the set $\{g_{a\gamma\gamma}, g_{aWW}, g_{a\gamma Z}\}$ contribute to the one-loop corrected effective coupling $g^{\text{eff}}_{a\gamma\gamma}$. Using the  gauge-invariance relations Eq.~(\ref{relationcouplingc2}), we choose to express the final result   in terms of just two of them, e.g.  the set $\{g_{a\gamma\gamma}, g_{aWW}\}$, plus fermionic couplings:  
\begin{equation} 
\begin{aligned}\label{gagammaeff}
    g_{a\gamma \gamma}^{\text{eff}} = \, & g_{a\gamma\gamma}  \left\{ 1 + \frac{\alpha_{em}}{6 \pi}\, A^{Z/\gamma\rightarrow \gamma}
      \right\} 
     + \frac{2 \alpha_{em}}{\pi} g_{aWW}  B_2 \left( \frac{4 M_W^2}{p^2} \right) - \frac{\alpha_{em}}{\pi f_a}\sum_{\text{f}} c_\text{f} \, Q_\text{f}^2 N_C B_1 \left( \frac{4 m_\text{f}^2}{p^2} \right) \,,
\end{aligned}
\end{equation}
where here and all through the rest of the paper (unless stated otherwise) the sum over fermions denotes  all possible individual fermion flavours, $\text{f}=u,c,t,d,s,b,e,\mu,\tau$, 
 and $p$ denotes the 4-momentum of the ALP, $N_C$ is the number of colours for a given fermion $\text{f}$ (i.e. 3 for quarks and 1 for leptons), $Q_\text{f}$ is its electric charge. The functions $B_1$ and $B_2$ have already been defined in Ref.~\cite{Bauer:2017ris}  as:    
\begin{equation}
    B_1 (\tau) = 1 - \tau f^2 (\tau) \,, \qquad B_2(\tau) = 1 - (\tau - 1) f^2 (\tau) \,, \qquad \mbox{with } f (\tau) =  \begin{cases} \arcsin \frac{1}{\sqrt{\tau}} & \mbox{for } \tau \ge 1 \\ \frac{\pi}{2} + \frac{i}{2} \ln{\frac{1 + \sqrt{1 - \tau}}{1- \sqrt{1 - \tau}}} & \mbox{for } \tau < 1 \end{cases}\,.
    \label{ftau}
\end{equation}
 The function $A^{Z/\gamma\rightarrow \gamma}$ encodes pure  leg radiative corrections stemming from diagramas \textbf{D1}, \textbf{D2} and \textbf{D5} in Fig.~\ref{correctionexternallegs} 
 (with the virtual gauge boson  attached to $a$ being  either a photon or a $Z$ boson, while the $W$ boson runs in the closed gauge loops), 
\begin{equation}
\begin{aligned}
\label{gammagammaren}
A^{Z/\gamma\rightarrow \gamma}=&
    1- 2 \sum_\text{f}  Q_\mf^2 N_C \log \left(  \frac{\Lambda^2}{m_\text{f}^2} \right) + \frac{21 }{2 } \log \left(  \frac{\Lambda^2}{M_W^2} \right)  \,.
\end{aligned}
\end{equation}
This computation 
has been carried out in dimensional regularization, trading next the  $1/\epsilon$  UV-divergent terms for an  energy cutoff $\Lambda$ via the $\overline{\text{MS}}$ prescription $ 1/{\epsilon} - \gamma_E + \log \left( 4 \pi \mu^2 \right) \to \log \Lambda^2$.  This leg correction  
correspond to the SM one-loop redefinition of $\alpha_{em}$.  
Indeed, were the hatted basis of gauge operators to be used -- in which $\alpha_1$ and $\alpha_2 $ enter explicitly in the operators definition (see Eqs.~(\ref{chats}) and (\ref{gXX0-II})), the one-loop corrections   would read
\begin{equation} 
\begin{aligned} \label{gagammaeffhat}
    \hat{g}_{a\gamma \gamma}^{\text{eff}} = \,  \hat{g}_{a\gamma\gamma} 
      + \frac{2 \alpha_{em}}{\pi s_w^2 } \hat{g}_{aWW}  B_2 \left( \frac{4 M_W^2}{p^2} \right) - \frac{4}{f_a}\sum_{\text{f}}c_\text{f}  \,Q_\text{f}^2 N_C B_1 \left( \frac{4 m_\text{f}^2}{p^2} \right) \,,
\end{aligned}
\end{equation}
 a result which in the on-shell ALP limit reduces straightforwardly to that in Ref.~\cite{Bauer:2017ris}.  
 
Eq.~(\ref{gagammaeff}) 
could be rewritten if wished in terms of  $\{c_{\tilde{B}}, c_{\tilde{W}}  \}$ (and $c_\text{f}$) applying Eq.~(\ref{gaZZ}) for  $g_{aWW}$ and the last equality in Eq.~(\ref{gagamma_cWcB}) for   $g_{a\gamma \gamma}$ (and analogously for Eq.~(\ref{gagammaeffhat})) .
 
For an on-shell ALP ($p^2=m_a^2$) the one-loop corrected decay width is simply given by
\begin{equation}
 \Gamma (a \to \g \g) = \frac{m_a^3 |g_{a\g\g}^{\text{eff}}|^2}{64 \pi} \,,
\end{equation}
or the equivalent expression in the hat basis with the replacement $g_{a\g\g}^{\text{eff}}\rightarrow \alpha_{em}/4\pi \, \hat{g}_{a\g\g}^{\text{eff}}$.
We show next some limits of the exact results  above for the functions $B_1$ and $B_2$, for an off-shell ALP, which are of interest in particular experimental contexts.

\subsubsection{\tpdf{$g^{\text{eff}}_{a \gamma\gamma}$}{gaAA} for high, intermediate and low ALP $p^2$  }

\begin{itemize}

\item For $p^2 \to \infty$ ($p^2 \gg (m_\mf^2 , M_Z^2 , M_W^2)$),  only the anomaly contribution remains from fermion coupling insertions. These contributions and those from  $g_{aWW}$ insertions reduce to, respectively, 
\begin{flalign}
 B_1 = 1 \,,\qquad  B_2 = - \frac{1}{4} \left( \log \left( \frac{M_W^2}{p^2} \right) + i \pi \right)^2 \,. 
\end{flalign}

\item For intermediate values of $p^2$ ($m_\mf^2 \ll p^2 \ll (M_Z^2, M_W^2) \ll m_t^2$), i.e. smaller than the top and all gauge bosons masses but larger than all other fermion masses, it results
\begin{flalign}
 B_1 = \begin{cases}
          1 \,, & \text{for light fermion insertions: } m_\mf^2 \ll p^2 \ll M_Z^2 \,, \\
          0 \,, & \text{for top quark insertion: } p^2 \ll M_Z^2 \ll m_t^2 \,,
         \end{cases} \qquad\qquad B_2=0\,. 
\end{flalign}
\item For $p^2 \to 0$,  i.e. smaller than all fermion masses, both functions vanish $ B_1 =  B_2=0$.
\end{itemize}

\subsection{ALP anomalous coupling to gluons} 
\label{sec-gagg}
The Feynman diagrams which induce one loop corrections to the effective anomalous coupling of an ALP to two gluons, $g_{agg}$, are depicted by diagrams {\textbf A}, {\textbf B}, {\textbf C} and {\textbf D} of Fig.~\ref{correctionsgammaZ}    
with  $V=V'=g$ (which implies that all virtual gauge bosons are also gluons).    Only the 
ALP-quark couplings $c_\text{f}$, and $g_{agg}$ itself, can contribute at one-loop to the $g^{\text{eff}}_{gaa}$ amplitude,
\begin{equation} \label{gagluon}
\begin{aligned}
    g_{a g g}^{\text{eff}} = \, & g_{a g g}  \left\{ 1 + \frac{\alpha_{s}}{12 \pi}\, G^{g g}
      \right\} - \frac{\alpha_{s}}{2 \pi f_a}\sum_{\substack{\text{f} = u,  c,  t, \\ d,  s, b}}c_\text{f} \, B_1 \left( \frac{4 m_\text{f}^2}{p^2} \right) \,,
\end{aligned}
\end{equation}
where $B_1$ was defined in Eq.~\eqref{ftau},  and the function $G^{g g}$  encodes the corrections 
stemming from the vertex diagram   \textbf{A} in  Fig.~\ref{correctionsgammaZ} plus those from external leg corrections in diagrams \textbf{D1}, \textbf{D2} and \textbf{D5} of Fig.~\ref{correctionexternallegs}. 
 
We have performed the computation  of $g_{a g g}^{\text{eff}}$ in the $R_\xi$ gauge and using dimensional regularization. The latter 
respects gauge invariance and regulates both ultraviolet (UV) and infrared  (IR) divergences when present,  portraying both as poles in $1/\epsilon$ and thus mixing them. It is possible to separate UV and IR divergences, though, via the implementation as a previous step of any IR regularization procedure~\cite{HeinrichLectures} --e.g. setting the external gluons off-shell or using an effective gluon ``mass''~\footnote{It is meant here to simply replace the gluon propagator by a massive one. 
This is not a gauge invariant procedure and it thus leaves finite terms which are $\xi$-dependent and in consequence physically meaningless, but it allows to identify properly the UV divergences (and with this information restart the whole procedure using only dimensional regularization). It is of course possible to give a mass to the gluon in a gauge invariant way by ``Higgsing'' QCD: this would add the contribution of the would-be gluonic Goldstone bosons, and we checked that all the $\xi$-dependence  would cancel then. Nevertheless, this Higgsed theory does not recuperate QCD in the massless  gluon limit: for instance, the beta function is modified by the contribution of the extra scalar degrees of freedom present. 
}-- so as to identify first the UV divergences, and  then using this information on the complete pure dimensional regularization result. We obtain,
\begin{align}\label{Ggg-dimensionalreg}
 G^{gg} = \, & - 2 \sum_{\substack{\text{f} = u, c,  t, \\ d,  s,  b}} \left( \frac{1}{\epsilon_{\text{UV}}} - \gamma_E + \log \left( \frac{4 \pi \mu^2_{\text{UV}}}{m_\mf^2} \right) \right)  + 33 \left( \frac{1}{\epsilon_{\text{UV}}} - \gamma_E + \log \left( \frac{4 \pi \mu^2_{\text{UV}}}{p^2} \right)  \right) \nonumber \\
 &- 9 \left( \frac{1}{\epsilon_{\text{IR}}} - \gamma_E + \log \left( - \frac{4 \pi \mu^2_{\text{IR}}}{p^2} \right)  \right)^2\,
  - 33 \left( \frac{1}{\epsilon_{\text{IR}}} - \gamma_E + \log \left( \frac{4 \pi \mu^2_{\text{IR}}}{p^2} \right)  \right) + 36 + \frac{3 \pi^2}{2} \,,
\end{align}
where $\epsilon_{\text{UV}}$ ($\epsilon_{\text{IR}}$) and $\mu^2_{\text{UV}}$  ($\mu^2_{\text{IR}}$) account respectively for the UV (IR) divergence and renormalization scale. 
This result can be rewritten in terms of UV and IR cutoffs via the $\overline{\text{MS}}$ prescription 
\begin{align}\label{prescription}
 \frac{1}{\epsilon_{\text{UV}}} - \gamma_E + \log \left( 4 \pi \mu^2_{\text{UV}} \right) \to \log \Lambda^2 \,, \\
 \frac{1}{\epsilon_{\text{IR}}} - \gamma_E + \log \left( 4 \pi \mu^2_{\text{IR}} \right) \to \log \lambda^2 \,,
\end{align}
where $\Lambda$  and $\lambda$ denote respectively the UV and IR energy cut-offs, leading to
\begin{equation}\label{Ggg}
 G^{gg} = 33 \log \left( \frac{\Lambda^2}{p^2} \right) - 2 \sum_{\substack{\text{f} = u,  c,  t, \\ d,  s,  b}} \log \left( \frac{\Lambda^2}{m_\mf^2} \right) - 33 \log \left( \frac{\lambda^2}{p^2} \right) - 9 \left( \log \left( \frac{\lambda^2}{p^2} \right)  + i \pi \right)^2 + 36 + \frac{3 \pi^2}{2} \,.
\end{equation}
When computing the probability for a given physical processes, the unphysical dependence on IR divergences will cancel with that stemming from soft and/or collinear gluon bremsstrahlung.  In turn, the UV-divergent terms in this equation lead to the beta function for $c_{\tilde G}$ in Eq.~(\ref{betacG}).

\subsection{ALP anomalous coupling to \tpdf{$Z$}{Z} plus photon} 
\label{gagammaZ-approx}

The effective $g_{a\gamma Z}$ coupling  receives one-loop corrections  from the fermion-ALP couplings $c_\text{f}$ and from the complete set of electroweak couplings $\{g_{a\gamma\gamma}, g_{aWW},  g_{aZZ}, g_{a\gamma Z}\}$. The relevant Feynman diagrams are those in   Figs. \ref{correctionsgammaZ} and \ref{correctionexternallegs} (except diagram \textbf{E}), with the external vector bosons being either photon or $Z$,  and with $V\ne V'$. 
In consequence, the gauge boson running in closed gauge loops can only be the $W$ boson, while  the virtual boson attached to $a$ in diagrams  \textbf{D1}, \textbf{D2} and \textbf{D5}  is either $Z$ or $\gamma$, and   $V'=\gamma$ in diagrams \textbf{D3} and \textbf{D4}.

The results are shown to become $\xi$-independent --as they must-- only when the gauge-electroweak parameter space is reduced to three couplings, using  Eq.~(\ref{relationcouplingc2}). Applying the latter again, the electroweak set can be further reduced to two anomalous electroweak operators, that we choose to be the set $\{g_{a\gamma Z}, g_{aWW} \}$. The total result
can be summarized as  
\begin{equation} \label{gagamaZ_eff}
\begin{aligned}
    g_{a\gamma Z}^{\text{eff}} =   &\, g_{a \gamma Z} \,  \left\{  1+ \frac{\alpha_{em}}{12 \pi}\, \Big(A^{Z/\gamma\rightarrow \gamma}\, +  \frac{1}{ c_w^2 s_w^2} \,A^{Z/\gamma\rightarrow Z}\Big)
 \right\} + \frac{\alpha_{em}}{\pi} \frac{c_w}{s_w} g_{a WW} \, A^{WW}\,   
   + \frac{\alpha_{em}}{\pi c_w s_w }\sum_{\text{f}} \frac{c_{\text{f}}}{f_a} A^{\text{f}} \,,
\end{aligned}
\end{equation}
where the exact expressions for all the functions in this equation can be found in App.~\ref{app-agammaZ-coupling-complete},  for an off-shell ALP and on-shell external SM particles. They are defined as follows: 
\begin{itemize}
\item $A^{Z/\gamma\rightarrow \gamma}$  gathers the external leg corrections with a photon as final particle, (Fig.~\ref{correctionexternallegs} \textbf{D1}-\textbf{D5}   with $V=\gamma$ and $V'=Z$). Its expression was given in Eq.~(\ref{gammagammaren}).
\item $A^{Z/\gamma\rightarrow Z}$ encodes the external leg corrections with  $Z$ as final particle 
(Fig.~\ref{correctionexternallegs} \textbf{D1}-\textbf{D5}   with $V=Z$ and $V'=\gamma$). It can be expanded as
\begin{equation}
\label{AZgammatoZ_total}
A^{Z/\gamma\rightarrow Z}=  A_{\text{ferm}}^{Z/\gamma\rightarrow Z} \, + \, A_{\text{Higgs}}^{Z\rightarrow Z}\,+\,  A_{\text{gauge}}^{Z/\gamma\rightarrow Z}  \,, 
\end{equation}
 where 
 \begin{itemize}
\item $A_{\text{ferm}}^{Z/\gamma\rightarrow Z}$ accounts for the SM fermion loop corrections,   Fig.~\ref{correctionexternallegs} \textbf{D5}, see Eq.~(\ref{renZgammatoZ_ferm}).  

\item $A_{\text{Higgs}}^{Z\rightarrow Z}$ encodes Higgs corrections to external legs 
in  Fig.~\ref{correctionexternallegs} \textbf{D3} and \textbf{D4}, see Eq.~(\ref{renZtoZ_Higgs}). 
\item  $A_{\text{gauge}}^{Z/\gamma\rightarrow Z}$ gathers the gauge boson corrections to external legs 
in  Fig.~\ref{correctionexternallegs} \textbf{D1} and \textbf{D2}
(with $W$ bosons running in the loop),  plus the $g_{a \gamma Z}$ component of the corrections stemming from $g_{a\gamma\gamma}$ and $g_{aZZ}$ insertions in  Fig.~\ref{correctionexternallegs} \textbf{D1}-\textbf{D5}, projected on  the parameter space $\{g_{a \gamma Z}, g_{aWW}\}$, see Eq.~(\ref{renZgammatoZ_gauge}).
\end{itemize}
\item  $A^{WW}$ contains the contributions 
from direct vertex insertions of $g_{aWW}$ in 
diagrams  \textbf{A} and \textbf{B}
of Fig.~\ref{correctionsgammaZ},
plus the $g_{a WW}$ component of the corrections stemming from $g_{a\gamma\gamma}$ and $g_{aZZ}$ insertions in Fig.~\ref{correctionexternallegs} \textbf{D1}-\textbf{D5}   projected on  the parameter space $\{g_{a \gamma Z}, g_{aWW}\}$, see Eq.~(\ref{AWW}). 
\item $A^{\text{f}}$ encodes the fermion triangle correction from diagram \textbf{C} in Fig.~\ref{correctionsgammaZ}, see Eq.~(\ref{Af}).

\end{itemize}
  Eq.~(\ref{gagamaZ_eff}) can be rewritten in terms of  $\{c_{\tilde{B}}, c_{\tilde{W}}  \}$ (and $c_\text{f}$) applying Eq.~(\ref{gaZZ}) for  $g_{aWW}$ and the last equality in Eq.~(\ref{gagammaZ_cWcB}) for   $g_{a\gamma Z}$.

An example of physical process to which the exact results can be directly applied in case $m_a<M_Z$ is given by the decay width of a $Z$ boson to photon plus ALP,  
 \begin{equation}
  \Gamma (Z \to a \g) = \frac{M_Z^3 |g_{a \g Z}^{\text{eff}}|^2}{384 \pi} \left( 1 - \frac{m_a^2}{M_Z^2} \right)^{3} \,,
 \end{equation}
 while for $m_a>M_Z$, the ALP decay width into $Z$ plus photon reads
\begin{equation}
  \Gamma (a \to \g Z) = \frac{m_a^3 |g_{a \g Z}^{\text{eff}}|^2}{128 \pi} \left( 1 - \frac{M_Z^2}{m_a^2} \right)^{3} \,.
 \end{equation}

We illustrate next the results obtained above for a generic off-shell ALP in some particular limits of practical interest.

\subsubsection{\tpdf{$g^{\text{eff}}_{a \gamma Z}$}{gaAZ} for high ALP \tpdf{$p^2$}{p2}}

For $p^2 \to \infty$, ($p^2 \gg (m_\mf^2 , M_Z^2 , M_W^2)$), the anomaly contribution yields:
\begin{flalign}
 A^\mf = 2 N_C Q_\mf^2 s_w^2 \,, &&
\end{flalign}
while the correction proportional to $g_{aWW}$ is given by
\begin{flalign}
 A^{WW} = \frac{42 M_W^2 + M_Z^2}{12 M_W^2} \log \left( \frac{\Lambda^2}{M_W^2} \right) - \sum_\mf \frac{N_C Q_\mf (T_{3,\mf} - 2 Q_\mf s_w^2)}{3 c_w^2} \log \left( \frac{\Lambda^2}{m_\mf^2} \right) - \left( \log \left( \frac{M_W^2}{p^2} \right) + i \pi \right)^2 \,,&&
\end{flalign}
where  the  terms proportional to $\log \Lambda$ are kept, because  consistency of the EFT expansion requires $p^2<\Lambda^2$.

\subsubsection{\tpdf{$g^{\text{eff}}_{a \gamma Z}$}{gaAZ} for intermediate  and low ALP \tpdf{$p^2$}{p2}}
Both for  $m_\mf^2 \ll p^2 \ll (M_Z^2, M_W^2) \ll m_t^2$,  where $\mf$ refers to all fermion mass but the top one,  and for $p^2 \to 0$ ($p^2 \ll (m_\mf^2 , M_Z^2)$), i.e. smaller than all fermion masses (which can apply for instance to  $Z$ decay to ALP + photon),  the contribution of fermionic ALP couplings to 
 $g^{\text{eff}}_{a \gamma Z}$ is well approached by 
\begin{flalign}
 A^\mf = \begin{cases}
          2 N_C Q_\mf^2 s_w^2 \,, & \text{for light fermions: } m_\mf^2 \ll M_Z^2 \,, \\
          \frac{3 Q_t}{2} \,, & \text{for the top quark: } m_t^2 \gg M_Z^2 \,,
         \end{cases}
 &&
\end{flalign}
while the correction proportional to $g_{aWW}$ reads
\begin{flalign}
 A^{WW} = \frac{42 M_W^2 + M_Z^2}{12 M_W^2} \log \left( \frac{\Lambda^2}{M_W^2} \right) -  \sum_{\text{f}} \frac{N_C Q_\mf (T_{3,\mf} - 2 Q_\mf s_w^2 )}{3 c_w^2} \log \left(  \frac{\Lambda^2}{m_\text{f}^2} \right) +...\,, &&
\end{flalign}
where dots stand for constant terms.

\subsection{  ALP anomalous coupling to \tpdf{$ZZ$}{ZZ}} 
\label{sec-ZZ-approx}
The effective $g_{aZ Z}$ coupling  receives  corrections induced by  three of the four electroweak gauge couplings: the set $\{g_{aWW}, g_{aZZ} , g_{a\gamma Z}\}$, plus $\bf{c}_{\text{f}}$ fermion corrections. All Feynman diagrams in  Figs. \ref{correctionsgammaZ}  and \ref{correctionexternallegs} contribute with $V=V'=Z$. Using  Eq.~(\ref{relationcouplingc2}),   the contributions resulting from electroweak gauge insertions can be projected on a two-dimensional space of couplings, which we choose to be 
here $\{g_{a Z Z}, g_{aWW} \}$. The total effective coupling  $g_{a Z Z}$ can then be expressed as 
\begin{equation} \label{gaZZ_eff}
\begin{aligned}
    g_{a Z Z}^{\text{eff}} = \, &  g_{a ZZ} \left\{1  + \frac{\alpha_{em}}{6 \pi c_w^2 s_w^2} \, (A^{Z/\gamma \rightarrow Z} +  B^{\text{Higgs}})\right\}
    + \frac{\alpha_{em}}{\pi} \frac{c_w^2}{s_w^2} g_{a WW} B^{WW}
     + \frac{\alpha_{em}}{\pi c_w^2 s_w^2 }  \sum_{\text{f}} \frac{c_{\text{f}}}{f_a}B^{\text{f}}\,,
\end{aligned}
\end{equation}
 where the complete expressions for the functions in this expression can be found in App.~\ref{app-aZZ-coupling-complete}. They correspond to: 
 \begin{itemize}
 \item $A^{Z/\gamma\rightarrow Z}$ encodes corrections to the external legs (diagrams \textbf{D1}-\textbf{D5} in Fig.~\ref{correctionexternallegs}), 
 see Eq.~(\ref{renZgammatoZ_ferm}) for the exact result. 
 \item $B^{\text{Higgs}}$ stems from the vertex insertion of $g_{aZZ}$ with a  Higgs particle exchanged between the two $Z$ bosons (diagram \textbf{E} in Fig.~\ref{correctionsgammaZ}), see Eq.~(\ref{BHiggs}).
 \item   $B^{WW}$ collects the contributions proportional to $g_{aWW}$ resulting from direct vertex insertions of $g_{a WW}$  in   Fig.\ref{correctionsgammaZ}  \textbf{A} and \textbf{B}, plus the $g_{a WW}$ component of the contributions seeded by the insertion of $g_{a \gamma Z}$  in the external legs and then projected onto the parameter space $\{g_{a Z Z}, g_{aWW} \}$, see Eq.~(\ref{BWW}). 
\item Finally,  the function $B^{\text{f}}$ encodes the contributions from vertex insertions of the fermionic couplings $c_{\text{f}}$ (Fig.~\ref{correctionsgammaZ} {\bf C}), see Eq.~(\ref{Bf}).  
\end{itemize}
Eq.~(\ref{gaZZ_eff}) can be rewritten in terms of  $\{c_{\tilde{B}}, c_{\tilde{W}}  \}$ (and $c_\text{f}$) applying Eq.~(\ref{gaZZ}) for  $g_{aWW}$ and the last equality in Eq.~(\ref{gaZZ_cWcB}) for   $g_{aZZ}$.

The results in this subsection can be applied to a variety of transitions in which the ALP may be on-shell or off-shell. For instance, for $m_a> 2M_Z$ the one-loop corrected ALP decay width into two $Z$ bosons is simply given by
 \begin{equation}
  \Gamma (a \to Z Z) = \frac{m_a^3 |g_{aZZ}^{\text{eff}}|^2}{64 \pi} \left( 1 - \frac{4 M_Z^2}{m_a^2} \right)^{3/2} \,.
 \end{equation}

 We present next for illustration the limit of the complete results in App.~\ref{app-aZZ-coupling-complete} in the particular case of high ALP four-momentum squared, which can be of  interest for instance for non-resonant collider ALP searches.\footnote{For intermediate ALP momentum ($m_\mf^2 \ll p^2 \ll (M_Z^2, M_W^2) \ll m_t^2$) and low four-momentum ($p^2 \ll (m_\mf^2 , M_Z^2)$) the transition is not kinematically possible with the gauge bosons on-shell.}

\subsubsection{\tpdf{$g^{\text{eff}}_{a Z Z}$}{gaZZ} for high ALP \tpdf{$p^2$}{p2}}
For $p^2 \to \infty$ ($p^2 \gg (m_\mf^2 , M_Z^2 , M_W^2)$),  only the anomaly contribution remains from the insertion of ALP-fermions couplings, 
\begin{flalign}
 B^\mf = - N_C Q_\mf^2 s_w^4 \,, 
\end{flalign}
while the contribution proportional to $g_{aWW}$ simplifies to
\begin{flalign}
 B^{WW} = \frac{42 M_W^2 + M_Z^2}{12 M_W^2} \log \left( \frac{\Lambda^2}{M_W^2} \right) - \sum_\mf \frac{N_C Q_\mf (T_{3,\mf} - 2 Q_\mf s_w^2)}{3 c_w^2} \log \left( \frac{\Lambda^2}{m_\mf^2} \right) - \frac{1}{2} \left( \log \left( \frac{M_W^2}{p^2} \right) + i \pi \right)^2\,, &&
\end{flalign}
and that proportional to $g_{aZZ}$ corrected by Higgs boson exchange between external legs vanishes,  $B^{\text{Higgs}} = 0$.

\subsection{ ALP anomalous coupling to \tpdf{$W^+W^-$}{WW}} 
\label{sec-aWW-approx}
All four couplings in the ensemble $\{g_{a\gamma\gamma}, g_{aWW}, g_{aZZ} , g_{a\gamma Z}\}$    induce one-loop corrections to the effective $g^{\text{eff}}_{aWW}$ coupling. All Feynman diagrams in  Figs. \ref{correctionsgammaZ}  and \ref{correctionexternallegs} contribute. The complete 
results can be found  in App.~\ref{sec-aWW-complete}. Using Eq.~(\ref{relationcouplingc2}),  the total result can be expressed for instance as a function of  
$\{g_{a WW}, g_{a\gamma\gamma} \}$ plus  fermionic couplings, 
\begin{equation}\label{gaWWeff}
 \begin{aligned}
 g_{a W W}^{\text{eff}} = \, &  g_{a WW} \Bigg\{1+   \frac{\alpha_{em}}{24 \pi s_w^2} (A^{W\rightarrow W} + C^{WW}+ C^{\text{Higgs}} ) \Bigg\}   + \frac{\alpha_{em}}{2 \pi} g_{a \gamma\gamma}\,C^{\gamma\gamma} + \frac{\alpha_{em}}{\pi  s_w^2}\sum_{\text{f}} \frac{c_{\text{f}}}{f_a}C^{\text{f}} \,,
\end{aligned}
\end{equation}
 where:
 \begin{itemize}
 
\item $A^{W\rightarrow W}$ contains two sources of one-loop external-leg SM corrections to the insertion of $g_{a WW}$ itself:   fermionic and Higgs corrections,
 \begin{equation}\label{AWW}
 A^{W\rightarrow W}= A^{W\rightarrow W}_{\text{ferm}}+A^{W\rightarrow W}_{\text{Higgs}}\,,
\end{equation}
 with only fermion doublets contributing  to $A^{W\rightarrow W}_{\text{ferm}} $, see diagram {\bf D5} in Fig.~\ref{correctionexternallegs} and Eq.~(\ref{renAWtoW_ferm}), and the Higgs-dependent term $A^{W\rightarrow W}_{\text{Higgs}}$ stemming from diagrams {\bf D3} and {\bf D4} in Fig.~\ref{correctionexternallegs}, see Eq.~(\ref{renAWtoW_Higgs}).

\item $C^{WW}$ accounts for corrections proportional to  $g_{a W W}$,  and gathers one-loop SM corrections on the external legs (Fig.~\ref{correctionexternallegs} {\bf D1} and {\bf D2}) together with vertex ones (Fig.~\ref{correctionsgammaZ} {\bf A} and {\bf B}) (see Eq.~(\ref{Cgauge}) for the complete expression): 
\begin{itemize}
\item The leg corrections and those from the vertex diagram {\bf B}  are directly seeded by the insertion of $g_{aWW}$.
\item The contributions originated from  diagram {\bf A} correspond to the combination of direct vertex insertions of  $g_{aWW}$, plus  the $g_{aWW}$ component  of the contributions seeded by   $\{ g_{aZZ} , g_{a\gamma Z}\}$ insertions  projected onto the $\{ g_{a\gamma \gamma}, g_{aWW}\}$ parameter space.
\end{itemize} 

\item $C^{\text{Higgs}}$ is a pure vertex correction resulting from the direct insertion of $g_{a WW}$ with the Higgs boson  exchanged between the two $W$ legs (diagram {\bf E} in Fig.~\ref{correctionsgammaZ}), see  Eq.~(\ref{CHiggs}).

\item The vertex function $C^{\gamma\gamma}$ corresponds to Fig.~\ref{correctionsgammaZ} {\bf A}, combining the results from the direct insertion of $g_{a\gamma\gamma}$ { \it and}  the $g_{a\gamma\gamma}$ component of the contributions seeded by $\{  g_{a\gamma Z}, g_{aZZ}\}$  insertions projected onto the $\{ g_{\gamma\gamma }, g_{aWW}\}$ parameter space, see Eq.~(\ref{Cgammagamma}). 
\item  Finally, the vertex function $C^{\text{f}}$  accounts for the fermionic triangle contributions (Fig.~\ref{correctionsgammaZ} {\bf C}), induced by fermionic couplings $c_\text{f}$ insertions, see Eq.~(\ref{Cferm}).
 
\end{itemize}  
Eq.~(\ref{gaWWeff}) can be rewritten in terms of  $\{c_{\tilde{B}}, c_{\tilde{W}}  \}$ (and $c_\text{f}$) applying Eq.~(\ref{gaZZ}) for  $g_{aWW}$ and the last equality in Eq.~(\ref{gagamma_cWcB}) for   $g_{a\gamma\gamma}$.

 We present next for illustration the  high ALP four-momentum squared limit of  the functions in Eq.~(\ref{gaWWeff}). 
\subsubsection{\tpdf{$g^{\text{eff}}_{a WW}$}{gaWW} for high ALP \tpdf{$p^2$}{p2}}

In the limit $p^2 \to \infty$ ($p^2 \gg (m_\mf^2 , M_Z^2 , M_W^2)$),  the fermionic contribution to the anomaly  vanishes. The same holds in this limit for the correction proportional to $g_{a\g\g}$ as well as that stemming from Higgs boson-exchange between the external $W$ bosons, 
\begin{equation}
 C^\mf = C^{\g\g} = C^{\text{Higgs}} =0\,.
\end{equation}
The only non-vanishing contributions  in this limit are those  proportional to $g_{a WW}$ itself and stemming from $A^{W\rightarrow W}$ and $C^{WW}$. The function $A^{W\rightarrow W}$  is independent of $p^2$: in consequence, it is not  further simplifed from the  relatively cumbersome complete expressions in Eqs.~(\ref{renAWtoW_ferm}) and Eq.~(\ref{renAWtoW_Higgs}), see Eq.~(\ref{AWW}). The function $C^{WW}$ simplifies to
\begin{flalign}\label{CWWapprox}
  C^{WW} = 43 \log \left( \frac{\Lambda^2}{M_W^2} \right) - 12 \left( \log \left( \frac{M_W^2}{p^2} \right) + i \pi \right)^2  
 - 12 s_w^2 \log \left( \frac{\lambda^2}{M_W^2} \right) \left( 1 + i \pi + \log \left( \frac{M_W^2}{p^2} \right) \right) ... \,, &&
 \end{flalign}
where $\Lambda$ is the UV cutoff (this logarithmic dependence cannot be disregarded in front of that in $p^2$ for EFT consistency), and $\lambda$ denotes the IR cutoff. The computation has been carried out entirely in dimensional regularization, with the $1/\epsilon$ terms  traded next for energy cutoffs via a protocol alike to that used for $g^{\text{eff}}_{agg}$ --Eq.~(\ref{Ggg-dimensionalreg})--   and the prescription in Eq.~(\ref{prescription}). The $\log \Lambda$ dependence contained in $C^{WW}$ combined with that in the leg correction $A^{W\rightarrow W}$ determines the beta function for $c_{\tilde{W}}$ in Eq.~(\ref{betacW}).
 
The two first terms in  Eq.~(\ref{CWWapprox}) are the leading contributions for large enough $p^2$. 
The third term   exhibits a logarithimic dependence on the IR cutoff which is instead physically irrelevant and can be disregarded, as it will exactly cancel for any physical observable against the contributions from soft and/or collinear photon brehmsstrahlung.  The latter may also contribute additional finite terms to be combined with the finite and $p^2$ independent terms in $C^{WW}$  (see the exact expression in Eq.~(\ref{Cgauge}) in App.~\ref{sec-aWW-complete}), encoded here by dots.

\vspace{0.5cm}
 For intermediate  ($m_\mf^2 \ll p^2 \ll (M_Z^2, M_W^2) \ll m_t^2$) and low ($p^2 \ll (m_\mf^2 , M_Z^2)$) ALP four-momentum,  the ALP-$WW$ transition is again not kinematically possible for gauge bosons on-shell.

\subsection{ ALP fermionic couplings} 
\label{sec-fermion-couplings}
\begin{figure}[h]
\centering
\includegraphics{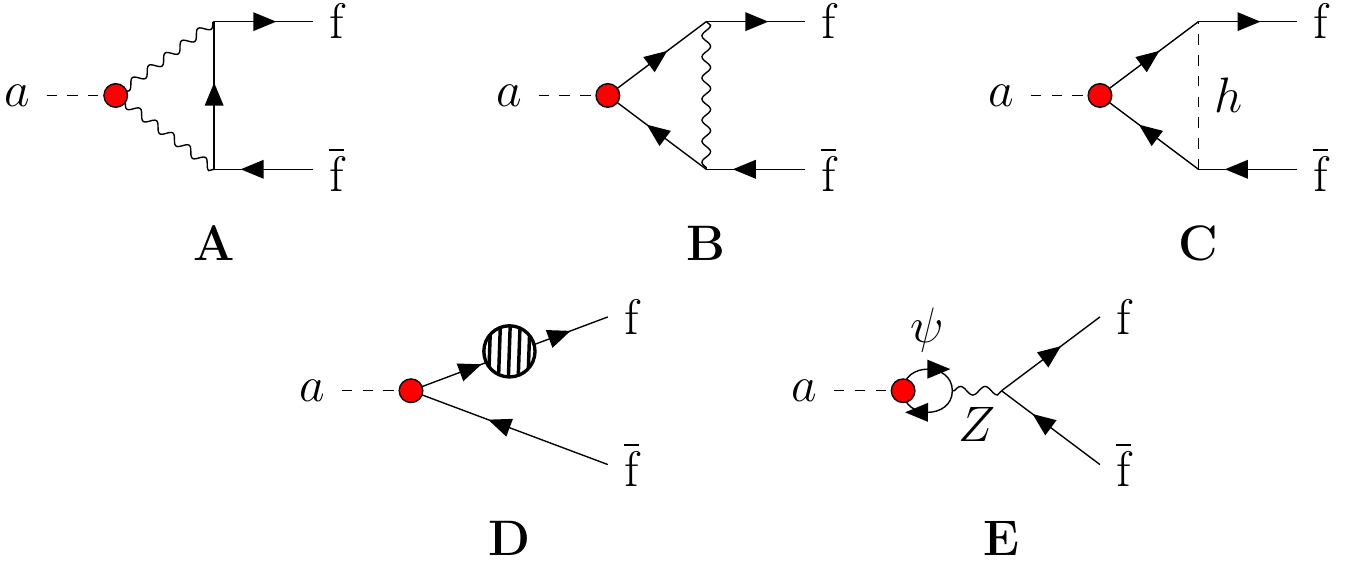}
\caption{One-loop diagrams contributing to $c_{\text{f}}$  at one-loop (plus the corresponding diagrams with Goldstone bosons). The wavy lines denote gauge bosons: gluons, photons, $W$ and $Z$ bosons.}
\label{correctionsferm}
\end{figure}

\begin{figure}[h]
\centering
\includegraphics{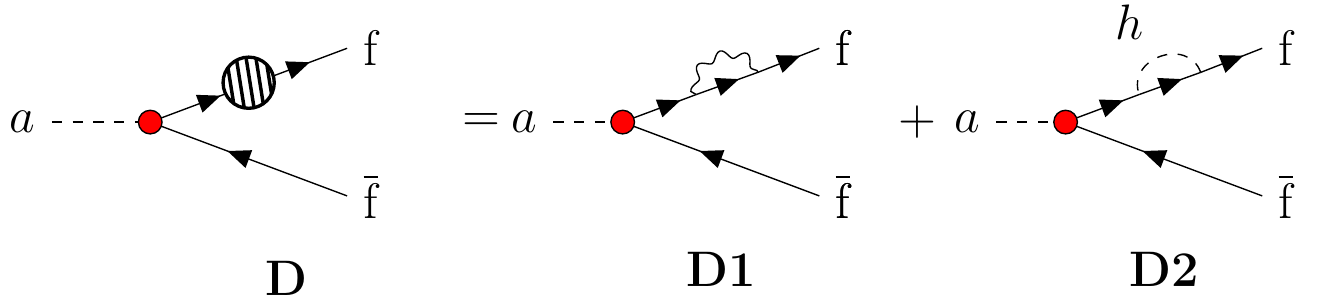}
\caption{One-loop diagrams contributing to the correction to the external fermion legs. Diagrams with Goldstone bosons are included.}
\label{correctionexternallegsferm}
\end{figure}

The one-loop corrections to the effective ALP-fermion-fermion couplings are depicted in Figs.~\ref{correctionsferm} and \ref{correctionexternallegsferm}, where the internal wavy lines denote either the gluon in the case of the gluon-ALP coupling $g_{agg}$ (only possible for quark final states) or electroweak gauge bosons. 
Contrary to the case for all previous effective couplings described, the individual contributions seeded
by each 
of the electroweak couplings in the set $\{g_{a\gamma\gamma}, g_{aWW}, g_{aZZ} , g_{a\gamma Z}\}$ are separately gauge invariant. In other words, the $\xi$-independence of the results holds already at the level of each one of those contributions, that is, prior to their projection onto a reduced parameter space of electroweak gauge couplings.
For this reason, we will present  those contributions individually. If wished, the reader can trivially project those results in the two-coupling $\{c_{\tilde{W}}, c_{\tilde{B}}\}$ parameter space, or on any other parameter space (e.g. $\{g_{a\gamma \gamma}, {g_{aWW}}\}$), using the gauge-invariance  relations in  Eq.~(\ref{relationcouplingc2}).

The results can be summarized as
\begin{equation}
 \begin{aligned}
\frac {c_{\text{f}}^{\text{eff}} }{f_a}= & \frac {c_{\text{f}}}{f_a} \left\{  1+ \frac{\alpha_{em}}{2 \pi}\, D^{c_\text{f}} + \frac{\alpha_{s}}{3 \pi}\, D^{c_\text{f}}_g \right\}  +   \frac{\alpha_{em}}{2 \pi f_a}\,
 \big\{ c_{\text{f'}} \,   D^{c_{\text{f'}}} +   \sum_{\psi}  {c_{\psi}} \, \, D^{c_\psi}_{\text{mix}}  \big\}\\
 +  &\frac{\alpha_{em}}{2 \pi}\left\{   g_{a \gamma \gamma}\, D^{\gamma\gamma} +    g_{a\gamma Z} \,D^{\gamma Z} +     g_{aZZ} \,D^{ZZ}+    g_{aWW} \,D^{WW} 
  \right\}  +  \frac{\alpha_{s}}{3 \pi}\left\{   g_{a g g}\, D^{g g}\right\}\,,
 \end{aligned}
 \label{cfeff}
\end{equation}
where the sum over fermions runs over all possible flavours, $\psi=u,c,t,d,s,b,e,\mu,\tau$, and 
the terms in the second line account --respectively-- for vertex insertions of the phenomenological ALP electroweak couplings  $\{g_{a \gamma \gamma}, g_{a\gamma Z},  g_{aZZ},  g_{aWW} \}$ plus the anomalous gluon coupling $g_{a gg}$: they all stem from diagram {\bf A} in Fig.~\ref{correctionsferm}, and  each term is separately gauge invariant. The complete expressions for the functions $D^{\gamma\gamma}$, $D^{\gamma Z}$, $D^{ZZ}$ and $D^{WW}$ can be found in Eqs.~(\ref{Dgg})-(\ref{DWW}) of App.~\ref{app-fermion-couplings-complete}.   
The first line in Eq.~(\ref{cfeff}) encodes instead insertions of: 
\begin{itemize}
\item The fermionic coupling $c_\text{f}$ itself accompanied by one-loop exchange of a gluon, encoded in $D^{c_\text{f}}_g $, or by the one-loop exchange of either a photon, a $Z$, a $W$ or a Higgs boson, i.e. 
\begin{equation}
D^{c_\text{f}} = D^{c_\text{f}}_\gamma +D^{c_\text{f}}_Z+ D^{c_\text{f}}_W+ D^{c_\text{f}}_h\,,
\label{Dcf}
\end{equation}
where $D^{c_\text{f}}_W$ is a pure leg correction from $W$ exchange (Fig.~\ref{correctionexternallegsferm} {\bf D1}), and it is $\xi$-independent by itself, see Eq.~(\ref{DcfW}).  In contrast, in to order get results in an explicitly  gauge invariant formulation, the one-loop corrections 
due to  photon or  $Z$ exchange --encoded respectively in $D^{c_\text{f}}_\gamma$ and  $D^{c_\text{f}}_Z$-- require the combination of the vertex diagram {\bf B} in Fig.~\ref{correctionsferm} and  the leg correction in Fig.~\ref{correctionexternallegsferm} {\bf D1}, see Eqs.~(\ref{Dcfgamma}) and (\ref{DcfZ}).  Similarly, manifest gauge invariance of the Higgs-exchange corrections --encoded in $D^{c_\text{f}}_h$-- results after combining the vertex correction in Fig.~\ref{correctionsferm} {\bf C} and the leg corrections in Fig.~\ref{correctionexternallegsferm} {\bf D2}, see Eq.~(\ref{DcfHiggs}).
\item  The  contribution from $c_\text{f'}$, where $\text{f'}$ denotes  the  $SU(2)_L$ flavour partner of fermion $\text{f}$, encoded in the function $D^{c_\text{f'}}$ given in Eq.~(\ref{Dcfprime}). It corresponds to the vertex correction due to $W$ exchange  in Fig.~\ref{correctionsferm} {\bf B}, which is gauge invariant by itself.
\item All possible fermionic contributions to the mixed $a$-$Z$ correction in Fig.~\ref{correctionsferm} {\bf E}, which are encoded through the functions $D^{c_\psi}_{\text{mix}}$, which are also separately gauge-invariant, see the complete result in Eq.~(\ref{exact-mix}).
\end{itemize} 
The results can be applied to a variety of physical transitions with an ALP on- or off-shell. For instance, for ALP decay into a fermionic $\text{f}\bar{\text{f}}$ channel when  $m_a> 2 \, m_\text{f}$, the one-loop corrected width is simply obtained from 
\begin{equation}
  \Gamma (a \to \mf \, \bar \mf) = \frac{N_C m_a m_\mf^2 |c_\mf^{\text{eff}}|^2}{8 \pi f_a^2} \sqrt{ 1 - \frac{4 m_\mf^2}{m_a^2} } \,.
 \end{equation}

For simplicity and for illustration purposes, we present next in this subsection some useful limits of the exact functions in App.~\ref{app-fermion-couplings-complete}, for a generic off-shell ALP.

\subsection{\tpdf{${c_{\text{f}}^{\text{eff}} }$}{cf eff}  for high ALP $p^2$}
For non-resonant searches at the LHC and other colliders, and/or for very heavy ALPs, the limit  $m_\mf^2 \ll (M_Z^2,M_W^2,M_H^2) \ll m_t^2 \ll p^2$ is of physical interest, where $m_{\text{f}}$ refers to all fermion masses but the top one.  In this subsection we set $m_\mf=0$  except in divergent terms.

\subsubsection{Limit of light external fermions for ${\text{f}=u,d,s,b,e,\mu}$}
Let us first consider the contribution of gauge-anomalous couplings to ${c_{\text{f}}^{\text{eff}} }$. 
 For instance, 
$D^{gg}$  encodes the $g_{agg}$ contribution with gluons running in the internal loop of diagram {\bf A} in Fig.~\ref{correctionsferm}, which in this limit reduces to
\begin{flalign}
\label{Dgluon-high-p2}
 D^{g g} =  \Bigg\{ 3 \log \left( \frac{\Lambda^2}{m_\mf^2} \right) - 4 - \frac{2 \pi^2}{3} - \frac{1}{2} \left( \log \left( \frac{m_\mf^2}{p^2} \right) + i \pi \right)^2 \Bigg\} \,. &&
\end{flalign}
Analogously, 
$D^{\gamma\gamma}$  accounts for the $g_{a\gamma\gamma}$ insertion with two photons running in the internal loop of diagram {\bf A} in Fig.~\ref{correctionsferm}, with an expression very close to that of  $D^{g g}$ which in this limit reduces to
\begin{flalign} \label{Dgamma-high-p2}
D^{\gamma\gamma} =  \frac{ Q_\text{f}^2}{2 }  D^{g g} \,, &&
\end{flalign}
while the $D^{\gamma Z}$ term stems from that same  diagram  with one photon and one $Z$ boson in the internal loop,
\begin{flalign}
&
D^{\gamma Z}
  = \frac{ Q_\text{f} \,(T_{3,\text{f}} - 2 Q_\text{f} s_w^2)}{16 c_w s_w}  \Bigg\{  12 \log \left( \frac{\Lambda^2}{M_Z^2} \right) - 19 - \frac{2 \pi^2}{3} - 2 \left( \log \left( \frac{M_Z^2}{p^2} \right) + i \pi \right)^2
  \\
  &
  - \log \left( \frac{m_\mf^2}{M_Z^2} \right) \left[ 6 + 4 i \pi +  4 \log \left( \frac{M_Z^2}{p^2} \right)\right] \Bigg\} \,. &&
 \end{flalign}
Similarly, 
the same  diagram  in Fig.~\ref{correctionsferm} {\bf A}  although with  two internal $Z$ bosons results in
 \begin{flalign}
& D^{Z Z}
  = \frac{ 1 }{8 c_w^2 s_w^2}  \Bigg\{ (T_{3,\text{f}}^2 - 2 T_{3,\text{f}} Q_\text{f} s_w^2 + 2 Q_t s_w^4) \left(  6 \log \left( \frac{\Lambda^2}{M_Z^2} \right) - 11 \right)
  \\
  & + 4 T_{3,\mf}^2 \left(1 + \log \left( \frac{M_Z^2}{p^2} \right) + i \pi \right) + 4 Q_\mf^2 s_w^2 (T_{3,\mf} - Q_\mf s_w^2) \left( \log \left( \frac{M_Z^2}{p^2} \right) + i \pi \right)^2 \Bigg\} \,, &&
\end{flalign}
while $D^{WW}$  corresponds to that same diagram, albeit with internal W bosons,
\begin{flalign}
D^{W W}
  = \begin{cases}
  \frac{ 1 }{16 s_w^2}  \Bigg\{ 6 \log \left( \frac{\Lambda^2}{p^2} \right) - 2 \log \left( \frac{M_W^2}{p^2} \right) - 7 + 4 i \pi \Bigg\} \,, & \begin{aligned} &\text{for leptons and quarks} \\
   & \text{except top and bottom} \,, \end{aligned} \\
\frac{ 1 }{16 s_w^2}  \Bigg\{  6 \log \left( \frac{\Lambda^2}{p^2} \right) - 2  \log \left( \frac{m_t^2}{p^2} \right) - 9 + 4 i \pi \Bigg\} \,,  & \text{for the bottom quark} \,.
    \end{cases} &&
  \end{flalign}
  
For the contributions resulting from the 
insertions of ALP fermionic couplings, the one-loop gluon corrections (vertex plus legs), and the analogous one-loop photon corrections lead  in this limit to, respectively,  
\begin{flalign} \label{Dcfgg_highp2}
 D^{c_\text{f}}_g = - 2 \Bigg\{ 1 -  \frac{\pi^2}{6} + \log \left( \frac{\lambda^2}{m_\mf^2} \right) \left( 1 + i \pi + \log \left( \frac{m_\mf^2}{p^2} \right) \right)  + \frac{1}{2} \left( \log \left( \frac{m_\mf^2}{p^2} \right) + i \pi \right)^2 \Bigg\} \,, &&
\end{flalign}
\begin{flalign}
D^{c_{\text{f}}}_\gamma=  \frac{ Q_{\text{f}}^2 } {2}\,  D^{c_\text{f}}_g 
\,, &&
\label{Dcfgamma_highp2}
\end{flalign}
where $\lambda$ is an infrared cutoff which encodes  the IR-divergent contributions to the $1/\epsilon$ dimensional regularization terms via the prescription in Eq.~(\ref{prescription}), following the same protocol used for the gluonic IR divergences in Eq.~(\ref{Ggg-dimensionalreg}).
and the photonic ones in Eq.~(\ref{CWWapprox}). Those unphysical IR logarithmic dependences will again exactly cancel in physical transitions against those from the phase space integral  terms stemming from tree-level soft and/or collinear gluon and photon bremsstrahlung.

  In turn, $Z$ exchange (vertex plus legs) is free from IR divergences and leads to 
\begin{flalign}
 D^{c_{\text{f}}}_Z = - \frac{Q_\mf s_w^2 ( T_{3,\text{f}}  -  Q_\text{f} s_w^2 )}{2  c_w^2 s_w^2}  \Bigg\{   \frac{2 \pi^2}{3} + \left( \log \left( \frac{M_Z^2}{p^2}\right) + i \pi \right)^2 \Bigg\} \,. &&
\end{flalign}

The ${c_{\text{f}}^{\text{eff}} }$ component resulting from one-loop $W$-exchange corrections to ALP fermion-coupling insertions unfolds as explained as two $\xi$-independent contributions: i) the leg correction from the insertion of $c_{\text{f}}$  in Fig.~\ref{correctionexternallegsferm} {\bf D1}, encoded in $D^{c_{\text{f}}}_W$, which in this particular limit vanishes,  
 and ii)  the vertex correction  induced by the insertion of  the $SU(2)$ flavour-partner coupling $c_{\text{f}'}$ in Fig.~\ref{correctionsferm} {\bf B}, encoded in $D^{c_{\text{f'}}}$:
\begin{flalign}
 D^{c_{\mfp}}
 = \begin{cases}
0 \,, &  \begin{aligned} & \text{for leptons and quarks} \\
 & \text{except top and bottom} \,, \end{aligned} \\
- \frac{ m_{t}^2}{8 M_W^2 s_w^2}  \Bigg\{ \log \left( \frac{\Lambda^2}{p^2} \right) + \log \left( \frac{M_W^2}{p^2} \right) + \frac{7}{2} + 2 i \pi \Bigg\} \,, & \text{for the bottom quark} \,.
   \end{cases} \,. &&
\end{flalign}

The one-loop Higgs corrections to $c_{\text{f}}$ insertions also vanish in this limit, $D^{c_{\text{f}}}_h  = 0$.
 Finally,  the mixed one-loop contribution to $c_{\text{f}}^{\text{eff}}$ from diagram {\bf E} in Fig.~\ref{correctionsferm}  receives contributions from all possible ALP fermionic couplings --quarks and leptons, and it is also $\xi$-independent by itself. Its expression is particularly simple even in the exact case (see Eq.~(\ref{exact-mix}) in App.~\ref{app-complete-results}), while in the present limit all contributions vanish but for that with the top quark running in the loop,
\begin{flalign}
 D^{c_{t}}_\text{mix}  = -  \frac{3 T_{3,\text{f}} \, m_t^2 }{ 2 s_w^2 M_W^2} 
\Bigg\{ \log \left( \frac{\Lambda^2}{p^2} \right) + 2 +i \pi \Bigg\} \,, &&
\end{flalign}
where $T_{3,\text{f}}$ denotes the third component of weak isospin for the external flavour $\text{f}$.
The logarithimic dependence was already obtained in Ref.~\cite{Feng:1997tn}. This result shows that, in the limit  under study, the top-coupling contribution can be the dominant one on the quest for signals of ALP couplings to light fermions, because the contributions are proportional to the mass of the fermion running in the loop and independent of the external flavour. In fact,  this conclusion extends as well to the exact result in Eq.~(\ref{exact-mix}). This may be very relevant for instance on the searches for ALP couplings to electrons in XENON and other experiments, see Sec.~\ref{sec-LHC-probes}.
 
\subsubsection{Limit of light internal fermions for external $\text{f}=t$}
The analogous high ALP $p^2$ results when the external fermion is the top, i.e. the contributions to ${c_{t}^{\text{eff}} }$ neglecting light fermion masses, are reported next. 

Let us consider first the impact of the insertions of ALP gauge anomalous couplings. In the case of the ALP-photon and ALP-gluon couplings, $g_{agg}$ and $g_{a\gamma\gamma}$, the corresponding functions  $D^{gg}$ and $D^{\gamma\gamma}$ are exactly as those in  Eqs.~(\ref{Dgluon-high-p2}) and  (\ref{Dgamma-high-p2}) albeit with the replacement $m_{\text{f}}\rightarrow m_t$. For the other anomalous couplings, the results simplify to
\begin{flalign}
D^{\gamma Z}
  = \frac{ Q_t \,(T_{3,t} - 2 Q_t s_w^2)}{4 c_w s_w}  \Bigg\{ 3 \log \left( \frac{\Lambda^2}{m_t^2} \right) - 4 \Bigg\} \,, &&
  \end{flalign}
\begin{flalign}
&
D^{Z Z}
  = \frac{ 1 }{4 c_w^2 s_w^2}  \Bigg\{ (T_{3,t}^2 - 2 T_{3,t} Q_t s_w^2 + 2 Q_{t} s_w^4) \left( 3 \log \left( \frac{\Lambda^2}{m_t^2} \right) - 4 \right) 
  \\
  &
  + 2 Q_t^2 s_w^2 (T_{3,t} - Q_t s_w^2) \Bigg[ \frac{2 \pi^2}{3} + \frac{1}{2} \left( \log \left( \frac{m_t^2}{p^2} \right) + i \pi \right)^2 \Bigg] + 2 T_{3,t}^2 \left( \log \left( \frac{m_t^2}{p^2} \right) + i \pi \right) \Bigg\} \,, &&
\end{flalign}
\begin{flalign}
D^{W W}
  = \frac{ 1 }{8 s_w^2}  \Bigg\{ 3 \log \left( \frac{\Lambda^2}{p^2} \right) - \log \left( \frac{m_t^2}{p^2} \right) - 3 + 3 i \pi \Bigg\} \,. &&
  \end{flalign}
  
In turn, the one-loop gluon and photon contributions  to ${c_{t}^{\text{eff}} }$ stemming from  ALP-fermion couplings, i.e.  $D^{c_\text{f}}_g$   and $D^{c_\text{f}}_\gamma$,   are respectively identical to those found above for the light external fermion limit  
in Eqs.~(\ref{Dcfgg_highp2}) and  (\ref{Dcfgamma_highp2}).  The rest of the one-loop boson corrections to insertions of ALP-fermion couplings reads in this limit:

\begin{equation} \begin{aligned}
&
 D^{c_{t}}_Z = \frac{1}{2  c_w^2 s_w^2}  \Bigg\{ - \frac{m_t^2 T_{3,t}^2}{M_Z^2} \left(  \log \left( \frac{\Lambda^2}{p^2} \right) + 2 + i \pi \right) - ( T_{3,t}^2  + 4  T_{3,t}  Q_t s_w^2 - 4 Q_t^2 s_w^4 ) \log \left( \frac{m_t^2}{M_Z^2}\right)
 \\
 &
 - Q_t s_w^2 ( T_{3,t}  -  Q_t s_w^2 ) \left[ \frac{\pi^2}{3} +2 \log \left( \frac{m_t^2}{M_Z^2}\right) \left( \log \left( \frac{m_t^2}{p^2}\right) + i \pi \right) - \left( \log \left( \frac{m_t^2}{p^2}\right) + i \pi \right)^2 \right] \Bigg\} \,, &&
\end{aligned} \end{equation}
\begin{flalign}
 D^{c_{t}}_W
 = - \frac{m_{t}^2}{8 M_W^2 s_w^2}  \Bigg\{ \log \left( \frac{\Lambda^2}{m_{t}^2} \right) + 1 + i \pi \Bigg\} \,, &&
\end{flalign}
\begin{flalign}
 D^{c_{t}}_h  = - \frac{m_{t}^2}{8 \pi s_w^2 M_W^2} \Bigg\{ \log \left( \frac{\Lambda^2}{M_H^2} \right) +  \log \left( \frac{p^2}{M_H^2} \right) - 2 - i \pi \Bigg\} \,, &&
\end{flalign}
\begin{flalign}
 D^{c_{\psi}}_{\text{mix}}  = -  \frac{3 m_t^2 }{4 s_w^2 M_W^2} \Bigg\{ \log \left( \frac{\Lambda^2}{m_{t}^2} \right) + 2 +i \pi \Bigg\} \,, &&
\end{flalign}
while  $D^{c_{\text{f'}}}_W = 0 $ with   $\text{f'}=b$ in this particular case.
  
\subsection{${c_{\text{f}}^{\text{eff}} }$  for intermediate ALP $p^2$ and light fermions}

We explicit now the limits for an ALP with a  low   $p^2$,  smaller or equal than all  SM boson gauge boson masses  but larger than the mass squared of all light fermions $m_\mf^2 \ll p^2 \ll M_Z^2 , M_W^2, M_H^2$
 with $\text{f} =u,d,c,s,b,e,\mu, \tau$. This limit is of interest for instance when considering decays of a light ALP to leptons or light fermions, such as those searched for in rare decays.  
 
The contribution stemming from the insertions of  $g_{agg}$  and $g_{a\gamma\gamma}$, i.e. the functions $D^{gg}$  and $D^{\gamma\gamma}$,  are again exactly as those in Eqs.~(\ref{Dgluon-high-p2}) and (\ref{Dgamma-high-p2}).
For the other anomalous couplings, the results simplify in this limit to
 \begin{flalign}
 \label{DgammaZ-lightALP}
D^{\gamma Z}
  = \frac{ Q_\text{f} \,(T_{3,\text{f}} - 2 Q_\text{f} s_w^2)}{8 c_w s_w}  \Bigg\{ 6 \log \left( \frac{\Lambda^2}{M_Z^2} \right) - 13 \Bigg\} \,, &&
  \end{flalign}
\begin{flalign}
D^{Z Z}
  = \frac{ (T_{3,\text{f}}^2 - T_{3,\text{f}} 2 Q_\text{f} s_w^2 + 2 Q_{\text{f}} s_w^4)}{8 c_w^2 s_w^2}  \Bigg\{ 6 \log \left( \frac{\Lambda^2}{M_Z^2} \right) - 19 \Bigg\} \,, &&
  \end{flalign}
\begin{flalign}
\label{DW-lightALP}
D^{W W}
  = \frac{ 1 }{16 s_w^2}  \Bigg\{ 6 \log \left( \frac{\Lambda^2}{M_Z^2} \right) - 19 \Bigg\} \,. &&
  \end{flalign}
The results  for  $D^{\gamma\gamma}$  and Eqs.~(\ref{DgammaZ-lightALP})-(\ref{DW-lightALP}) have been addressed previously in Ref.~\cite{Bauer:2017ris} for an on-shell ALP ($p^2=m_a^2$); our results are in agreement with those, except for a minor factor in $D^{\gamma\gamma}$, Eq.~(\ref{Dgamma-high-p2}).

\noindent  We consider next the impact of inserting ALP fermionic couplings. Their contributions vanish in this particular limit for the following functions: 
\begin{flalign}
 D^{c_{\text{f}}}_Z =  D^{c_{\text{f}}}_W = D^{c_{\text{f}}'}_W =D^{c_{\text{f}}}_h  = 0 \,, &&
\end{flalign}
while  the gluon and photon corrections $D^{c_\text{f}}_g$   and $D^{c_\text{f}}_\gamma$ coincide  with those in Eqs.~(\ref{Dcfgg_highp2}) and  (\ref{Dcfgamma_highp2}). Finally, the $a$-$Z$ mixing corrections read in this limit~\cite{Feng:1997tn}
\begin{flalign} \label{mixingtope}
 D^{c_{t}}_\text{mix}  = -  \frac{3 T_{3,\text{f}} \, m_t^2 }{ 2 s_w^2 M_W^2} \Bigg\{ \log \left( \frac{\Lambda^2}{m_t^2} \right) \Bigg\} \,. &&
\end{flalign}

\section{Gauge invariance at one-loop level}
\label{sec-gauge-inv-one-loop}

This section analyzes the modifications to the tree-level gauge invariance relations in Eq.~(\ref{relationcouplingc2}) and (\ref{gaBB}), which result from rewriting the only two independent parameters of the electroweak sector $g_{aWW}$ and $g_{aBB}$ (i.e. $c_{\tilde{W}}$ and $c_{\tilde{B}}$, see Eqs.~(\ref{gaZZ}) and (\ref{gaBB})) in terms of the measured phenomenological couplings, e.g. 
\begin{equation}
 \label{relationcouplingc3}
    \begin{aligned}
      &  g_{aWW}= g_{a\gamma\gamma} + \frac{c_w}{2 s_w} g_{a \gamma Z} \,, \\
 & g_{aBB} =  c_w^2 g_{a \gamma\gamma} + s_w^2 g_{a ZZ} - c_w s_w g_{a \gamma Z} \,.
    \end{aligned}
\end{equation}

Radiative corrections which include mass effects 
(spontaneously) break the explicit gauge invariance of the original Lagrangian in Eq.~(\ref{general-NLOLag-lin}) and Table~\ref{tab.basis}. In other words, corrections proportional to the Higgs vev $v$ are to be expected, which can be summarized as contributions to both the original $SU(2)_L\times U(1)_Y$-invariant operators and to additional effective couplings which are not invariant under the electroweak (and custodial) symmetry.  The results can then be encoded 
as the strength of the following set of four effective couplings 
 \begin{equation}
 \{a \, B^{\mu\nu}  \tilde {B}_{\mu\nu},\, a\,W^{\mu\nu}  \tilde {W}_{\mu\nu}, \,a \, B^{\mu\nu}  \tilde {W}^3_{\mu\nu}, \,a\,W_3^{\mu\nu}  \tilde {W}_{3\,\mu\nu}\}\,, \label{set-radiative-corrections}
 \end{equation}
where the last two are new and do not respect electroweak and custodial symmetries, while the first two were already present in the original gauge-invariant Lagrangian Eq.~(\ref{general-NLOLag-lin}). The radiative corrections to the  $aW^1_{\mu \nu} \tilde{W}^{1 \mu \nu}$ coupling must equal  exactly  those for the $aW^2_{\mu \nu} \tilde{W}^{2 \mu \nu}$ coefficient because of electric charge conservation:  $g^{\text{eff}}_{a W W}$ will encode them as well as the identical ones for 
the $aW^3_{\mu \nu} \tilde{W}^{3 \mu \nu}$ interaction, while the ``excess'' will be accounted for in the coefficient for a $a\,W_3^{\mu\nu}  \tilde {W}_{3}$ interaction denoted $\Delta_{WW}$. In turn, $\Delta_{ BW}$ will encode  corrections of the form $a \, B^{\mu\nu}  \tilde {W}^3_{\mu\nu}$, i.e. 
\begin{equation} \label{Leff-non-GI}
\mathcal{\delta L}_a^\text{total}  \supset \frac{1}{4} \Delta_{B W} a B_{\mu \nu} \tilde{W}^{3 \mu \nu} +\frac{1}{4} \Delta_{WW} a W^3_{\mu \nu} \tilde{W}^{3 \mu \nu}\,.
\end{equation}
The  two new effective couplings 
can be expressed as 
the following combinations of radiatively-corrected phenomenological parameters:
\begin{equation} \label{invrelat}
    \begin{aligned}
    \Delta_{WW} = & \bar{s}_w^2 \,g^{\text{eff}}_{a \gamma\gamma} + \bar{c}_w^2 \,g^{\text{eff}}_{a ZZ} + \bar{c}_w \bar{s}_w \,g^{\text{eff}}_{a \gamma Z}- g^{\text{eff}}_{a WW} \,, \\
    \Delta_{ BW} = & 2 \bar{c}_w \bar{s}_w ( g^{\text{eff}}_{a \gamma \gamma} - g^{\text{eff}}_{aZZ} ) + (\bar{c}_w^2 - \bar{s}_w^2) g^{\text{eff}}_{a \gamma Z} \,.
    \end{aligned}
\end{equation}
It is straightforward to compute the exact values of $ \Delta_{WW} $ and $ \Delta_{BW} $ from the results for the effective couplings in Sect.~\ref{complete-one-loop-main} and App.~\ref{app-complete-results} and the expression for $\bar{c}_w$   in Eq.~(\ref{cwbar}). 
The tree-level closed gauge-invariance relations in Eq.~(\ref{relationcouplingc2}) will be modified in consequence. 

\begin{figure}[h]
\centering
\includegraphics{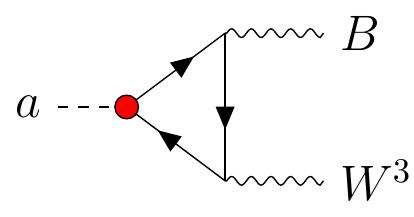}
\caption{Illustration of fermionic one-loop contributions which induce an effective coupling $a \,B^{\mu\nu}  \tilde {W}^3_{\mu\nu}$.}
\label{inducedcouplingBW}
\end{figure}

\subsection*{Gauge invariant ancestors of radiatively corrected couplings}
As stated above, the operators $a B_{\mu \nu} \tilde{W}^{3 \mu \nu}$ and  $a W^3_{\mu \nu} \tilde{W}^{3 \mu \nu}$ are neither custodial nor $SU(2)_L$ invariant.  
Nevertheless, there must be a fully gauge-invariant formulation of any possible correction to the effective Lagrangian and its corrections, because electroweak gauge symmetry is unbroken in nature.
Indeed,  in generic EFTs, one-loop corrections are expected to give contributions to higher order terms in the EFT expansion.
Both in the SMEFT and in the linear ALP EFT, these contributions are always finite, i.e. all UV-divergences are reabsorbed order-by-order in the EFT expansion.
Well-known examples in the SM are the magnetic and electric dipole moments in the SM, whose gauge invariant version corresponds to operators with mass dimension six and above.

Higher order radiative corrections, and in particular mass dependent ones (which are equivalent to multiple Higgs insertions that then take a vev) can imply that a full tower of operators may be needed to formulate those corrections 
in a gauge invariant way. 
The putative $SU(2)_L\times U(1)_Y$-invariant ancestors of the four gauge anomalous couplings in the Lagrangian $\mathcal{ L}_a^\text{total}+\mathcal{\delta L}_a^\text{total}$, i.e. Eqs.~(\ref{general-NLOLag-lin})  and Eq.~(\ref{Leff-non-GI}), 
can be formulated alike to those  in Ref.~\cite{Helset:2020yio} for CP-conserving Higgs couplings. For our ALP set, we expect $v$-dependent radiative corrections encoded in the gauge invariant operators 
\begin{align}
\label{eq.BB_n}
&(5+2n): \quad a \, (\Phi^\dagger \Phi)^{n}  \, B^{\mu\nu}  \tilde {B}_{\mu\nu}\,,   \\ 
\label{eq.WW_n}
& (5+2n): \quad a\,(\Phi^\dagger \Phi)^{n}\,   W^{\mu\nu}  \tilde {W}_{\mu\nu}\,, \\ 
\label{eq.BW_n}
& (7+2n): \quad a\,(\Phi^\dagger \Phi)^{n}\,(\Phi^\dagger \vec{\sigma}\Phi)\,    \vec{W}^{\mu\nu}  \tilde {B}_{\mu\nu} \longrightarrow a \, B^{\mu\nu}  \tilde {W}^3_{\mu\nu} \,, \\
& (9+2n): \quad a\,(\Phi^\dagger \Phi)^{n}\,(\Phi^\dagger \vec{\sigma^a}\Phi) \,(\Phi^\dagger \vec{\sigma^b}\Phi)\,    \vec{W}_a^{\mu\nu}  \tilde {W}_{b\,\mu\nu} \longrightarrow a \, W_3^{\mu\nu}  \tilde {W}^3_{\mu\nu} \,,
\label{GI-radiative-operators}
\end{align}
where  $n$ is integer, $n \ge0$. In the last two lines  it is indicated that those two towers of operators lead --after spontaneously symmetry breaking-- to the custodial and $SU(2)_L$ non-invariant couplings $a \, B^{\mu\nu}  \tilde {W}^3_{\mu\nu}$ and $ a\, W_3^{\mu\nu}  \tilde {W}^3_{\mu\nu}$ postulated earlier: note that their mass dimension 
is at least seven and nine,  respectively,  and that they vanish for $v=0$. 
In contrast,  the couplings  in the first two lines can receive mass-independent one-loop corrections even for $n=0$, as computed in the previous section. An important consequence of this is that loops induced by dimension-5 ALP operators can give UV-divergent contributions to the structures in~\eqref{eq.BB_n} and~\eqref{eq.WW_n} for $n=0$, while contributions to all other structures must be finite. We find that this is indeed the case for $\Delta_{WW}$ and $\Delta_{BW}$.

A pertinent question is the scale that would weight down those higher-dimension operators. Only one inverse power of $f_a$ is possible, because ALP insertions must enter as powers of $a/f_a$, and only one ALP insertion is considered here. The remaining scale dependence must then correspond to either another BSM scale (not considered here) or simply to SM mass parameters when only SM radiative corrections are present as in the present work, i.e. to powers of the electroweak scale. These SM corrections should generate coefficient contributions proportional in addition to the SM sources of custodial breaking.

\subsection{Gauge invariance relations among effective electroweak couplings at one-loop}
\label{results-gauge-invariance-loop}
It is easy to verify that  $ \Delta_{ WW}=\Delta_{ BW}=0$ in the massless limit, i.e. for $v=0$,\footnote{Moreover,  the radiative correction to $g_{aBB}$ is proportional as expected to  $\sum_\psi y^2_\psi$, where $y_\psi$ denote the fermion hypercharges.} and the one-loop corrections to the anomalous gauge couplings  satisfy the tree-level gauge invariance relations Eq.~(\ref{relationcouplingc2}). 
Instead, when mass corrections are taken into account, non-zero values for    $\Delta_{ WW}$ and $\Delta_{ BW}$ do emerge. As an example, our results show that the contributions stemming from ALP-fermion coupling insertions --see Fig.~\ref{inducedcouplingBW}-- are finite and take the general form 
\begin{equation}
\begin{aligned}
\Delta_{ BW} & =Y_L \, (F_L(m_1) - F_L(m_2))+Y_{R1} \,F_R(m_1)-Y_{R2} \,F_R(m_2) \,,
\end{aligned}
\end{equation}
where $m_{i=1,2}$ denote fermion masses of $SU(2)_L$ fermion partners and the functions $F_{R}(m)$ and $F_L(m)$  cancel in the massless fermion limit, $F_{L,R}(0)=0$.  In other words,  a non-vanishing $\Delta_{ BW}$ coupling requires as expected that the sources of custodial breaking be at play: different fermion hypercharges and non-degenerate fermion partners running in the loop.

More in general,  it follows from the analysis above that the tree-level gauge invariance relations in Eq.~(\ref{relationcouplingc2}) are to be substituted by the one-loop corrected ones, which we choose to parametrize as:
\begin{equation}
 \label{relationcouplingc2-one-loop}
\begin{aligned}
&  g^{\text{eff}}_{aWW}= g^{\text{eff}}_{a\gamma\gamma} + \frac{\bar{c}_w}{2 \bar{s}_w} g^{\text{eff}}_{a \gamma Z}  - \frac{\bar{c}_w}{2 \bar{s}_w} \Delta_{BW} - \Delta_{WW}\,, \\
& g^{\text{eff}}_{a ZZ} = g^{\text{eff}}_{a \gamma \gamma} + \frac{\bar{c}_w^2 - \bar{s}_w^2}{2 \bar{c}_w \bar{s}_w} g^{\text{eff}}_{a \gamma Z}  - \frac{1}{2 \bar{c}_w \bar{s}_w} \Delta_{BW} \,,
\end{aligned}
\end{equation}
where ${\bar{c}_w}$ was defined in Eq.~(\ref{cwbar}). 
These one-loop corrections gauge invariance relations may impact on the limits inferred 
for  a given coupling from the experimental bounds on another couplings known at present with higher precision (e.g.  the bounds on the ALP-$ZZ$ anomalous coupling 
obtained from the experimental limits on the ALP-$\gamma \gamma$ coupling 
in certain mass regimes~\cite{Bauer:2017ris,Alonso-Alvarez:2018irt,Gavela:2019wzg}).

\subsubsection*{Limit  \boldmath{$m_\mf^2, M_Z^2 , M_W^2, M_H^2 \ll p^2 \le m_t^2$}  }
Because   $\Delta_{ WW}$ and $\Delta_{ BW}$ vanish for $v=0$,  they vanish in the limit $p^2 \to \infty$. The contribution of the top quark may thus dominate for large $p^2$ close to $m_t^2$.
That is,  the contribution of  the top-ALP coupling ${\bf{c}}_{t}$ may dominate in the limit in which all SM particle masses but the top one are neglected with respect to the ALP $p^2$: 
\begin{flalign}
  \Delta_{BW} \sim - c_{t} \,\frac{\alpha_{em}}{\pi c_w s_w} \frac{m_t^2}{p^2} \Bigg\{ 6 + 6 i \sqrt{1- \frac{4 m_t^2}{p^2}} f \left( \frac{4 m_t^2}{p^2} \right) - 4 f \left( \frac{4 m_t^2}{p^2} \right)^2 \Bigg\} \,, &&
\end{flalign}
\begin{flalign}
 \Delta_{WW} \sim  c_{t} \,\frac{3 \alpha_{em}}{2 \pi s_w^2} \frac{m_t^2}{p^2} \Bigg\{ 1 + m_t^2 \mathcal{C} \left( 0 , 0 , p^2 , m_t , 0 , m_t \right) \Bigg\} \,, &&
\end{flalign}
where the function   $f(\tau)$  was defined in Eq.~(\ref{ftau}) and $\mathcal{C}$ is defined in  Eqs.~(\ref{functionC})-(\ref{functionCmasslessALP}) of App.~\ref{app-aZZ-coupling-complete}. Notice that these expressions vanish in the limit $p^2 \to \infty$, as they must.

\section{Some phenomenological consequences of loop-induced ALP couplings}
\label{sec-LHC-probes}

High-precision measurements may be increasingly able to probe loop corrections to tree-level effective couplings. Currently, sensitivity to loop-induced couplings is particularly interesting when a tree-level coupling is suppressed and the loop contributions dominate. 

In this section we are going to explore two examples of such situations: high-energy gluon-initiated production of an electroweak ALP, and very precise low-energy searches for ALPs which rely on couplings to electron-positron. In both cases, we focus on the loop effects of the ALP coupling to top quarks.

\subsection{LHC probes for heavy ALPs}
\begin{figure}[ht!]
\centering
\includegraphics[scale=0.2]{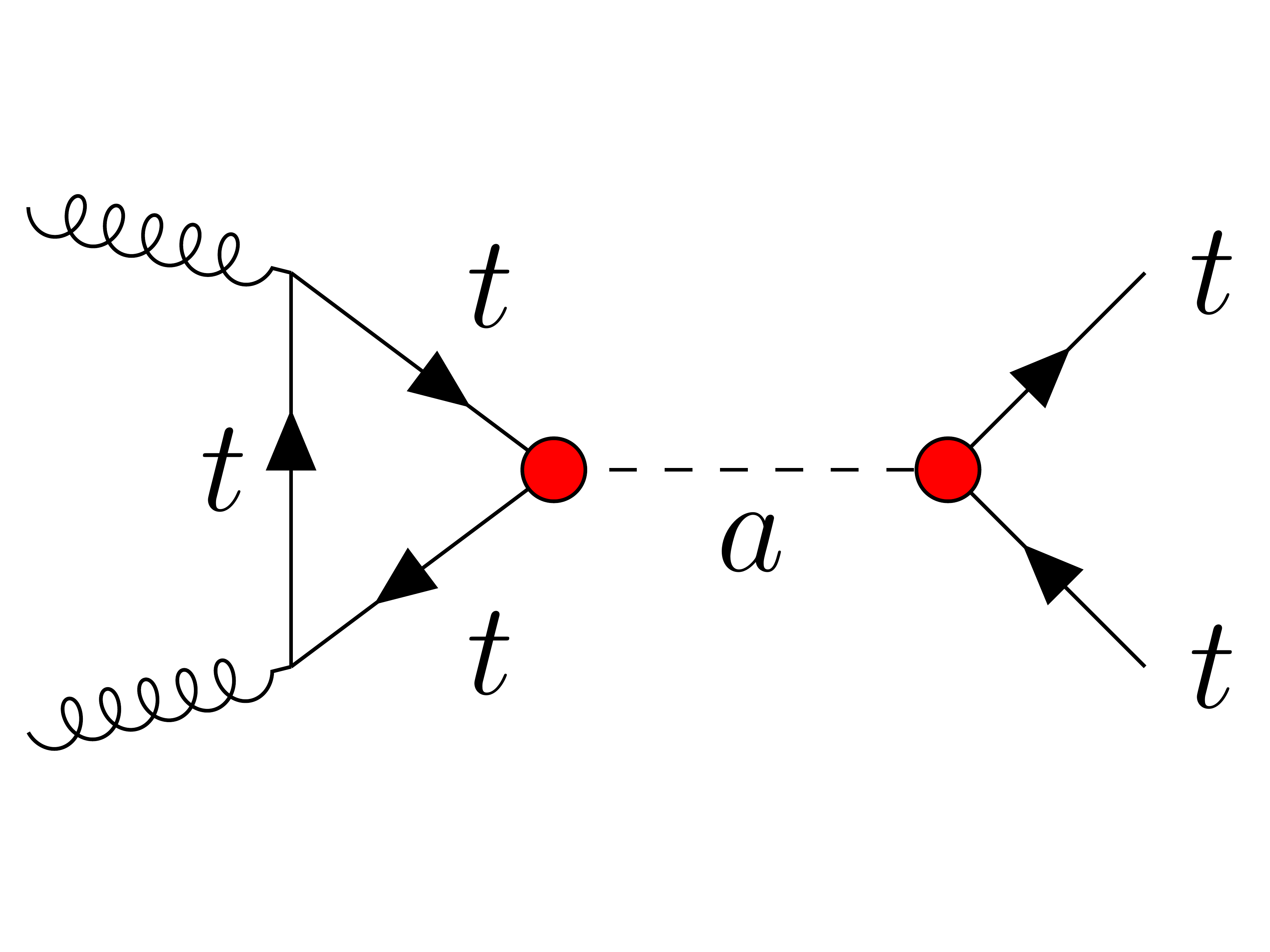}
\caption{One-loop contribution to ALP production via gluon-fusion, and its decay to a pair of tops. }
\label{axionttbar}
\end{figure}

In the Lagrangian Eq.~(\ref{general-NLOLag-lin}), we provided the ALP with couplings to the whole SM: the electroweak bosons, gluons and fermions. This is a rather general coupling structure, yet ALPs may have restrictions on how they communicate at tree-level to the SM. For example, ALPs could originate from a UV  sector participating  in the  mechanism of electroweak symmetry breaking, coupled to the $SU(2)_L\times U(1)_Y$ sector and not to the $SU(3)_c$ one, e.g. in Composite Higgs models where an additional heavy CP-odd state arises as a partner to the Composite Higgs~\cite{No:2015bsn, Croon:2015naa}.  This is just an example of theories with vanishing or very suppressed tree-level effective ALP-gluon coupling ($g_{a g g }$ in Eq.~(\ref{gagluon})): an electroweak ALP.

These  models of electroweak ALPs would be hard to probe at the LHC, as  protons are mainly made of light quarks and gluons. 
Then, the  leading contribution  to gluon-fusion cross-section could correspond to integrating out tops. For definiteness, let us consider exclusively the ALP-top diagonal coupling $c_t$ defined in Eq.~(\ref{updiagonal}),
\begin{equation}
\label{eq:ct}
\Lag \supset c_t\,\dfrac{\de_\mu a}{2f_a} \, \left(\bar{t}\gamma^\mu \gamma_5 t \right)\,.
\end{equation}
The ALP production would then be mediated by a top running in the gluon loop and could be constrained, for instance, in $gg\to a\to t\bar t$ processes, as illustrated in Fig.~\ref{axionttbar}. In this process, the ALP could be either resonant or non-resonant~\cite{Gavela:2019cmq}, depending on its mass.\footnote{
A competing channel, that takes place at tree level, is $p  p \to t  \bar t  a$ with $a \to t \bar t$. However, the phase-space suppression for this channel is stronger than the loop suppression in $gg\to t\bar t$. For example, for $m_a=1$ TeV, 
$\sigma(g g \to a)/\sigma(p p \to t\bar t a) \simeq 2 \times 10^4 $.
}

For definiteness, here we consider ALPs with $m_a>2m_t$, such that the top-antitop pair can be resonant. This allows us to derive constraints from existing searches for resonances in $t\bar t$ final states, that are at a very mature stage in the LHC collaborations. This is true in particular at high-mass, where the fully hadronic topology can be accessed using jet substructure techniques. As an illustration of how LHC probes could be used to search for heavy ALPs, we re-interpret the recent ATLAS analysis~\cite{ATLAS:2020lks} to set bounds on  $c_t/f_a$.

\begin{figure}[t!]
\centering
\includegraphics[width=10cm]{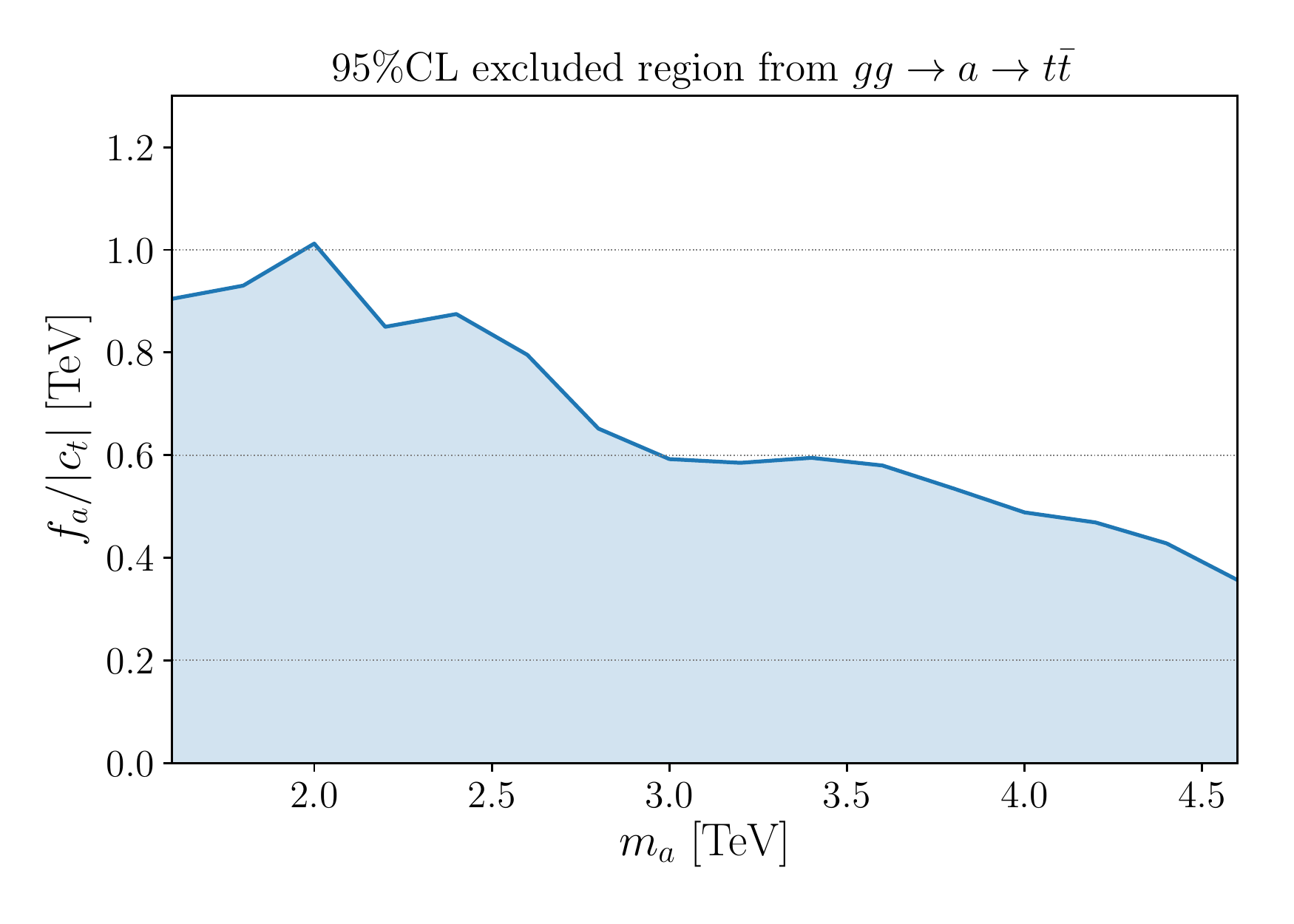}
\caption{Limits on $f_a/|c_t|$, as a function of the ALP mass, extracted from the all-hadronic $t\bar t$ resonance search by ATLAS of  Ref.~\cite{ATLAS:2020lks}. }
\label{fatoplimit}
\end{figure}

We simulate separately the pure $gg\to a\to t\bar t$ signal and the component stemming from the interference of this process with SM $gg\to t\bar t$ production. Expressing $g_{agg}^{\text{eff}}$ as a function of $c_t$ as in Eq.~\eqref{gagluon}, the former scales with $(c_t/f_a)^4$ and the latter with $(c_t/f_a)^2$. The simulation is performed generating $10^5$ events in each channel with {\tt MadGraph5\_aMC@NLO}~\cite{Alwall:2011uj}, using an in-house UFO implementation of the Lagrangian in Eq.~\eqref{general-NLOLag-lin}. Variations of $g_{agg}^{\text{eff}}$ with $p^2$ are neglected, as they only induce a few \% correction to the numerical value of the gluon coupling. The imaginary part stemming from expanding the $B_1$ loop function is also subdominant and can be safely neglected in the simulation.

We perform a very simple analysis at parton level, without decaying the top quarks and without performing full parton shower and detector simulations. To partially compensate for this, a gaussian smearing with a $6\%$ width is applied to the simulated top-antitop invariant mass ($m_{t\bar t}$) distribution,   and the latter is multiplied by a $m_{t\bar t}$-dependent suppression factor estimated from Fig.~2 in Ref.~\cite{ATLAS:2020lks}, that accounts for the tagging efficiencies.\footnote{We assume that the efficiency for the ALP detection does not differ significantly from that for a $Z'$. A more detailed analysis would require simulating both particles and comparing how the fat-jet tagging  efficiency varies depending on the coupling properties of the resonance. This dependence has been often found to be subdominant in previous studies, see e.g. Refs.~\cite{Gouzevitch:2013qca,Kasieczka:2021xcg}.} 
The acceptance correction is implemented by applying, at the generator level, the cuts reported in Ref.~\cite{ATLAS:2020lks} on the top quarks pseudo-rapidities $\eta_{t,\bar t}$ and transverse momenta $p_{T}(t,\bar t)$, and on their rapidity and azimuthal-angle separations, $\Delta y_{t\bar t}$ and $\Delta \phi_{t\bar t}$ respectively.

The distribution obtained (summing signal and interference components) is compared to the difference between measured and predicted number of events in the $m_{t\bar t}$ spectra reported in Ref.~\cite{ATLAS:2020lks}, that is available on~\href{https://www.hepdata.net/record/ins1795076?version=1}{HEPdata}. We implement a basic test statistics constructing a $\chi^2$ as
\begin{equation} \label{chisquared}
\chi^2(c_t/f_a) = \sum_k \frac{1}{\sigma_k^2}
\left[\frac{c_t^4}{f_a^4} \, a_k + \frac{c_t^2}{f_a^2}\, i_k + b_k - d_k\right]\,,
\end{equation}
where the index $k$ runs over the bins of the $m_{t\bar t}$ distributions for the 1- and 2-$b$-tagged signal regions, $a_k$ ($i_k$) is the number of events estimated for the pure ALP signal (ALP-SM interference) in the $k$-th bin with $c_t/f_a=\unit[1]{TeV^{-1}}$.  In this equation, $b_k$ ($d_k$) is the number of expected background events (observed events) reported by the ATLAS Collaboration. Finally, the uncertainty $\sigma_k$ is estimated by summing in quadrature the total systematic uncertainty reported by ATLAS, the statistical error $\sqrt{d_k}$ on the measured data points and the statistical uncertainty associated to our Monte Carlo simulation. As a conservative choice, bins with 0 observed events are removed from the analysis, as in this case a $\chi^2$ statistics cannot be applied. We repeat this analysis for various values of $m_a$ in the range from $1.6$ to $\unit[4.6]{TeV}$ and extract, for each value, a 95\%CL upper limit on $c_t/f_a$.

The results of this naive re-interpretation are shown in  Fig.~\ref{fatoplimit}. The limits on $f_a$ obtained lie at the boundaries of a good effective description of the ALP Lagrangian as, for $|c_t|=1$, the bound on $f_a$ is mostly below $m_a$. 
On the other hand, in a strongly interacting regime where $|c_t|\simeq 4\pi$ (as could be the case of a Composite Higgs model), the limits on $f_a$ improve by an order of magnitude and result well above $m_a$.
A dedicated analysis, potentially extended to the leptonic and semi-leptonic channels, could improve these bounds significantly.

\subsection{Limits on the couplings to top quarks for light ALPs }

Another interesting use of loop-induced ALP couplings appears when a tree-level coupling is very well measured and can provide a good constraint on loop-induced couplings, assuming no substantial cancellations happen between tree and loop-induced couplings.
Among these, the loop-induced ALP-electron diagonal coupling $c_e$ (defined in Eq.~\eqref{leptdiagonal}),  
\begin{equation}
\label{eq:ce}
\Lag \supset c_e\,\dfrac{\de_\mu a}{2f_a} \, \left(\bar{e}\gamma^\mu \gamma_5 e \right)\,, 
\end{equation}
is particularly interesting as electrons are found in stable matter. Astrophysical objects like red giants or precise non-collider experiments such as Dark Matter Direct Detection experiments provide an excellent handle on that coupling.   
Here we consider the current limits on the axion-electron coupling collected in Ref.~\cite{ciaran_o_hare_2020_3932430}, that include results from  Red Giants~\cite{Capozzi:2020cbu}, Solar neutrinos~\cite{Gondolo:2008dd} and LUX~\cite{LUX:2017glr}, which are derived for solar axions and extend to very low ALP masses, as well as from 
Edelweiss~\cite{EDELWEISS:2018tde},  
PandaX~\cite{PandaX:2017ock}, SuperCDMS~\cite{SuperCDMS:2019jxx}   
and XENON-1T~\cite{XENON:2019gfn,XENON:2020rca,VanTilburg:2020jvl}, that cover the region $\unit[100]{eV}\lesssim m_a \lesssim \unit[100]{keV}$ assuming the ALP to be the main DM constituent.\footnote{Note that --strictly speaking-- the bounds extracted from DM searches only apply in scenarios where the ALP is stable and can be produced with the correct relic abundance. Verifying the latter condition for the particular ALP scenario considered here is beyond the scope of this work.} The most stringent bounds are those from red giants and from DM direct detection at XENON-1T, and give $| c_{e}^{\text{eff}}|\, (m_e/f_a) \lesssim  10^{-13}$.

\begin{figure}[t]\centering
 \includegraphics[width=16cm]{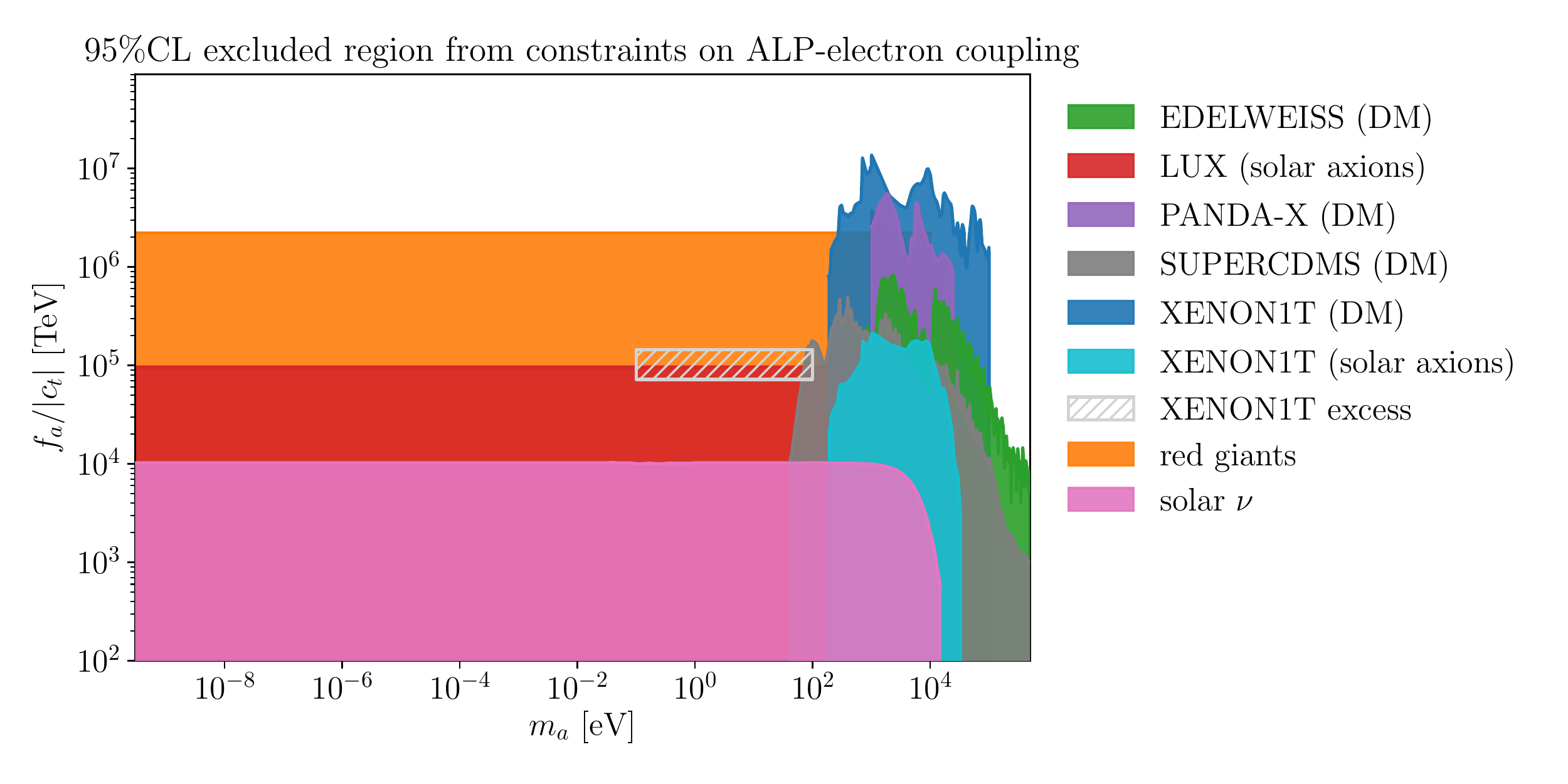}
 \caption{Limits on $f_a/|c_t|$ as a function of the ALP mass, extracted rescaling existing constraints on the ALP-electron coupling, taken from Ref.~\cite{ciaran_o_hare_2020_3932430}. The grey hatched box marks the region roughly compatible with the excess observed by XENON1T~\cite{XENON:2020rca}.}\label{fig:bounds_ee}
\end{figure}

These limits can be translated into limits on the diagonal ALP-top coupling, using the one-loop  contributions computed in Sec.~\ref{sec-fermion-couplings}, corresponding  to  diagram {\bf E} in Fig.~\ref{correctionsferm}, see also Refs.~\cite{Feng:1997tn,Arias-Aragon:2020shv}.  Note that in DM direct detection experiments the typical energy range is the keV, hence our expressions must be taken in the limit of low-momentum exchange  in the detector between the ALP and the electron, i.e. below the electron mass. In this case one finds the  log-enhanced expression found in Eq.~(\ref{mixingtope}), namely:
\begin{equation} \label{cefromctop}
 c_{e}^{\text{eff}} \simeq  2.48 \, c_t \,  \alpha_{em} \,  \log \left( \frac{\Lambda^2}{m_t^2}  \right) \,.
\end{equation}  
For consistency, the cutoff of the loop integrals $\Lambda$ 
should be  of the same order as $f_a$. As the $\Lambda$ dependence is logarithmic we will use  
$\Lambda=\unit[10^6]{TeV}$   in this equation, to extract the bounds on $f_a/c_t$ shown in Fig.~\ref{fig:bounds_ee},\footnote{ The value $\unit[10^6]{TeV}$ was chosen a posteriori, so as to match the limits on $f_a$. } from which it follows
\begin{equation}\label{ctfromredgiants}
f_a/|c_t| > \unit[2.2\times10^6]{TeV}
\end{equation}
in the entire range considered.
  If the ALP is assumed to be DM, XENON1T bounds apply, leading to the stronger constraint
 \begin{equation} \label{ctfromDM}
 f_a/|c_t|>\unit[1.4\times 10^7]{TeV}\,.
 \end{equation}

XENON1T recently observed an excess in their data, which could have been explained by solar axions coupled to electrons and/or photons in the mass range $m_a \sim 0.1-100$ eV for the QCD axion~\cite{XENON:2020rca,Gao:2020wer}. Thus, instead of a limit, in this case XENON1T would identify a finite preferred region in the plane $(g_{a\gamma\gamma},c_e/f_a)$. Unfortunately, this interpretation of a QCD axion is in conflict with the data from red giants. Nevertheless and for the sake of the exercise, one can consider what would be the preferred value for $f_a/c_t$ if  that XENON1T excess was taken at face value. Using then Eq.~\eqref{cefromctop}, and the one-loop corrections to $g_{a\g\g}$ computed in Sec.~\ref{sec-photon-couplings} which correspond to diagram \textbf{C} in Fig.~\ref{correctionsgammaZ}, it follows that the induced value of $g_{a\g\g}$ is strongly suppressed for the ALP mass range considered here ($p^2 = m_a^2 \ll m_t^2$): $g_{a\g\g}^{\text{eff}} \lesssim 10^{-8} \alpha_{em} c_t /f_a \ll c_e^{\text{eff}} / f_a$. 
In this limit, the results from XENON1T could be interpreted as a preferred range for $c_e^{\text{eff}}$ independent of  $g_{a\g\g}$. This broadly includes values $2\times 10^{-12}\lesssim |c_e^{\text{eff}}| (m_e/f_a) \lesssim 4 \times 10^{-12}$. The projection of this interval in terms of  $f_a/c_t$ is shown as a grey-hatched region in Fig.~\ref{fig:bounds_ee}.

\vspace{0,3cm}

Finally, note that the type of analysis carried out in this subsection can be also applied to the flavour-diagonal ALP-bottom coupling $c_b$ (defined in Eq.~\eqref{downdiagonal}). Numerically, limits on $f_a/c_b$ can be approximately estimated rescaling by $m_b^2/m_t^2$~\cite{Arias-Aragon:2020shv} those on $f_a/c_t$  in Eqs.~(\ref{ctfromredgiants}) and (\ref{ctfromDM}) above, leading respectively to $f_a/|c_b| >  \unit[1.0\times10^3]{TeV}$ and  $f_a/|c_b| >  \unit[6.5\times10^3]{TeV}$, for $\Lambda= \unit[10^3]{TeV}$.

\section{Conclusions}
The search for axions and ALPs is intensifying in both the energy and the precision frontiers. The vastly different energy regions explored range from those typical of astrophysics and low-energy laboratory experiments to collider energies. At the same time, increasingly precise probes are targeted e.g.  photon and/or invisible channels in rare hadron decays and other low-energy channels. The point is well past in which the estimation of one-loop effects in the couplings of ALPs to SM particles is needed to  explore optimally BSM physics through the detection of pseudo-Goldstone bosons signals. From the theoretical point of view,  effective Lagrangian formulations allow to pursue this quest in a very model-independent way.

In this work, we have first clarified the relations among alternative --complete and non-redundant-- CP-even bases for the $d=5$ ALP linear effective Lagrangian. In doing so, we derived the exact relations between bases which differ in their choices of fermionic operators constructed with left-handed and right-handed currents and/or chirality-flip couplings. We identified  the precise combinations of gauge anomalous couplings involved in trading different bases. This includes the relations stemming from the anomalous global $B+L$ current and the conserved $B-L$ one. 
Although we then chose to work  on a complete and non-redundant basis containing gauge anomalous operators plus all possible right-handed fermionic currents and certain couplings made out of left-handed currents, 
 the relations obtained will allow easy translation of the results to other bases.
   
Furthermore, illustrative practical checks of bases equivalences were as well performed. For instance, the purely bosonic operator $\O_{a\Phi}$ 
can be written either as a combination of right-handed fermion currents or as a combination of left-handed fermion currents, right-handed ones and gauge anomalous couplings: it is explicitly shown how  all anomalous corrections vanish at one-loop level, as they should. 
   
In a second step, we have computed  the complete one-loop corrections --thus including all divergent and finite terms-- to all possible CP-even couplings of an ALP to SM fields,
for a generic off-shell ALP and on-shell SM particles.  Our results are formulated in the form of the effective one-loop interactions $\{g^{\text{eff}}_{agg}\,,g^{\text{eff}}_{a\gamma\gamma},\,g^{\text{eff}}_{aWW},\,  g^{\text{eff}}_{aZZ}\,,g^{\text{eff}}_{a\gamma Z}\, ,
c_{\text{f}}^{\text{eff}} \}$, 
where  the latter  is computed for all SM fermions 
--light and heavy--   but restricted to flavour diagonal external channels. Moreover three-generation CKM mixing is disregarded in the loop corrections.  Neutrino masses are disregarded as well.    Our computations thus carry  to novel territory previous studies restricted to on-shell ALPs and to certain channels and limits. All our computations have been performed in the covariant $R_\xi$ gauge, and the intermediate $\xi$-dependent steps made publicly available at 
\href{https://notebookarchive.org/2021-07-9otlr9o}{NotebookArchive}, together with the exact final gauge-invariant results. The latter are  shown as well  in App.~\ref{app-complete-results}, while in the main text limits relevant for high, intermediate and low energy experiments are extracted. Particular attention has been dedicated to the isolation of infrared divergences when present.  
  As a byproduct, the UV divergent terms of our computations also allowed to do a straightforward check of recent RG results in the literature in different bases.

An illustration of the reach of our results is the impact that any putative ALP coupling induces at one-loop on any other ALP interaction. For instance, we explored how, for heavy ALPs, the ALP-top coupling can be constrained by LHC measurements of top-pair final states, processes which are induced at one-loop by this coupling. These channels, pumped up by gluon fusion via a top loop, open up the possibility of studying many ALP final states with a sizeable cross-section, even when the tree-level coupling ALP-gluon would be zero. We also explored constraints on ALP-top interactions for light ALPs. In this case, the strictest limits are those derived from bounds on the ALP-electron coupling, extracted from astrophysical constraints and from DM Direct Detection searches~\cite{Feng:1997tn}.

An interesting point also clarified in this work is the one-loop modification of the electroweak tree-level gauge invariance relations. These are relevant as far as custodial symmetry breaking, i.e. mass and hypercharge differences, are relevant. We have determined these corrections, which will impact future one-loop extractions at LHC and other experiments of the sensitivity to a given ALP coupling from more precise data on another ALP coupling  (e.g. $g_{aWW}$ from data on  $g^{\text{eff}}_{a\gamma \gamma}$ or $g^{\text{eff}}_{a\gamma Z}$, and similar analyses).

A plethora of experimental channels should be explored using the results of this paper. 
 Future directions include  the one-loop complete results with all external particles off-shell and also flavour non-diagonal channels. A related interesting task is  the computation of box and other diagrams for certain physical processes,  which is mandatory to cancel all infrared divergences in processes involving $g_{agg}$, $g_{aWW}$ and $ c_{\text{f}}^{\text{eff}}$.  Finally, the analysis of the ALP bases should be extended to include CP violation in the ALP couplings. These and other exciting developments lie ahead in the BSM path to  uncover novel pseudo-Goldstone boson physics.

\acknowledgments
We thank Gonzalo Alonso, Josemi No, Jorge Fernandez de Troconiz, Arturo de Giorgi, Luca Merlo and Pablo Quilez for illuminating discussions.  
The work of J.B. was e supported by the Spanish MICIU through the National Program FPU (grant number FPU18/03047). J.B. and M.B.G. acknowledge partial financial support by the Spanish MINECO through the Centro de excelencia Severo Ochoa Program under grant SEV-2016-0597, by the Spanish ``Agencia Estatal de Investigac\'ion''(AEI) and the EU ``Fondo Europeo de Desarrollo Regional'' (FEDER) through the project PID2019-108892RB-I00/AEI/10.13039/501100011033. All authors acknowledge the European Union's Horizon 2020 research and innovation programme under the Marie Sk\l odowska-Curie grant agreement No 860881-HIDDeN. VS acknowledges support from the UK STFC via Grant ST/L000504/1. 

\newpage
\appendix

\section{Standard Model equations of motion}\label{app.eom}
\label{EOM}
In this Appendix we report the SM EOM for the fermion and Higgs fields, that are relevant for the discussion in  Sect.~\ref{secEffLag} and App.~\ref{app.redefinitions}.
For chiral fermions, the EOM read 
\begin{align}
\label{eq.eom_Q}
 i\slashed{D} Q_L & = \Phi Y_d d_R+\tilde \Phi Y_u u_R\,,
 &
 i\slashed{D} u_R & = \tilde \Phi^\dag Y_u^\dag Q_L\,,
 &
 i\slashed{D} d_R & = \Phi^\dag Y_d^\dag Q_L\,,
 \\
 i\slashed{D} L_L & = \Phi Y_e e_R\,,
 &
 i\slashed{D} e_R & = \Phi^\dag Y_e^\dag L_L\,,
 \label{eq.eom_L}
\end{align}
where flavor index contractions are implicit.
For the conjugate fields they imply
\begin{align}
\label{eq.eom_Q_conjugate}
 -i \bar Q_L \overleftarrow{\slashed{D}} & = \bar d_R Y_d^\dag\Phi^\dag +\bar u_R Y_u^\dag\tilde \Phi^\dag\,,
 &
 -i \bar u_R \overleftarrow{\slashed{D}} & = \bar Q_L Y_u\tilde \Phi\,,
 &
 -i \bar d_R \overleftarrow{\slashed{D}} & = \bar Q_L Y_d\Phi \,,
 \\
 -i \bar L_L \overleftarrow{\slashed{D}} & = \bar e_R Y_e^\dag\Phi^\dag  \,,
 &
 -i \bar e_R \overleftarrow{\slashed{D}} & = \bar L_L Y_e\Phi \,.
 \label{eq.eom_L_conjugate}
\end{align}

\noindent
The EOM for the Higgs field reads
\begin{align}
\label{eq.eom_H}
\square \Phi_i &= -\left[\bar d_R Y_d^\dag (Q_L)_i 
+ (\bar Q_L i\sigma^2)_i Y_u u_R 
+ \bar e_R Y_e^\dag (L_L)_i \right] + \frac{m_h^2}{2} \Phi_i - 2\lambda (\Phi^\dag\Phi)\Phi_i\,,
   \\
\label{eq.eom_H_conjugate}
\square \Phi^\dag_i &= -\left[(\bar Q_L)_i Y_d d_R+ \bar u_R Y_u^\dag (i\sigma^2 Q_L)_i + (\bar L_L)_i Y_e e_R \right] + \frac{m_h^2}{2} \Phi^\dag_i - 2\lambda (\Phi^\dag\Phi)\Phi^\dag_i   \,,
\end{align}
where $i$ is a free $SU(2)_L$ index and we have taken $V(\Phi^\dag\Phi) = -(m_H^2/2) \Phi^\dag\Phi + \lambda (\Phi^\dag\Phi)^2$, where $m_H$ and $\lambda$ denote respectively the Higgs mass and self-coupling.

The use of fermion EOM is tantamount to chiral rotations of fermion fields, at the classical level. When considering loop effects as in this work, they must be supplemented by the contributions of the SM anomalous global currents, i.e. 
\begin{align}
 \de_\mu(\bar Q_{L}^i \g^\mu Q_{L}^i) &  \supset 
 \dfrac{g^{\prime 2}}{96\pi^2} B_{\mu \nu} \tilde{B}^{\mu \nu}
 + \dfrac{3g^2}{32\pi^2} W_{\mu \nu}^{\a} \tilde{W}^{\a \mu \nu}
 + \dfrac{g_s^2}{16\pi^2} G_{\mu \nu}^{a} \tilde{G}^{a \mu \nu}
 \,,
 \\
\de_\mu(\bar u_{R}^i \g^\mu u_{R}^i) & \supset 
-\dfrac{g^{\prime 2}}{12\pi^2}B_{\mu \nu} \tilde{B}^{\mu \nu}
-\dfrac{g_s^2}{32\pi^2}G_{\mu \nu}^{a} \tilde{G}^{a \mu \nu}
\,,
\\
\de_\mu(\bar d_{R}^i \g^\mu d_{R}^i) & \supset 
-\dfrac{g^{\prime 2}}{48\pi^2}B_{\mu \nu} \tilde{B}^{\mu \nu}
-\dfrac{g_s^2}{32\pi^2}G_{\mu \nu}^{a} \tilde{G}^{a \mu \nu}
\,,
\\
\de_\mu(\bar L_{L}^i \g^\mu L_{L}^i) & \supset 
\dfrac{g^{\prime 2}}{32\pi^2}B_{\mu \nu} \tilde{B}^{\mu \nu}
+\dfrac{g^2}{32\pi^2}W_{\mu \nu}^{\a} \tilde{W}^{\a \mu \nu}
\,,
\\
\de_\mu(\bar e_{R}^i \g^\mu e_{R}^i) & \supset 
- \dfrac{g^{\prime 2}}{16\pi^2}B_{\mu \nu} \tilde{B}^{\mu \nu}
\,,
\end{align}
where we are not summing over the index $i$.

\section{Field redefinitions and operator basis reduction}\label{app.redefinitions}
In this appendix we consider the ALP-dependent field redefinitions that are required in order to relate and reduce the operator basis:
\begin{align}\label{eq.field_redefinitions}
\Phi &\mapsto \exp\left[i x_\Phi \frac{a}{f_a}\right]\Phi\,,
&
\mf &\mapsto \exp\left[i \x_{\mf} \frac{a}{f_a}\right]\mf\,,
\end{align}
where in flavour space $\mf=\{Q_L,u_R,d_R,L_L,e_R\}$ are vectors and $\x_\mf=\x_\mf^{ij}$ are tensors. For notation simplicity, the subindices $\{L,R\}$  will be omitted, i.e. $\mf=\{Q_L,u_R,d_R,L_L,e_R\}\equiv\{Q,u_,d_,L,e\}$. We take all rotation parameters $x_\Phi$ and $\x_\mf$ to be real, consistent with the assumption of CP conservation of the ALP couplings (the only CP-violation present is that of the SM contained in CKM, i.e. in the Yukawa matrices). Moreover, due to the hermicity of the Lagrangian it is only the symmetric component of the matrices $\x_\mf$ that contributes to a variation in it. Then, from now on we assume $\x_\mf$ to be symmetric, i.e. $\x_\mf^{ij} = \x_\mf^{ji}$, so $( \x_\mf + \x_\mf^T )/ 2 = \x_\mf$.
Discussing the basis reduction in terms of field redefinitions rather than via the direct use of EOMs makes their impact on the $\O_{\tilde X}$ operators more transparent: because the fermion rotations are chiral,  contributions to the latter are generated through the axial anomaly.

The general procedure for reducing the operator basis is as follows. The rotations  in~\eqref{eq.field_redefinitions} are first applied to $\Lag_{SM}$ (Eq.~\eqref{eq.LSM}), and an expansion at $O(1/f_a)$ is performed next.
The net shift resulting from the most general rotation reads~\cite{Brivio:2017ije}
\begin{equation}\label{eq.L_rotated}
\begin{aligned}
\Delta \Lag_{SM}&= - x_\Phi \, \O_{a\Phi}
- \sum_{\mf=Q,u,d,L,e}  \x_\mf \O_{\mf} +\Big[
\left( \x_L Y_e  - Y_e \x_e - x_\Phi Y_e \right)\O_{e\Phi} 
\\
& + \left( \x_Q Y_d - Y_d \x_d -x_\Phi Y_d \right)\O_{d\Phi} + \left( \x_Q Y_u - Y_u \x_u + x_\Phi Y_u \right)\O_{u\Phi}
+\hc \Big]
\\
& + \frac{g'^{2}}{32 \pi^2} \O_{\tilde{B}} \tr \left[ \frac{1}{3} \x_Q - \frac{8}{3} \x_
 u - \frac{2}{3} \x_d + \x_L - 2 \x_e \right] 
 + \frac{g^{2}}{32 \pi^2} \O_{\tilde{W}} \tr \left[ 3 \x_Q + \x_L \right] 
 \\
 & + \frac{g^{2}_s}{32 \pi^2} \O_{\tilde{G}} \tr \left[ 2 \x_Q - \x_u - \x_d \right] \,,
\end{aligned}
\end{equation}
where the anomalous operators $\O_{\tilde X}$ are defined in Table~\ref{tab.basis}, $\O_\mf$ are the chirality-conserving fermionic operators defined in~Eqs~\eqref{ferm}~\eqref{lept}, 
$\O_{\mf\Phi}$ are the chirality-flip ones defined in Eq.~(\ref{ferm-Yuk-1}), and finally 
$\O_{a\Phi}$ is defined in Table~\ref{tab.basis_bosonic}.
The trace in the last two lines of Eq.~(\ref{eq.L_rotated}) is over flavor indices, while in the first two lines the implicit contraction of flavour index of the effective coefficients and operators respects the convention in Eq.~(\ref{convention}), e.g.
$$\left(\x_L Y_e - Y_e\x_e - x_\Phi Y_e\right)\O_{e\Phi} = 
\sum_{i,j}\left(\x_L Y_e - Y_e\x_e - x_\Phi Y_e\right)_{ij}\O^{ij}_{e\Phi}
\,,\quad \text{etc.}\,,
$$
while the expressions inside parenthesis are matrix products, i.e. $(\x_L Y_e)_{ij}= \sum_k (\x_L)_{ik} ( Y_e)_{kj}$.  

At this point, one is free for instance to choose  $x_\Phi$ and $\x_\mf$ so that the terms in $\Delta\Lag_{SM}$ cancel off against redundant operators in $\Lag_a^{\rm total}$. Or to choose values for combination of indices so as to remove one or all of the anomalous coefficients $c_{\tilde X}$. 
It is not hard to verify that each field transformation is equivalent --up to shifts to the anomalous bosonic operators-- to the application of the EOM of the corresponding field, provided in App.~\ref{app.eom}.  In what follows, some specific applications of Eq.~\eqref{eq.L_rotated} are developed.

\subsection{Relation between $\O_{a\Phi}$ and fermionic operators}\label{app.Oaphi}
 Eq.~\eqref{eq.L_rotated} indicates that in order to remove $\O_{a\Phi}$ one needs to fix $x_\Phi = c_{a\Phi}$. This $\Phi$ rotation 
 comes at the price of introducing a set of chirality-flip operators~\cite{Georgi:1986df,Gavela:2014vra,Brivio:2017ije}, i.e.
\begin{equation}
 \O_{a\Phi} = Y_u \O_{u\Phi} - Y_d \O_{d\Phi} - Y_e \O_{e\Phi}+ \hc \,.
\end{equation}
As $\O_{a\Phi}$ is a purely bosonic operator, the flavor structure of the fermionic operators in this equation 
necessarily reflects the SM flavour structure. In other words, it follows the MFV ansatz~\cite{Chivukula:1987py,Hall:1990ac,D'Ambrosio:2002ex}, where a $U(3)^5$ global flavor symmetry is present in the Lagrangian but for the Yukawa couplings, which are treated as spurions.

The combination of chirality-flip operators obtained can  be traded next for chirality-preserving ones (plus in some cases shifts in the $\O_{\tilde X}$ operator coefficients) by fixing the quantities $\x_\mf$ such that the coefficients of $\O_{e\Phi}, \O_{u\Phi}, \O_{d\Phi}$ in Eq.~\eqref{eq.L_rotated} cancel. This is equivalent to applying the transformations of the fermion fields in Eqs.~\eqref{eq.Q_to_uHdH_anomalous}--\eqref{eq.e_to_eH_anomalous} below.
For instance, it is possible to map $\O_{a\Phi}$ onto just 3 operators out of the whole set $\{\O_u, \O_d, \O_e , \O_Q , \O_L\}$, plus $\O_{\tilde X}$ operators. For example, the mapping onto the set $\{\O_u, \O_d, \O_e\}$ is achived choosing $\x_e^{ij}=\x_d^{ij}=-\x_u^{ij}=\d^{ij}\, c_{a\Phi}$, and leads to   Eq.~(\ref{Axion-fermion-from-bosonic-1}), that is a combination of  only right-handed fermionic currents. As it can be easily checked from Eq.~\eqref{eq.L_rotated}, the contributions to gauge anomalous operators $\O_{\tilde X}$ cancel exactly in this case, which does not necessarily generalize to other choices.  
 
 For instance, one could alternatively map onto the set $\{\O_Q,\O_u,\O_L\}$ by choosing $\x_L^{ij}=\x_Q^{ij}=\x_u^{ij}/2=\delta^{ij}\, c_{a\Phi}$, which leads to
\begin{align}
 \O_{a\Phi} &=
 - \tr \left( \O_L +  \O_Q + 2 \O_u \right) + \frac{1}{8\pi^2}\left(g^{2}\O_{\tilde W} - g^{\prime 2} \O_{\tilde B}\right) n_g\,.
\end{align}
This result does not mean that $\O_{a\Phi}$ is anomalous! In fact, we have explicitly checked that when the product $c_{a\Phi} \O_{a\Phi}$ is considered at $\mathcal{O}(\alpha)$,  the contribution from the $\O_{\tilde X}$ terms on the last bracket are compensated exactly by the anomalous contributions stemming from the insertion in Fig. \ref{correctionsgammaZ} diagram \textbf{C} of the operators in the first bracket ($\O_u$, $\O_Q$ and $\O_L $), and only  $\mathcal{O} (m_\mf^2) $ finite terms of remain from the loop contribution. When instead the same computation is performed using the expression for $\O_{a\Phi}$ in Eq.~(\ref{Axion-fermion-from-bosonic-1}), i.e. as combination of the right-handed set $\{\O_u, \O_d, \O_e\}$, the anomalous contributions they induce cancel each other and only the same $\mathcal{O} (m_\mf^2)$ terms are present, as they should.

\subsection{Relations among fermionic operators}
\label{relations-flip-non-flip}
Collecting the terms proportional to $\x_\mf$ in Eq.~\eqref{eq.L_rotated} one can infer relations among the fermionic operators. Writing explicitly the flavor indices $i,j$, it follows that the relations between  chirality preserving and chirality-flip operators (plus anomalous couplings) read
fermion structures can be 
\begin{align}
\label{eq.Q_to_uHdH_anomalous}
 \O_Q^{ij} &= \left[\O_{d\Phi}^{ik}\, (Y_{d})_{jk} +\O_{u\Phi}^{ik}\, (Y_{u})_{jk} + (\O_{d\Phi}^\dag)^{kj}\, (Y_{d}^\dag)_{ki} +(\O_{u\Phi}^\dag)^{kj}\, (Y_{u}^\dag)_{ki}\right] 
 \\
 & \, \, \, \, \, \, +\left[
  \frac{g^{\prime 2}}{96\pi^2}\O_{\tilde B}
 +\frac{3g^{2}}{32\pi^2}\O_{\tilde W}
 +\frac{g_s^{2}}{16\pi^2}\O_{\tilde G}\right]\, \d^{ij}
 \,,
 \\
 \O_u^{ij} &= \left[- \O_{u\Phi}^{kj}\, (Y_u)_{ki} - (\O_{u\Phi}^\dag)^{ik}\, (Y_u^\dag)_{jk}\right]
 -\left[
  \frac{g^{\prime 2}}{12\pi^2}\O_{\tilde B}
 +\frac{g_s^{2}}{32\pi^2}\O_{\tilde G}\right]\, \d^{ij}
 \,,
 \\
 \O_d^{ij} &= \left[- \O_{d\Phi}^{kj}\, (Y_d)_{ki} - (\O_{d\Phi}^\dag)^{ik}\, (Y_d^\dag)_{jk}\right]
 -\left[
  \frac{g^{\prime 2}}{48\pi^2}\O_{\tilde B}
 +\frac{g_s^{2}}{32\pi^2}\O_{\tilde G}\right]\, \d^{ij}\,,
 \\
 \O_L^{ij} &= \left[  \O_{e\Phi}^{ik}\, (Y_e)_{jk} + (\O_{e\Phi}^\dag)^{kj}\, (Y_e^\dag)_{ki}\right]
 +\left[
  \frac{g^{\prime 2}}{32\pi^2}\O_{\tilde B}
 +\frac{g^{2}}{32\pi^2}\O_{\tilde W}\right]\, \d^{ij}
 \,,
 \\
 \O_e^{ij} &= \left[- \O_{e\Phi}^{kj}\, (Y_e)_{ki} - (\O_{e\Phi}^\dag)^{ik}\, (Y_e^\dag)_{jk}\right]
 -
  \frac{g^{\prime 2}}{16\pi^2}\O_{\tilde B}\, \d^{ij}
 \,,
 \label{eq.e_to_eH_anomalous}
\end{align}
where a sum over $k$ is understood. Combining them, the relations between chirality-conserving operators and chirality-flip ones are determined. 

The equations above showed how to express the chirality-conserving couplings as combinations of chirality-flip ones.  What about the inverse relation?  
 It is clear  from the counting of degrees of freedom shown earlier that the latter cannot be achieved in all generality. Indeed, only very particular combinations -- relatively weighed by Yukawa factors-- of a given chirality-flip operator can be extracted from Eqs.~(\ref{eq.Q_to_uHdH_anomalous})-(\ref{eq.e_to_eH_anomalous}), and written in terms of chirality-conserving plus anomalous couplings: this reduces in practice their $n_g^2$ degrees of freedom per fermionic operator (which sums to a total of $3 n_g^2$ fermionic parameters in the ALP Lagrangian) to an active number of $n_g(5n_g+3)/2 - 1$ fermionic parameters in total.

\subsection{Purely fermionic bases: removing anomalous operators}\label{app.removing_anomalous}
Finally, one could ask whether the anomalous operators $\O_{\tilde X}$ could be removed altogether from the basis, trading them for fermionic structures.
In order to do this, one needs to impose
\begin{align}\label{eq.noOB_condition}
 \frac{g^{\prime2}}{32\pi^2}\tr\left[\frac13\x_Q-\frac83 \x_u - \frac23 \x_d + \x_L - 2\x_e\right] &= - c_{\tilde B}\,,
 \\
 \label{eq.noOW_condition}
 \frac{g^{2}}{32\pi^2}\tr\left[3\x_Q+\x_L\right] &= - c_{\tilde W}\,.
 \\
 \label{eq.noOG_condition}
 \frac{g_s^{2}}{32\pi^2}\tr\left[2\x_Q-\x_u-\x_d\right] &= - c_{\tilde G}\,.
\end{align}
It is not difficult to show explicitly that the anomalous bosonic operators cannot be completely replaced by purely chirality-conserving fermionic ones.~\footnote{This is as expected on physical grounds, given the non-invariance of anomalous gauge couplings under the shift symmetry.}  
Indeed, it follows from Eq.~(\ref{eq.L_rotated})  that the conditions to remove all chirality-flip terms are  
\begin{equation}\label{eq.no_chiralityflip_condition}
\begin{aligned}
 \x_Q Y_u - Y_u \x_u &= 0\,,
 \\
 \x_Q Y_d - Y_d \x_d &= 0\,,
 \\
 \x_L Y_e - Y_e \x_e &= 0\,,
\end{aligned}
\end{equation}
and it is not possible to satisfy simultaneously these equations and the conditions in Eqs.~\eqref{eq.noOB_condition}--\eqref{eq.noOG_condition}.
Nevertheless, it is sufficient to relax two of the conditions in~\eqref{eq.no_chiralityflip_condition} in order for the system to be solvable. This implies that any solution of  Eqs.~(\ref{eq.Q_to_uHdH_anomalous})-(\ref{eq.e_to_eH_anomalous}) always involve chirality-flip terms. One example is:
 \begin{align}\label{eq.removing_OB}
\O_{\tilde B} &= - \frac{16\pi^2}{g^{\prime 2}n_g} \left[ \tr  \O_e + \left(Y_e \O_{e\Phi}+\hc\right)\right]\,,
 \\
\O_{\tilde W} &= \frac{32\pi^2}{g^{2}n_g} \left[ \tr \left( \O_L + \frac12\O_e \right) - \left(\frac{Y_e}{2} \O_{e\Phi}+\hc\right)\right]\,,
 \\
\O_{\tilde G} &= \frac{32\pi^2}{g_s^{2}n_g}\left[\tr \left( - \O_d + \frac{\O_e}{3} \right) - \left(Y_d \O_{d\Phi} - \frac{Y_e}{3} \O_{e\Phi}+\hc\right)\right]\,.
\label{eq.removing_OG}
\end{align}

A final comment on the non-equivalence of anomalous couplings and shift-invariant fermionic ones is pertinent in the case of the gauge hypercharge, i.e. the operator $\O_{\tilde{B}}$. 
As it is well known,  the pure gauge anomalous couplings can be written as total derivatives of non-gauge invariant quantities, $X_{\mu\nu} \tilde X^{\mu\nu}  = \de_\mu K_X^\mu$, a term that for pure $U(1)$ gauge Lagrangians does not contribute to the action because the gauge configurations die sufficiently fast at infinity, unlike for non-abelian groups. In this sense, it can appear at first sight surprising that the equations above show that the fermionic equivalent of  $\O_{\tilde B}$  does include chirality-flip (and thus not-shift invariant) terms.  Nevertheless,  in the presence of fermions it is the combination of $\O_{\tilde B}$ and $\O_{\tilde W}$ in Eq.~(\ref{eq.B+L-1}) the one  which is shift-invariant, because it corresponds to the non-conservation of the anomalous $B+L$  global $U(1)$, while the combination of $\O_{\tilde B}$ and $\O_{\tilde W}$ with opposite sign is endowed with a non shift-invariant nature, see Eq.~(\ref{eq.B-L-1}).

\section{Complete --finite and divergent-- corrections to effective couplings }
\label{app-complete-results}
We gather here the exact expressions for the one-loop corrections to the set of ALP-SM couplings $\{g_{a \gamma Z}, g_{aZZ}, g_{a WW}\}$ at $\mathcal{O}(1/f_a)$, for a generic off-shell ALP and on-shell external SM fields. These couplings were introduced and developed only in certain limits in Sect.~\ref{complete-one-loop-main}, while the complete expressions are presented below.

\subsection{ ALP-\tpdf{$Z$}{Z}-photon anomalous coupling}
\label{app-agammaZ-coupling-complete}
The results for the one-loop corrected $g^{\text{eff}}_{a \gamma Z}$ have been introduced in Sec.~\ref{gagammaZ-approx}, where the results were also presented in certain limits. We collect in this appendix the exact expressions for the functions defined in that section (the complete expression for $A^{Z/\gamma\rightarrow \gamma}$   was already given in Eq.~(\ref{gammagammaren})). All descriptions presented there for the origin of each term  apply here as well.
The intermediate $\xi$-dependent steps, together with the final $\xi$-independent expressions, can be found in 
\href{https://notebookarchive.org/2021-07-9otlr9o}{NotebookArchive}. The gauge invariant complete results are as follows:
\begin{equation} 
\begin{aligned} \label{renZgammatoZ_ferm}
A_{\text{ferm}}^{Z/\gamma\rightarrow Z}= &
    -2 \sum_\text{f} N_C \Bigg\{ \left( \frac{T_{3,\mf}^2}{2} - T_{3,\mf} Q_\mf s_w^2 + Q_\mf^2 s_w^4 \right) \times \\
    & \times \left( \log \left(  \frac{\Lambda^2}{m_\text{f}^2} \right) + \frac{2}{3} + \frac{M_Z^2 - 2 m_\text{f}^2}{M_Z^2 - 4 m_\text{f}^2 } \mathcal{DB} (M_Z^2 , m_\mf , m_\mf ) \right) \\
    & + \frac{m_\text{f}^2}{M_Z^2} \left( \frac{T_{3,\mf}^2}{2} + 2 T_{3,\mf} Q_\mf s_w^2 - 2 Q_\mf^2 s_w^4 \right) \Bigg( 1 - \frac{2  m_\text{f}^2}{M_Z^2 - 4 m_\text{f}^2 } \mathcal{DB} (M_Z^2 , m_\mf , m_\mf ) \Bigg) \\
    & - c_w^2 Q_\mf (T_{3,\mf} - 2 Q_\mf s_w^2 ) \Bigg[ \log \left(  \frac{\Lambda^2}{m_\text{f}^2} \right)
     \\
    & + \frac{12 m_\text{f}^2 + 5 M_Z^2}{3 M_Z^2} + \frac{2 m_\text{f}^2 + M_Z^2}{M_Z^2} \mathcal{DB} (M_Z^2 , m_\mf , m_\mf ) \Bigg] \Bigg\} \,, &&
  \end{aligned}
\end{equation}
where $T_{3,\mf}$ denotes the weak isospin of fermion $f$.

The function $A_{\text{Higgs}}^{Z\rightarrow Z}$ for the Higgs corrections to external legs is given by 
\begin{equation} \label{renZtoZ_Higgs}
\begin{aligned}
A_{\text{Higgs}}^{Z\rightarrow Z}= &
     \frac{1}{4} \Bigg\{ \frac{M_Z^4 -3 M_Z^2 M_H^2 + M_H^4}{M_Z^4} + \frac{12 M_Z^6 - 18 M_Z^4 M_H^2 + 9 M_Z^2 M_H^4 - 2 M_H^6}{4 M_Z^6} \log \left( \frac{M_H^2}{M_Z^2} \right) \\
   &  - \frac{36 M_Z^6 - 32 M_Z^4 M_H^2 + 13 M_Z^2 M_H^4 - 2 M_H^6}{2 M_Z^4 (M_H^2 - 4  M_Z^2)} \mathcal{DB} (M_Z^2 , M_Z , M_H ) \Bigg\} \,,\\
\end{aligned}
\end{equation}

while the gauge corrections to external legs  proportional to $g_{a \gamma Z}$ are gathered in
\begin{equation} \label{renZgammatoZ_gauge}
\begin{aligned}
A_{\text{gauge}}^{Z/\gamma\rightarrow Z}= & - \frac{1}{2}
     \left\{ \frac{42 M_W^4 + M_Z^4}{2 M_Z^4} \log \left(  \frac{\Lambda^2}{M_W^2} \right) + \frac{M_W^4}{4 M_Z^4} \log \left( \frac{M_W^2}{M_Z^2} \right)  \right. \\
    & +\frac{180 M_W^6 + 153 M_W^4 M_Z^2 - 12 M_W^2 M_Z^4 - 5 M_Z^6}{3 M_Z^6} \\
    & \left.  + \frac{120 M_W^6 + 108 M_W^4 M_Z^2 + 2 M_W^2 M_Z^4 + M_Z^6}{4 M_Z^6} \mathcal{DB} (M_Z^2 , M_W , M_W ) \right\} \,. &&
\end{aligned}
\end{equation}
The function contributions $A^{WW}$ which encodes the contributions proportional to $g_{aWW}$  reads 
\begin{flalign}
A^{WW} \equiv &  
      \Bigg\{ \frac{42 M_W^2 + M_Z^2}{12 M_W^2} \log \left( \frac{\Lambda^2}{M_W^2} \right) + \frac{36 M_W^4 + 93 M_W^2 M_Z^2 + 2 M_Z^4}{9 M_W^2 M_Z^2} \nonumber \\
    & + \frac{24 M_W^4 + 38 M_W^2 M_Z^2 + M_Z^4}{12 M_W^2 M_Z^2} \mathcal{DB} (M_Z^2 , M_W , M_W )  \nonumber \\
    &  - \frac{ 4 (4 M_W^2 - p^2)}{p^2 - M_Z^2 } \left( f^2 \left( \frac{4 M_W^2}{p^2} \right) - f^2 \left( \frac{4 M_W^2}{M_Z^2} \right)\right) \label{AWW} \\
    & - \frac{1}{3 c_w^2} \sum_{\text{f}} N_C Q_\mf (T_{3,\mf} - 2 Q_\mf s_w^2 ) \Bigg[ \log \left(  \frac{\Lambda^2}{m_\text{f}^2} \right)
    \nonumber \\
    & + \frac{12 m_\text{f}^2 + 5 M_Z^2}{3 M_Z^2} + \frac{2 m_\text{f}^2 + M_Z^2}{M_Z^2} \mathcal{DB} (M_Z^2 , m_\mf , m_\mf ) \Bigg] \Bigg\} \,, \nonumber &&
\end{flalign}
while the complete result for the function $A^{\text{f}}$ which encodes the fermion triangle correction is given by 
\begin{flalign} \label{Af}
A^{\text{f}}= 
 Q_\text{f} N_C \left\{ 2 Q_\text{f} s_w^2 + \frac{4 \left( T_{3,\text{f}} - 2 Q_\text{f} s_w^2 \right) m_\text{f}^2}{p^2 - M_Z^2} \left(f\left( \frac{4 m_\text{f}^2}{p^2} \right)^2 - f\left( \frac{4 m_\text{f}^2}{M_Z^2} \right)^2  \right) \right\} \,, &&
\end{flalign}
where the function $f(\tau)$ has been defined in Eq.~(\ref{ftau}) and the function $\mathcal{DB}(p^2,m_1,m_2)$ corresponds to function \texttt{DiscB} in \texttt{Package-X} and is defined as
\begin{equation} \label{functionDB}
\mathcal{DB} (p^2 , m_1 , m_2 ) \equiv \frac{ \sqrt{\rho(p^2,m_1^2,m_2^2)}}{p^2} \log \left( \frac{m_1^2 + m_2^2 - p^2 + \sqrt{\rho(p^2,m_1^2,m_2^2)}}{2 m_1 m_2} \right) \,,
\end{equation}
which is symmetric under $m_1 \leftrightarrow m_2$ and can be simplified in some specific cases:
\begin{align}
    \bullet & \qquad \mathcal{DB} (M^2 , M , m ) = \frac{m^2}{M^2} \sqrt{1 - \frac{4 M^2}{m^2}} \log \left( \frac{m^2  + \sqrt{m^4 - 4 M^2 m^2}}{2 M m} \right) \,,
    \\
    \bullet & \qquad \mathcal{DB} (p^2 , m , m ) = 2  i \sqrt{1 - \frac{4 m^2}{p^2}} f \left( \frac{4 m^2 }{p^2} \right) \,, \label{functionDBlimit}
\end{align}
and the function $\rho$ is the K\"{a}ll\'en function, that is defined as 
\begin{equation}\label{functionrho}
\rho(a,b,c) \equiv a^4 + b^4 +c^4 -2 a^2 b^2 - 2 b^2 c^2 - 2 c^2 a^2\,.
\end{equation}

\subsection{ALP-\tpdf{$ZZ$}{ZZ} anomalous coupling}
\label{app-aZZ-coupling-complete}
The results for the one-loop corrected $g^{\text{eff}}_{a Z Z}$ have been introduced in Sec.~\ref{sec-ZZ-approx}, where the results were also presented in certain limits. We collect in this appendix the exact expressions for the functions defined in that section. All descriptions presented there for the origin of each term  apply here as well.
The intermediate $\xi$-dependent steps, as well as the complete $\xi$-independent final expressions, can be found in \href{https://notebookarchive.org/2021-07-9otlr9o}{NotebookArchive}. The gauge invariant complete results, presented in the $\{g_{a Z Z}, g_{aWW} \}$  subspace of anomalous electroweak couplings ,are as follows:

The function $A^{Z/\gamma\rightarrow Z}$ which encodes corrections to the external legs  were given in Eq.~\eqref{AZgammatoZ_total} and~\eqref{renZgammatoZ_ferm}--\eqref{renZgammatoZ_gauge}.
  The function $B^{\text{Higgs}}$ accounting for the  the vertex insertion of $g_{aZZ}$ corrected at one-loop by  Higgs  exchange between the two $Z$ bosons reads
 \begin{equation} \label{BHiggs}
\begin{aligned}
B^{\text{Higgs}} = & \, 3 \Bigg\{ - \frac{2 M_Z^2}{4 M_Z^2 - p^2 }  \mathcal{DB} (M_Z^2 , M_Z , M_H )
+ \frac{2 M_Z^2}{4 M_Z^2 - p^2} \mathcal{DB} (p^2 , M_Z , M_Z ) \\
    & \, + M_Z^2 \left( \frac{2 M_H^2}{4 M_Z^2 - p^2} - 1 \right) \mathcal{C}(M_Z^2,M_Z^2,p^2,M_Z,M_H,M_Z) + \frac{M_H^2}{4 M_Z^2 - p^2}  \log \left( \frac{M_H^2}{M_Z^2} \right) \Bigg\}\,.
\end{aligned}
\end{equation}
The contributions proportional to $g_{aWW}$ encoded in $B^{WW}$ are given by 
\begin{equation}\label{BWW}
\begin{aligned}
B^{WW} = & \Bigg( \left\{ \frac{42 M_W^2 + M_Z^2}{12 M_W^2} \log \left(  \frac{\Lambda^2}{M_W^2} \right) + \frac{36 M_W^4 + 75 M_W^2 M_Z^2 + 2 M_Z^4}{9 M_W^2 M_Z^2} \right. \\
    & + \frac{24 M_W^4 + 38 M_W^2 M_Z^2 + M_Z^4}{12 M_W^2 M_Z^2} \mathcal{DB} (M_Z^2 , M_W , M_W ) \\
    & - \frac{ M_Z^4}{ M_W^2 (p^2 - 4 M_Z^2 )} \left( \mathcal{DB} (p^2 , M_W , M_W ) - \mathcal{DB} (M_Z^2 , M_W , M_W )  \right) \\
    & \left. +\left( (4 M_W^2 - p^2) + \frac{M_Z^4 (p^2 - 2 M_Z^2)}{2 M_W^2 ( p^2 - 4 M_Z^2)} \right) \mathcal{C}(M_Z^2,M_Z^2,p^2,M_W,M_W,M_W) \right\} \\
    & -  \frac{1}{3c_w^2} \sum_{\text{f}} N_C Q_\mf (T_{3,\mf} - 2 Q_\mf s_w^2 ) \Bigg\{ \log \left(  \frac{\Lambda^2}{m_\text{f}^2} \right) \\ 
    & + \frac{12 m_\text{f}^2 + 5 M_Z^2}{3 M_Z^2} + \frac{2 m_\text{f}^2 + M_Z^2}{M_Z^2} \mathcal{DB} (M_Z^2 , m_\mf , m_\mf ) \Bigg\} \Bigg)\,. &&
\end{aligned}
\end{equation}
Finally,  the function $B^{\text{f}}$ which encodes vertex insertions of fermionic couplings $c_{f}$ reads 
\begin{equation}\label{Bf}
\begin{aligned}
B^{\text{f}}=&
    - N_C \Bigg\{ Q_\mf^2 s_w^4 +  T^2_{3,f} \frac{2 m_\text{f}^2}{(4 M_Z^2 - p^2)} \left( \mathcal{DB} (p^2 , m_\mf , m_\mf ) - \mathcal{DB} (M_Z^2 , m_\mf , m_\mf )  \right) + \\
    & + \frac{2 m_\text{f}^2}{(4 M_Z^2 - p^2)} \left[M_Z^2 (T_{3,\mf} - 2 Q_\mf s_w^2)^2 + p^2 Q_\mf s_w^2 (T_{3,\mf} -  Q_\mf s_w^2) \right] \mathcal{C}\left(M_Z^2,M_Z^2,p^2,m_\text{f},m_\text{f},m_\text{f} \right) \Bigg\} \,,
\end{aligned}\end{equation}
where the function $\mathcal{C}(q_1^2 , q_2^2 , p^2 , m_1 , m_2 , m_3)$  is the $\text{C}_0$ Passarino-Veltman function~\cite{Passarino:1978jh} and is defined by:
\begin{equation}
\begin{aligned}\label{functionC}
& \mathcal{C} (q_1^2 , q_2^2 , p^2 , m_1 , m_2 , m_3) \equiv \\
& \int_0^1 \text{d}x \int_0^x \text{d}y \frac{1}{(x-y)y q_1^2 - (x-y) (x-1) q_2^2 -y (x-1) p^2 -y m_1^2 - (x-y) m_2^2 + (x-1) m_3^2} \,,
 \end{aligned} 
\end{equation}
which can be reduced to a combination of $f(\tau)$ and $\mathcal{DB}$ functions (see Eq.~\eqref{ftau} and Eqs.~\eqref{functionDB}-\eqref{functionDBlimit}) in the following cases:
\begin{align}
       \bullet & \qquad \mathcal{C}(0 , 0 , p^2 , m , m , m) = - \frac{2}{p^2} f \left( \frac{4 m^2}{p^2} \right)^2 \,, 
       \\
       \bullet & \qquad  \mathcal{C}(0 , M^2 , p^2 , m , m , m) = - \frac{2}{p^2 - M^2} \left[ f \left( \frac{4 m^2}{p^2} \right)^2 - f \left( \frac{4 m^2}{M^2} \right)^2 \right] \,, 
       \\
        \bullet & \qquad \mathcal{C}(M^2 , M^2 , 0 , m , m , m) = \frac{1}{ 4 m^2 - M^2} \mathcal{DB} (M^2 , m , m )  \,.  \label{functionCmasslessALP}
\end{align}

\subsection{ALP-\tpdf{$WW$}{WW} anomalous coupling}
\label{sec-aWW-complete} 
The results for the one-loop corrected $g^{\text{eff}}_{a WW}$ have been introduced in Sec.~\ref{sec-aWW-approx}, where the results were also presented in the high ALP $p^2$  limit. We collect in this appendix the exact expressions for the functions defined in that section. All descriptions presented there for the origin of each term  apply here as well.
The intermediate $\xi$-dependent steps, as well as the final $\xi$-independent results, can be found in \href{https://notebookarchive.org/2021-07-9otlr9o}{NotebookArchive}. The gauge invariant complete results, projected on the $\{g_{a \g\g}, g_{aWW} \}$  subspace of anomalous electroweak couplings are detailed next.

The function $A_{WW}$ results from the combination of fermionic and Higgs corrections, see Eq.~(\ref{AWW}). Only fermion doublets can contribute  to $A^{W\rightarrow W}_{\text{ferm}} $ (Fig.~\ref{correctionexternallegs} {\bf D5}):
 \begin{equation}\label{renAWtoW_ferm}
 \begin{aligned}
A^{W\rightarrow W}_{\text{ferm}}= 
  & \,  2 \sum_{\substack{\mf= u, c, t, \\ \nu_e, \nu_\mu, \nu_\tau}} N_C \Bigg\{ - \log \left(  \frac{\Lambda^2}{m_\text{f}^2} \right) - \frac{3 M_W^3 (m_\text{f}^2 + m^{2}_\text{f'}) + 6 (m_\text{f}^2 - m^{2}_\text{f'})^2 + 4 M_W^4}{6 M_W^2}  \\
    & + \frac{(m_\text{f}^2 - m^{2}_\text{f'})^3 - M_W^6}{2 M_W^6} \log \left( \frac{m_\text{f}^2}{m_\text{f'}^{2}} \right) + \mathcal{DB} (Mw^2 , m_\mf , m_\mfp ) \times \\
    &  \times \frac{M_W^6 (m_\text{f}^2 + m^{2}_\text{f'}) + 2 M_W^4 m_\text{f}^2 m_\text{f'}^{2} + M_W^2 (m_\text{f}^4 - m^{4}_\text{f'}) +(m_\text{f}^2 - m^{2}_\text{f'})^4 + M_W^8}{M_W^4 \, \rho(M_W^2,m_\text{f}^2,m_\text{f'}^{2})}   \Bigg\} \,, &&
\end{aligned}
\end{equation}
where $\Lambda$ is an UV cutoff (see Eq.~(\ref{prescription}), $m_\text{f}$ and $m_\text{f'}$ denote the masses of the two fermion mass eigenstates. 

The Higgs corrections to external legs gathered in $A^{W\rightarrow W}_{\text{Higgs}}$ 
(Fig.~\ref{correctionexternallegs} {\bf D3} and {\bf D4}) read
\begin{equation}\label{renAWtoW_Higgs}
 \begin{aligned}
A^{W\rightarrow W}_{\text{Higgs}} =  &    \frac{M_W^4 -3 M_W^2 M_H^2 + M_H^4}{M_W^4} +  \frac{12 M_W^6 - 18 M_W^4 M_H^2 + 9 M_W^2 M_H^4 - 2 M_H^6}{4 M_W^6} \log \left( \frac{M_H^2}{M_W^2} \right)  \\
    & - \frac{36 M_W^6 - 32 M_W^4 M_H^2 + 13 M_W^2 M_H^4 - 2 M_H^6}{2 M_W^4 (M_H^2 - 4 M_W^2)} \mathcal{DB} (M_W^2 , M_W , M_H ) \,. 
\end{aligned}
\end{equation}

The  gauge corrections  proportional to $g_{aWW}$ encoded by $C^{WW}$  are given by 
\begin{equation} \label{Cgauge}
 \begin{aligned}
C^{WW} =  &  \Bigg\{ 43 \log \left(  \frac{\Lambda^2}{M_W^2} \right) + \frac{236 M_W^4 + 33 M_W^2 M_Z^2 + 3 M_Z^4}{3 M_W^4} \\
    & + \left( \frac{36  M_W^6 - 34 M_W^4 M_Z^2 - M_W^2 M_Z^4 + 8 M_Z^6}{2 M_W^4 M_Z^2} - \frac{(24 M_W^6 - 30 M_W^4 M_Z^2 + 24 M_W^2 M_Z^4 - 6 M_Z^6) p^2}{2 M_W^4 M_Z^2 (4 M_W^2 - p^2)} \right) \times \\
    & \times \mathcal{DB} (M_W^2 , M_W , M_Z ) + \Bigg( \frac{48  M_W^8 +108 M_W^6 M_Z^2 - 60 M_W^4 M_Z^4}{4 M_W^6 M_Z^2}  \\
    &  \left. + \frac{- 17 M_W^2 M_Z^6 + 8 M_Z^8}{4 M_W^6 M_Z^2} - \frac{(24 M_W^6 - 54 M_W^4 M_Z^2 + 36 M_W^2 M_Z^4 - 6 M_Z^6) p^2 M_Z^2}{4 M_W^6 M_Z^2(4 M_W^2 - p^2)} \right) \log \left( \frac{M_W^2}{M_Z^2} \right) \\
    & + 12 s_w^2 (2 M_W^2 - p^2)  \mathcal{C}(M_W^2,M_W^2,p^2,M_W,\lambda,M_W) - 12 s_w^2 \log \left( \frac{\lambda^2}{M_W^2} \right) \\
    & + 6 c_w^2 \frac{16 M_W^4 +20 M_W^2 M_Z^2-6 M_W^4 -3 p^2 (4 M_W^2+M_Z^2) + 2 p^4}{ (4 M_W^2 - p^2)}  \mathcal{C}(M_W^2,M_W^2,p^2,M_W,M_Z,M_W) \Bigg\} \\
    &   
 + 6 (c_w^2 - s_w^2)  \, \Bigg\{  \frac{ 2 M_Z^2}{c_w^2 (4 M_W^2 - p^2)} \mathcal{DB} (p^2 , M_Z , M_Z ) \\
    & \left. +  \left( 2 (4 M_Z^2 - p^2) - \frac{M_Z^2 (2 M_Z^2 - p^2)}{c_w^2 ( 4 M_W^2 -p^2)} \right)  \mathcal{C}(M_W^2,M_W^2,p^2,M_Z,M_W,M_Z) \right\} \\
    & 
    + 24 s_w^2\, \Bigg\{  \frac{4 M_Z^2}{ M_W^2 p^2 (4 M_W^2 - p^2 ) } \mathcal{DB} (M_W^2 , M_W , M_Z ) \\
    & + \frac{2 M_Z^4}{p^2 (4 M_W^2 - p^2 ) }  \log \left( \frac{M_W^2}{M_Z^2} \right) - \frac{(p^2 - M_Z^2)^2}{p^2}  \mathcal{C}(M_W^2,M_W^2,p^2,0,M_W,M_Z)  \Bigg\} \,, \\
\end{aligned}
\end{equation}
where $\lambda$ is again an IR cutoff, which encodes  the IR contribution to the $1/\epsilon$ terms obtained in dimensional regularization via the prescription in Eq.~(\ref{prescription}), with a protocol alike to that for gluon corrections in Eq..~(\ref{Ggg-dimensionalreg}).

The vertex function $C^{\text{Higgs}}$ results from the direct  vertex insertion of $g_{a WW}$, with the Higgs particle  exchanged between the two $W$ legs (diagram {\bf E} in Fig.~\ref{correctionsgammaZ}): 
 \begin{equation} \label{CHiggs}
\begin{aligned}
C^{\text{Higgs}} = & \, 6 \Bigg\{ - \frac{2 M_W^2}{4 M_W^2 - p^2 } \mathcal{DB} (M_W^2 , M_W , M_H )  
+\frac{2 M_W^2}{4 M_W^2 - p^2} \mathcal{DB} (p^2 , M_W , M_W ) \\
    & \, + M_W^2 \left( \frac{2 M_H^2}{4 M_W^2 - p^2} - 1 \right) \mathcal{C}(M_W^2,M_W^2,p^2,M_W,M_H,M_W) + \frac{M_H^2}{4 M_W^2 - p^2}  \log \left( \frac{M_H^2}{M_W^2} \right) \Bigg\}\,.
\end{aligned}
\end{equation}
The vertex function $C^{\gamma\gamma}$ is given by 
\begin{equation} \label{Cgammagamma}
 \begin{aligned}
C^{\gamma\gamma} =
 &   - p^2 \mathcal{C}(M_W^2,M_W^2,p^2,0,M_W,0) + \frac{ M_Z^2}{c_w^2 (4 M_W^2 - p^2)} \mathcal{DB} (p^2 , M_Z , M_Z ) \\
    & +  \left( (4 M_Z^2 - p^2) - \frac{M_Z^2 (2 M_Z^2 - p^2)}{2 c_w^2 ( 4 M_W^2 -p^2)} \right)  \mathcal{C}(M_W^2,M_W^2,p^2,M_Z,M_W,M_Z) \\
    & -  \left( \frac{2 M_Z^2}{p^2} + \frac{M_Z^2}{c_w^2 (4 M_W^2 - p^2)} \right) \mathcal{DB} (M_W^2 , M_W , M_Z )  \\
    & - \left( \frac{M_Z^2}{c_w^2  p^2} - \frac{2 M_W^2 - M_Z^2}{2 c_w^4 (4 M_W^2 - p^2)} \right) \log \left( \frac{M_W^2}{M_Z^2} \right)  + \frac{2(p^2 - M_Z^2)^2}{p^2}  \mathcal{C}(M_W^2,M_W^2,p^2,0,M_W,M_Z)  \,.
 \end{aligned}
\end{equation}
Finally,  the fermionic triangle contributions induced by $c_\text{f}$ insertions (Fig.~\ref{correctionsgammaZ} {\bf C}) lead to 
\begin{flalign}
C^{\text{f}} = 
 &   - N_C \Bigg\{ \frac{m_\text{f}^2}{4 (4M_W^2-p^2)}\left(  \frac{m_\text{f}^2-m_\text{f'}^2}{M_W^2} - 1 \right) \log \left( \frac{m_\text{f}^2}{m_\text{f'}^2} \right) \nonumber \\
    & + \frac{ m_\text{f}^2}{2(4M_W^2 - p^2)} \left( \mathcal{DB} (p^2 , m_\mf , m_\mf ) -  \mathcal{DB} (M_W^2 , m_\mf , m_\mfp ) \right) \label{Cferm}  \\
    & + \frac{m_\text{f}^2 (M_W^2 - m_\text{f}^2 + m_\text{f'}^2)}{2(4M_W^2 - p^2)}  \mathcal{C}\left(M_W^2,M_W^2,p^2,m_\text{f},m_\text{f'},m_\text{f} \right) \Bigg\}\,.  \nonumber &&
 \end{flalign}

\subsection{ALP-fermion couplings}
\label{app-fermion-couplings-complete}
The results for the one-loop corrected {\boldmath{$c^{\text{eff}}_{\text{f}}$}} have been introduced and presented in Sec.~\ref{sec-fermion-couplings} in certain limits of interest.  We collect in this appendix the exact expressions for the functions defined in that section. All descriptions presented there for the origin of each term  apply here as well.
The intermediate $\xi$-dependent steps can be found in \href{https://notebookarchive.org/2021-07-9otlr9o}{NotebookArchive}. The gauge invariant complete results are as follows:
\begin{flalign}\label{Dgg}
 D^{g g} =  \Bigg\{ 3 \log \left( \frac{\Lambda^2}{m_\mf^2} \right) - 4 - p^2 \mathcal{C} \left( m_\mf^2 , m_\mf^2 , p^2 , 0 , m_\mf , 0 \right) \Bigg\} \,.
&&\end{flalign}
\begin{flalign}
D^{\gamma\gamma} =  \frac{ Q_\text{f}^2}{2 } D^{g g}  \,,
&&\end{flalign}
where the function $\mathcal{C}$ was defined in Eqs.~(\ref{functionC})-(\ref{functionCmasslessALP}), and 
\begin{flalign} \label{DgammaZ}
\begin{aligned}
 &D^{\gamma Z}
  = \frac{ Q_\text{f} \,(T_{3,\text{f}} - 2 Q_\text{f} s_w^2)}{16  c_w s_w}  \Bigg\{ 12 \log \left( \frac{\Lambda^2}{M_Z^2} \right) + 2 \frac{M_Z^2 - 8 m_\text{f}^2 }{m_\text{f}^2}  \\
 & + 2 \frac{2 m_\text{f}^2 (M_Z^2 + p^2) + M_Z^2 p^2}{m_\text{f}^2 p^2}  \mathcal{DB} (m_\mf^2 , m_\mf , M_Z ) \\
 & - \frac{12 m_\text{f}^4 p^2 - 2 m_\text{f}^2 M_Z^4 - M_Z^4 p^2}{m_\text{f}^4 p^2} \log \left( \frac{m_\text{f}^2}{M_Z^2} \right) - 4 \frac{(M_Z^2 - p^2)^2}{p^2} \mathcal{C}(m_\text{f}^2,m_\text{f}^2,p^2,0,m_\text{f},M_Z) \Bigg\} \,.
\end{aligned} &&\end{flalign}
\begin{flalign} \label{DZZ}
\begin{aligned}
 & D^{Z Z}= \frac{1}{8 c_w^2 s_w^2}  \Bigg\{ 2 \left( T_{3,\text{f}}^2 - 2 T_{3,\text{f}} \,Q_\text{f} s_w^2 + 2 Q_\text{f}^2 s_w^4 \right) \left( 3 \log \left( \frac{\Lambda^2}{M_Z^2} \right) + \frac{M_Z^2 - 4 m_\text{f}^2}{m_\text{f}^2} \right) \\
 & - \left( 2 (T_{3,\text{f}}^2 - 6 T_{3,\text{f}} \,Q_\text{f} s_w^2 + 6 Q_\text{f}^2 s_w^4) + \frac{2 M_Z^2 T_{3,\mf}^2}{m_\text{f}^2} -  \frac{M_Z^4}{m_\text{f}^4} (T_{3,\text{f}}^2 - 2 T_{3,\text{f}} \,Q_\text{f} s_w^2 + 2 Q_\text{f}^2 s_w^4) \right) \times \\
 & \times \log \left( \frac{m_\text{f}^2}{M_Z^2} \right) + 4 T_{3,\text{f}}^2 \mathcal{DB} (p^2 , M_Z , M_Z ) \\
 & + \frac{2 M_Z^2 (T_{3,\text{f}}^2 - 2 T_{3,\text{f}} \, Q_\text{f} s_w^2 + 2 Q_\text{f}^2 s_w^4 ) - 8 m_\text{f}^2 Q_\text{f} s_w^2 ( T_{3,\text{f}} - Q_\text{f} s_w^2) }{m_\text{f}^2 } \mathcal{DB} (m_\mf^2 , m_\mf , M_Z ) \\
 &  + 4 \left[ M_Z^2 ( T_{3,\text{f}} - 2 Q_\text{f} s_w^2 )^2 + p^2 Q_\mf s_w^2 ( T_{3,\text{f}} - Q_\text{f} s_w^2) \right] \mathcal{C}(m_\text{f}^2,m_\text{f}^2,p^2,M_Z,m_\text{f},M_Z) \Bigg\} \,,
\end{aligned} &&\end{flalign}
\begin{flalign} \label{DWW}
\begin{aligned}
 & D^{WW}  = \frac{1}{16 s_w^2}  \Bigg\{ 6 \log \left( \frac{\Lambda^2}{M_W^2} \right) - \frac{2 (3 m_\text{f}^2 + m_\text{f'}^2 - M_W^2 )}{m_\text{f}^2} \\
 & + 4 \mathcal{DB} (p^2 , M_W , M_W ) - \frac{m_\text{f}^4 + 2 m_\text{f}^2 m_\text{f'}^2 - ( m_\text{f'}^2 - M_W^2)^2}{m_\text{f}^4} \log \left( \frac{m_\text{f'}^2}{M_W^2} \right)  \\
 & + \frac{2 (m_\text{f}^2 - m_\text{f'}^2 + M_W^2)}{m_\text{f}^2} \mathcal{DB} (m_\mf^2 , m_\mfp , M_W ) - 4 \left( m_\text{f}^2 - m_\text{f'}^2 - M_W^2 \right) \mathcal{C}(m_\text{f}^2,m_\text{f}^2,p^2,M_W,m_\text{f'},M_W) \Bigg\} \,,
\end{aligned} &&\end{flalign}
where the function $\mathcal{DB}$ was defined in Eqs.~(\ref{functionDB})-\eqref{functionDBlimit}.

The contributions to $\bf{c_\text{f}^{\text{eff}}}$  from insertions of ALP fermionic couplings 
are given by 
\begin{flalign}
 D^{c_\text{f}}_g = - 2 \Bigg\{ 1 + \log \left( \frac{\lambda^2}{m_\mf^2} \right) + ( p^2 - 2 m_\mf^2 ) \mathcal{C} \left( m_\mf^2 , m_\mf^2 , p^2 , m_\mf , \lambda , m_\mf \right) \Bigg\} \,,
&&\end{flalign}
\begin{flalign}\label{Dcfgamma}
D^{c_\text{f}}_\gamma=    \frac{Q_\text{f}^2}{2}  D^{c_\text{f}}_g  \,,
&&\end{flalign}
\begin{flalign} \label{DcfZ}
\begin{aligned}
 & 
 D^{c_\text{f}}_Z = \frac{1}{4  c_w^2 s_w^2}  \Bigg\{ - \frac{2 m_\text{f}^2 T_{3,\text{f}}^2}{M_Z^2} \log \left( \frac{\Lambda^2}{M_Z^2} \right) + 4 (T_{3,\text{f}}^2 + T_{3,\text{f}} \, Q_\text{f} s_w^2 - Q_\text{f}^2 s_w^4) - \frac{4 m_\text{f}^2 T_{3,\text{f}}^2}{M_Z^2} \\
 & - \frac{4 M_Z^2}{m_\text{f}^2} (T_{3,\text{f}}^2 - 2 T_{3,\text{f}} \, Q_\text{f} s_w^2 + 2 Q_\text{f}^2 s_w^4) + \frac{1}{m_\text{f}^4 M_Z^2} \left[ T_{3,\text{f}}^2 (2 m_\text{f}^6 - m_\text{f}^4 M_Z^2 + 5 m_\text{f}^2 M_Z^4 - 2 M_Z^6) \right.\\
 & \left. -4 Q_\text{f} s_w^2 (T_{3,\text{f}}  - Q_\text{f} s_w^2) M_Z^2 (m_\text{f}^4 + m_\text{f}^2 M_Z^2 - M_Z^4) \right] \log \left( \frac{m_\text{f}^2}{M_Z^2} \right) \\
 & + \frac{2}{m_\text{f}^2 ( M_Z^2 - 4 m_\text{f}^2 )} \left[ T_{3,\text{f}}^2 ( -7 m_\text{f}^4 + 9 m_\text{f}^2 M_Z^2 - 2 M_Z^4 )  \right. \\
 & \left. -4 Q_\text{f} s_w^2 (T_{3,\text{f}}  - Q_\text{f} s_w^2) (m_\text{f}^4 + 3 m_\text{f}^2 M_Z^2 - M_Z^4)  \right] \mathcal{DB} (m_\mf^2 , m_\mf , M_Z ) \\
 & - \frac{2 m_\text{f}^2 T_{3,\text{f}}^2}{M_Z^2} \mathcal{DB} (p^2 , m_\mf , m_\mf ) \\
 & + 2 \left[ m_\text{f}^2 ( T_{3,\text{f}} - 2 Q_\text{f} s_w^2 )^2 + 2 p^2 Q_\mf s_w^2 ( T_{3,\text{f}} - Q_\text{f} s_w^2) \right] \mathcal{C}(m_\text{f}^2,m_\text{f}^2,p^2,m_\text{f},M_Z,m_\text{f}) \Bigg\} \,.
\end{aligned} &&\end{flalign}
\begin{flalign} \label{DcfW}
\begin{aligned}
 & 
 D^{c_\text{f}}_W
 = - \frac{1}{16 s_w^2}  \Bigg\{ \frac{2 m_\text{f}^2}{M_W^2} \log \left( \frac{\Lambda^2}{M_W^2} \right) + \frac{2 (m_\text{f}^4 + 2 m_\text{f}^2 m_\text{f'}^2 - 2 m_\text{f'}^4 - m_\text{f}^2 M_W^2 - 2 m_\text{f'}^2 M_W^2 + 4 M_W^4 )}{m_\text{f}^2 M_W^2} \\
 & - \frac{m_\text{f}^6 + 3 m_\text{f}^2 (m_\text{f}'^2 + M_W^2) - 2 (m_\text{f'}^6 - 3 m_\text{f'}^2 M_W^4 + 2 M_W^6)}{m_\text{f}^4 M_W^2} \log \left( \frac{m_\text{f'}^2}{M_W^2} \right) \\
 & + \frac{2}{m_\text{f}^2 M_W^2 \, \rho(m_\text{f}^2, m_\text{f'}^2, M_W^2)} \left[ m_\text{f}^8 - m_\text{f}^6 (m_\text{f'}^2 + M_W^2) - m_\text{f}^4 (3 m_\text{f'}^4 + 2 m_\text{f'}^2 M_W^2 - 3 M_W^4) \right. \\
 &  \left. + m_\text{f}^2 (5 m_\text{f'}^6 + m_\text{f'}^4 M_W^2 + m_\text{f'}^2 M_W^4 - 7 M_W^6) - 2 ( m_\text{f'}^2 - M_W^2 )^3 ( m_\text{f'}^2 + 2 M_W^2) \right] \mathcal{DB} (m_\mf^2 , m_\mfp , M_W ) \Bigg\} \,,
 \end{aligned} &&\end{flalign}
\begin{flalign} \label{Dcfprime}
 \begin{aligned}
 & 
 D^{c_{\text{f'}}}
 = 
  -  \frac{ m_\text{f'}^2}{16 s_w^2}  \Bigg\{ \frac{2 }{M_W^2} \log \left( \frac{\Lambda^2}{M_W^2} \right) - \frac{3 m_\text{f}^4 - 2 m_\text{f}^2 m_\text{f'}^2 + (m_\text{f'}^2 - M_W^2)^2}{m_\text{f}^4 M_W^2} \log \left( \frac{m_\text{f'}^2}{M_W^2} \right)  \\
 & + \frac{2 (2 m_\text{f}^2 + m_\text{f'}^2 - M_W^2)}{m_\text{f}^2 M_W^2}   - \frac{2 (m_\text{f}^2 - m_\text{f'}^2 + M_W^2) }{m_\text{f}^2 M_W^2} \mathcal{DB} (m_\mf^2 , m_\mfp , M_W )  \\
 & + \frac{2}{ M_W^2} \mathcal{DB} (p^2 , m_\mf , m_\mf ) + 4 \mathcal{C}(m_\text{f}^2,m_\text{f}^2,p^2,m_\text{f'},M_W,m_\text{f'}) \Bigg\} \,.
\end{aligned} &&\end{flalign}
\begin{flalign} \label{DcfHiggs}
\begin{aligned}
 & 
 D^{c_{\text{f}}}_h  = \frac{1}{16 \pi s_w^2} \Bigg\{ - \frac{2 m_\text{f}^2}{M_W^2} \log \left( \frac{\Lambda^2}{M_H^2} \right) + \frac{2 (2 m_\text{f}^2 - M_H^2)}{M_W^2} \\
 & - \frac{2 m_\text{f}^4 - 3 m_\text{f}^2 M_H^2 + M_H^4}{m_\text{f}^2 M_W^2} \log \left( \frac{m_\text{f}^2}{M_H^2} \right) + \frac{2 (m_\text{f}^2 - M_H^2)}{M_W^2} \mathcal{DB} (m_\mf^2 , m_\mf , M_H ) \\
 & + \frac{2 m_\text{f}^2}{M_W^2} \mathcal{DB} (p^2 , m_\mf , m_\mf ) + \frac{2 m_\text{f}^2 (M_H^2 - 4 m_\text{f}^2)}{M_W^2} \mathcal{C}(m_\text{f}^2,m_\text{f}^2,p^2,m_\text{f},M_H,m_\text{f}) \Bigg\} \,.
\end{aligned} &&\end{flalign}
\begin{flalign} 
\begin{aligned}
D^{c_\psi}_{\text{mix}} = -  \frac{T_{3,\text{f}} }{ s_w^2 M_W^2} 
N_C T_{3,\psi} m_\psi^2 \Bigg\{ \log \left( \frac{\Lambda^2}{m_\psi^2} \right) + 2 + \mathcal{DB} (p^2 , m_\psi , m_\psi ) \Bigg\} \,.
\end{aligned} \label{exact-mix}
&&\end{flalign}

\section{One-loop corrections to the weak angle}
\label{sec-cos-renorm} 
In Eq.~(\ref{cwbar}) we defined a quantity ${\bar{c}}_w$ as the ratio of two input observables: the $W$ and $Z$ masses, whose renormalized formulation was expressed in terms of $\Delta c_w$, see Eqs.~(\ref{cwbar}), and (\ref{Deltacwbar}). The exact $\Delta c_w$ expression can be split in three parts,
\begin{equation}
 \frac{\Delta c_w}{c_w} = \frac{\Delta c_w^{\text{gauge}}}{c_w} + \frac{\Delta c_w^{\text{Higgs}}}{c_w} + \frac{\Delta c_w^{\text{ferm}}}{c_w} \,,
\end{equation}
 which correspond respectively to  the gauge boson corrections to the self-energies,  the Higgs corrections and the fermions corrections:
\begin{equation}
\begin{aligned}
 & \frac{\Delta c_w^{\text{gauge}}}{c_w} = \frac{\a_{em}}{\pi} \Bigg\{ \frac{42 M_W^2 + M_Z^2}{48 M_W^2} \log \left( \frac{\Lambda^2}{M_W^2} \right) + \frac{288 M_W^6 + 696 M_W^4 M_Z^2 - 74 M_W^2 M_Z^4 -3 M_Z^6}{288 M_W^4 M_Z^2}
 \\
 &
 + \frac{80 M_W^4 - 14 M_W^2 M_Z^2 - M_Z^4}{192 s_w^2 c_w^2 M_W^4} \log \left( \frac{M_W^2}{M_Z^2} \right) + \frac{48 M_W^6 + 68 M_W^4 M_Z^2 -16 M_W^2 M_Z^4 - M_Z^6}{96 s_w^2 c_w^2 M_W^2 M_Z^6} \times
 \\
 &
 \times  \left[ M_Z^2 \mathcal{DB} (M_W^2 , M_W , M_Z ) -  M_W^2 \mathcal{DB} (M_Z^2 , M_W , M_W ) \right] \Bigg\} \,,
 \end{aligned}
\end{equation}
\begin{equation}
\begin{aligned}
 & \frac{\Delta c_w^{\text{Higgs}}}{c_w} = \frac{\a_{em}}{\pi} \Bigg\{ \frac{M_H^4 - 24 M_W^2 M_Z^2 }{96 M_W^4} + \frac{M_H^4 [ M_H^2 (M_W^2 + M_Z^2) - 6 M_W^2 M_Z^2]}{192 M_W^6 M_Z^2} \log \left( \frac{M_W^2}{M_H^2} \right) 
 \\
 & 
 - \frac{M_H^6 - 6 M_H^4 M_Z^2 + 18 M_H^2 M_Z^4 - 24 M_Z^4}{192 s_w^2 M_W^2 M_Z^4} \log \left( \frac{M_W^2}{M_Z^2} \right)
 \\
 &
 + \frac{M_H^4 - 4 M_H^2 M_Z^2 + 12 M_Z^4}{96 s_w^2 M_W^2 M_Z^2} \mathcal{DB} (M_Z^2 , M_Z , M_H )
 \\
 &
 - \frac{ M_H^4 - 4 M_H^2 M_W^2 + 12 M_W^4 }{96 s_w^2 M_W^4} \mathcal{DB} (M_W^2 , M_W , M_H ) \Bigg\} \,,
 \end{aligned}
\end{equation}
\begin{equation}
\begin{aligned}
 & \frac{\Delta c_w^{\text{ferm}}}{c_w} = \frac{\a_{em}}{\pi} \sum_{\substack{\mf = u,  c, t, \\ \nu_e,  \nu_\mu,  \nu_\tau}} \Bigg\{ \frac{4 (Q_\mf^2 + Q_\mfp^2) s_w^2 - 1}{24 c_w^2} \log \left( \frac{\Lambda^2}{m_\mf^2} \right)
 \\
 &
 + \frac{(m_\mf^2 - m_\mfp^2)^2}{48 s_w^2 M_W^4} +\frac{24 m_\mf^2 Q_\mf (2 Q_\mf s_w^2 - 1) + 24 m_\mfp^2 Q_\mfp (2 Q_\mfp s_w^2 + 1) + 5 M_Z^2 ( 4 s_w^2 ( Q_\mf^2 + Q_\mfp^2) - 1)}{72 M_W^2}
 \\
 &
 - \frac{(m_\mf^2 - m_\mfp^2 - M_W^2)^2 (m_\mf^2 - m_\mfp^2 + 2 M_W^2) - 2 M_W^4 M_Z^2 (8 Q_\mfp^2 s_w^4 + 4 Q_\mfp s_w^2 + 1)}{96 s_w^2 M_W^6} \log \left( \frac{m_\mf^2}{m_\mfp^2} \right)
 \\
 &
 + \frac{ (m_\mf^2 - m_\mfp^2 )^2 + M_W^2 (m_\mf^2 + m_\mfp^2) - 2 M_W^4 }{48 s_w^2 M_W^4} \mathcal{DB} (M_W^2 , m_\mf , m_\mfp )
 \\
 &
 + \frac{(2 m_\mf^2 + M_Z^2) ( 8 Q_\mf^2 s_w^4 - 4 Q_\mf s_w^2 +1) - 3 m_\mf^2}{48 s_w^2 M_W^2} \mathcal{DB} (M_Z^2 , m_\mf , m_\mf )
 \\
 &
 + \frac{(2 m_\mfp^2 + M_Z^2) ( 8 Q_\mf^2 s_w^4 + 4 Q_\mf s_w^2 +1) - 3 m_\mfp^2}{48 s_w^2 M_W^2} \mathcal{DB} (M_Z^2 , m_\mfp , m_\mfp )
 \Bigg\} \,,
 \end{aligned}
\end{equation}
where the funcions ${f(\tau)}$ and $ \mathcal{DB} (p^2 , m_1 , m_2 )$ were defined in Eq.~(\ref{ftau}) and Eqs.~\eqref{functionDB}-\eqref{functionDBlimit}. 

These $\Delta c_w$ corrections allow to express the tree-level phenomenological couplings $\{g_{a \g \g}, g_{a \g Z},  g_{a Z Z} \}$ as a combination of the 
two fundamental Lagrangian parameters $\{ c_{\tilde B}, c_{\tilde W} \}$ and  observable quantities, see Eqs.~(\ref{gagamma_cWcB})-(\ref{gaZZ_cWcB}).

\bibliographystyle{JHEP}
\bibliography{bibliography}
\end{document}